\documentclass[fleqn,usenatbib]{mnras}

\usepackage{newtxtext,newtxmath,ulem}
\usepackage[T1]{fontenc}
\usepackage{ae,aecompl}

\usepackage{graphicx}	

\usepackage{hyperref}
\usepackage[nolist,nohyperlinks]{acronym}
\usepackage{supertabular}
\usepackage{pdflscape}
\usepackage{float}
\usepackage{epstopdf}
\usepackage{multicol}
\usepackage{graphicx}
\usepackage{subcaption}
\usepackage{geometry}
\usepackage{changepage}
\usepackage{siunitx}
\usepackage{dcolumn}
\usepackage{longtable}
\usepackage{afterpage}
\usepackage{array}
\usepackage{varwidth}

\usepackage{lastpage}
\usepackage[nolist]{acronym}

\usepackage{graphicx}	
\usepackage{xspace}   
\usepackage{hyperref}
\usepackage{xcolor}
\usepackage{comment}
\usepackage{scalerel}
\usepackage{lineno}
\usepackage{float}

\newcommand{\Spitzer}{{\it Spitzer}}
\newcommand{\Einstein}{{\it Einstein}}
\newcommand{\ROSAT}{{\it ROSAT}}
\newcommand{\WISE}{{\it WISE}}
\newcommand{\xmm}{{\it XMM--Newton}}
\newcommand{\ergcm}[1]{erg~cm$^{-2}$ s$^{-1}$}

\newcommand{\aegean}{\textsc{Aegean}}

\newcommand{\HII}{{H\,{\sc ii}}\xspace}
\newcommand{\SII}{[S\,{\sc ii}]}
\newcommand{\OIII}{[O\,{\sc iii}]}

\newcommand{\D}{$^\circ$}

\newcommand{\Ha}{H$\alpha$}

\def\HII{\hbox{H\,{\sc ii}}}

\def\arcmin{\hbox{$^\prime$}}
\def\arcsec{\hbox{$^{\prime\prime}$}}

\newcommand{\TOTALNUMBERCHZEROPEAKFLUX}{108,330}
\newcommand{\TOTALNUMBERCHTWOPEAKFLUX}{30,322}
\newcommand{\TOTALNUMBERCHTHREEPEAKFLUX}{18,298}
\newcommand{\TOTALNUMBERCHFOURPEAKFLUX}{33,782}
\newcommand{\TOTALNUMBERCHFIVEPEAKFLUX}{34,929}
\newcommand{\TOTALNUMBERCHSIXPEAKFLUX}{32,587}
\newcommand{\TOTALNUMBERCHSEVENPEAKFLUX}{23,872}
\newcommand{\TOTALNUMBERCHTENPEAKFLUX}{29,337}
\newcommand{\TOTALNUMBERCHELEVENPEAKFLUX}{34,415}
\newcommand{\TOTALNUMBERCHTWELVEPEAKFLUX}{33,516}
\newcommand{\TOTALNUMBERCHTHIRTEENPEAKFLUX}{21,346}
\newcommand{\TOTALNUMBERCHFOURTEENPEAKFLUX}{9,880}
\newcommand{\TOTALNUMBERCHFIFTEENPEAKFLUX}{18,032}
\newcommand{\TOTALSPINDEX}{40,571}

\newcommand{\TOTALBCE}{517}
\newcommand{\MANUALBCE}{6}

\newcommand{\chg}[1]{{  #1}} 

\begin{acronym}[AWGN]
\acro{2MASS}{Two Micron All Sky Survey}
\acro{2MASX}{Two Micron All Sky Survey Extended Source Catalogue}
\acro{30Dor}[30 Dor]{30~Doradus}
\acro{AGN}{active galactic nuclei}
\acro{ATCA}{Australia Telescope Compact Array}
\acro{ATESP}{Australia Telescope ESO Slice Project}
\acro{ATNF}{Australia Telescope National Facility}
\acro{ATOA}{Australia Telescope Online Archive}
\acro{AT20G}{Australia Telescope 20 GHz Survey}
\acro{ASKAP}{Australian Square Kilometre Array Pathfinder}
\acro{BETA}{Boolardy Engineering Test Array}
\acro{BL}[BL Lac]{BL Lacertae Objects}
\acro{CASS}{CSIRO Astronomy and Space Science}
\acro{CABB}{Compact Array Broadband Back-end}
\acro{CHII}[{\sc CHii}]{Compact \textsc{Hii}}
\acro{CMB}{Cosmic Microwave Background}
\acro{CSIRO}{Australian Commonwealth Scientific and Industrial Research Organisation}
\acro{CSS}{Compact Steep Spectrum}
\acro{DS9}[\textsc{DS9}]{\textsc{SAOImage DS9}}
\acro{DSS}{Digital Sky Survey}
\acro{EM}{Electromagnetic}
\acro{EMU}{Evolutionary Map of the Universe}
\acro{ev}[eV]{electronvolt\acroextra{: 1 eV $\approx 1.6 \times 10^{-19}$ J}}
\acro{FITS}[\textsc{Fits}]{Flexible Image Transport System}
\acro{FRB}{Fast Radio Bursts}
\acro{FSRQ}{Flat Spectrum Radio Quasars}
\acro{FWHM}{Full Width at Half-Maximum}
\acro{GLEAM}{GaLactic Extragalactic All-sky MWA}
\acro{GPS}{Gigahertz Peak Spectrum}
\acro{HEMT}{High Electron Mobility Transistor}
\acro{HFP}[HFP]{High-Frequency Peaker}
\acro{HPBW}{Half Power Beam Width} 
\acro{HzRGs}{High Redshift Radio Galaxies}
\acro{pHFP}[pHFP]{Potential High Frequency Peaker}
\acro{HI}[H{\sc i}]{Neutral Atomic Hydrogen} 
\acro{HST}{\textit{Hubble Space Telescope}}
\acro{ICM}{intracluster medium}
\acro{IAU}{International Astronomical Union}
\acro{IFRSs}{Infrared Faint Radio Sources}
\acro{ISM}{Interstellar Medium}
\acro{IFRS}{Infrared Faint Radio Source}
\acro{IR}{Infrared}
\acro{JY}[Jy]{Jansky\acroextra{, 1 Jy = $10^{-26} \times \mathrm{W~ m}^{-2}~\mathrm{Hz}^{-1}$}} 
\acro{LLS}{Largest Linear Size}
\acro{LFAA}{Low-Frequency Aperture Array}
\acro{LMC}{Large Magellanic Cloud}
\acro{LSO}[LSOs]{Large Scale Objects}
\acro{MACHO}{Massive Astrophysical Compact Halo Objects}
\acro{MC}[MCs]{Magellanic Clouds}
\acro{mc}[MC]{Magellanic Cloud}    
\acro{MCELS}{Magellanic Cloud Emission Line Survey}
\acro{MW}{Milky Way}
\acro{MIRIAD}[\textsc{Miriad}]{Multichannel Image Reconstruction, Image Analysis and Display}
\acro{MIT}{Massachusetts Institute of Technology}
\acro{MOST}{Molonglo Observatory Synthesis Telescope}
\acro{MQS}{Magellanic Quasars Survey}
\acro{MRC}{Molonglo Reference Catalogue of Radio Sources}
\acro{MWA}{Murchison Widefield Array}
\acro{NRAO}{National Radio Astronomy Observatory}
\acro{NVSS}{NRAO VLA Sky Survey}
\acro{OPAL}{Online Proposal Applications \& Links}
\acrodefplural{ORC}{Odd Radio Circles}
\acro{ORC}{Odd Radio Circle}
\acro{OVV}{Optically Violent Variable Quasars}            
\acro{PAF}{Phased Array Feed}
\acro{pc}{parsec\acroextra{: 1 pc $\simeq 3.09 \times 10^{16}$ m}}
\acro{PMN}{Parkes-MIT-NRAO}
\acro{PNe}{Planetary Nebulae}
\acro{PWN}{Pulsar Wind Nebulae}
\acro{QSO}{Quasi-Stellar Object}
\acro{RA}{Right Ascension}
\acro{RFI}{Radio-Frequency Interference}
\acro{RMS}[rms]{Root Mean Squared}
\acro{SARAO}{South African Radio Astronomy Observatory}
\acro{SDSS}{Sloan Digital Sky Survey}
\acro{SED}{spectral energy distribution}
\acro{SI}[$\alpha$]{Spectral Index\acroextra{, $S \propto \nu^\alpha$}}
\acro{SKA}{Square Kilometre Array}
\acro{SMB}{Super Massive Blackholes}
\acro{SMC}{Small Magellanic Cloud}
\acro{SN}{Supernova}
\acro{SUMSS}{Sydney University Molonglo Sky Survey}
\acro{TOPCAT}[\textsc{Topcat}]{Tool for OPerations on Catalogues And Tables}
\acro{USS}{Ultra Steep Spectrum}
\acro{YSOs}{young stellar objects}
\acro{WBAC}{Wide-Band Analogue Correlator}
\acro{WIFES}[WiFeS]{Wide-Field Spectrograph}
\acro{WISE}{Wide-Field Infrared Survey Explorer}
\acro{VLBI}{Very Long Baseline Interferometry} 
\acro{VLSR}[\textbf{$v_{lsr}$}]{Velocity in the Line of Sight}
\acro{SMASH}{Survey of the MAgellanic Stellar History}
\acro{SNR}{Supernova Remnant}
\acrodefplural{SNR}{Supernova Remnants}
\acro{SD}{single-degenerate}
\acro{SUMSS}{Sydney University Molonglo Sky Survey}
\acro{SMBH}{Super Massive Black Hole}
\end{acronym}

\title[The MeerKAT 1.3 GHz Survey of the SMC]{The MeerKAT 1.3\,GHz Survey of the Small Magellanic Cloud}

\author[W. Cotton et al.]{W. D. Cotton,$^{1,2}$\thanks{E-mail: bcotton@nrao.edu}
M. D. Filipovi\' c,$^{3}$
F. Camilo,$^{2}$
R. Indebetouw,$^{4,1}$
R. Z. E. Alsaberi,$^{3}$ 
\newauthor
J. O. Anih,$^{5}$
M. Baker,$^{5}$
T. S. Bastian,$^{1}$
I. {Boji{\v c}i{\'c}},$^{3}$
E. Carli,$^{6}$
F. Cavallaro,$^{7}$
\newauthor
E. J. Crawford,$^{3}$
S. Dai,$^{3,8}$
F. Haberl,$^{9}$
L. Levin,$^{6}$
K. Luken,$^{3}$
C. M. Pennock,$^{10}$
\newauthor
N. Rajabpour,$^{3}$
B. W. Stappers,$^{6}$
J. Th. van Loon,$^{5}$
A. A. Zijlstra,$^{11}$
S. Buchner,$^{2}$
\newauthor
M. Geyer,$^{2,12}$
S. Goedhart,$^{2}$
M. Serylak$^{2,13,14}$ 
\\ \\
$^{1}$National Radio Astronomy Observatory, 520 Edgemont Road, Charlottesville, VA 22903, USA \\
$^{2}$South African Radio Astronomy Observatory (SARAO), 2 Fir Street, Black River Park, Observatory, Cape Town 7925, South Africa \\
$^{3}$Western Sydney University, Locked Bag 1797, Penrith South DC, NSW 2751, Australia \\
$^{4}$Department of Astronomy, University of Virginia, Charlottesville, VA 22904, USA \\
$^{5}$Lennard-Jones Laboratories, Keele University, Staffordshire ST5 5BG, UK\\
$^{6}$Jodrell Bank Centre for Astrophysics, Department of Physics and Astronomy, The University of Manchester, Manchester M13 9PL, UK\\
$^{7}$INAF-Osservatorio Astrofisico di Catania, Via S. Sofia 78, I-95123 Catania, Italy\\
$^{8}$CSIRO Astronomy and Space Science, PO Box 76, Epping, NSW 1710, Australia \\
$^{9}$Max-Planck-Institut f\"{u}r extraterrestrische Physik, Gie{\ss}enbachstra{\ss}e 1, D-85748 Garching, Germany \\
\chg{$^{10}$Institute for Astronomy, University of Edinburgh, Royal Observatory, Blackford Hill, Edinburgh EH9 3HJ, UK}\\
$^{11}$Jodrell Bank Centre for Astrophysics, The University of Manchester, Oxford Road M13 9PL, UK\\
$^{12}$High Energy Physics, Cosmology \& Astrophysics Theory (HEPCAT) Group, Department of Mathematics \& Applied Mathematics,\\ University of Cape Town, Cape Town 7700, South Africa\\
$^{13}$SKA Observatory, Jodrell Bank, Lower Withington, Macclesfield, SK11 9FT, UK\\
$^{14}$Department of Physics and Astronomy, University of the Western Cape, Bellville, Cape Town, 7535, South Africa\\
}

\date{Accepted 19 Jan. 2024. Received 15 Jan 2024; in original form 13 Nov 2023}
\pubyear{2023}
\hypersetup{draft}
\begin{document}
\label{firstpage}
\pagerange{\pageref{firstpage}--\pageref{lastpage}}
\maketitle

\newpage
\begin{abstract}
We present new radio continuum images and a source catalogue from the MeerKAT survey in the direction of the \ac{SMC}.
The observations, at a central frequency of 1.3\,GHz across a bandwidth of 0.8\,GHz, encompass a field of view $\sim$7\D$\times$7\D\ and result in images with resolution of 8 arcsec. The median broad-band Stokes~I image Root Mean Squared noise value is $\sim$11\,$\mu$Jy\,beam$^{-1}$.
The catalogue produced from these images contains \TOTALNUMBERCHZEROPEAKFLUX\ point sources and \TOTALBCE\ compact extended sources. 
We also describe a UHF (544--1088 MHz) single pointing observation.
We report the detection of a new confirmed \ac{SNR} (MCSNR~J0100--7211) with an X-ray magnetar at its centre and 10 new \ac{SNR} candidates. 
This is in addition to the detection of 21 previously confirmed \acp{SNR} and two previously noted \ac{SNR} candidates. Our new \ac{SNR} candidates have typical surface brightness an order of magnitude below those previously known, and on the whole they are larger.
The high sensitivity of the  MeerKAT survey also enabled us to detect the bright end of the \ac{SMC} \ac{PNe} sample -- point-like radio emission is associated with 38 of 102 optically known \ac{PNe}, of which 19 are new detections.
Lastly, we present the detection of three foreground radio stars amidst 11 circularly polarised sources, and a few examples of morphologically interesting background radio galaxies from which the radio ring galaxy ESO~029--G034 may represent a new type of radio object.

\end{abstract}



\section{Introduction}
\label{intro}

Over the past 50 years, several  radio surveys have greatly improved our knowledge of the neighbouring \ac{SMC} galaxy and its source populations. This is an exciting time for the study of nearby galaxies, especially in radio wavebands using the new generation of radio telescopes such as MeerKAT, the \ac{ASKAP} and the \ac{MWA}. These nearby Local Group galaxies offer ideal laboratories since they are close enough to be resolved, yet located at well-known distances \citep[see e.g.][]{2019Natur.567..200P}. 

New generations of Magellanic Cloud (MC) surveys across the entire electromagnetic spectrum provide a major opportunity to study different objects and processes in the evolution and elemental enrichment of the \ac{ISM}. The study of these in different domains, including radio, IR, optical and X-ray \citep{2013A&A...558A.101S}, allow a better understanding of objects such as \acp{SNR}, \ac{PNe}, \ac{YSOs}, \HII~regions and (super)bubbles and their environments. Also, the \ac{SMC}'s large-scale properties such as its magnetic field are essential to understand the kinematics and dynamics of this galaxy.

The focus of our new radio continuum study -- the \ac{SMC} -- is a gas-rich irregular and nearby dwarf galaxy interacting with the \ac{LMC} and \ac{MW} halo. The nearby distance of 62.44$\pm$0.47~kpc\footnote{For easier comparison with previous studies, we adopt a distance of 60~kpc for all our calculations in this paper as most of the previous studies used this value.} but with the sizeable line-of-sight depth reaching 7~kpc \citep{2020ApJ...904...13G} and the low Galactic foreground emission and absorption ($N_{\rm HI}\sim6\times10^{20}$\,cm$^{-2}$) enables the entire source population in the \ac{SMC} to be studied in various wavebands and at excellent spatial resolution. The \ac{SMC} also attracts particular attention because it has 5 times lower metallicity in its \ac{ISM} than the solar neighbourhood. At the same time, a vast amount of foreground and especially background sources pose for detailed investigation. 

For  these reasons, the \ac{SMC} has been the  target of many radio studies over the past 50 years. Starting in the 1970s, the \ac{SMC} has been surveyed by single-dish and interferometric radio telescopes. These large-scale observational campaigns have produced several catalogues of sources towards the \ac{SMC} \citep{1976MNRAS.174..393C,1976AuJPh..29..329M,1986A&A...159...22H,1990PKS...C......0W,1997A&AS..121..321F,1998PASA...15..280T,1998A&AS..130..421F,2002MNRAS.335.1085F,2004MNRAS.355...44P,2005MNRAS.364..217F,2006MNRAS.367.1379R,2007MNRAS.376.1793P,2011SerAJ.182...43W,2011SerAJ.183...95C,2011SerAJ.183..103W,2012SerAJ.184...93W,2012SerAJ.185...53W,2018MNRAS.480.2743F,2019MNRAS.490.1202J}.

The MeerKAT survey of the \ac{SMC} field presented here is a significant improvement on previous similar radio continuum studies. MeerKAT's high resolution and sensitivity have already been demonstrated in other studies \citep[e.g.,][]{2022ApJ...925..165H}. In this paper, we present point and compact extended source catalogues from the MeerKAT \ac{SMC} survey. These catalogues were obtained from images taken at the broad band of 1283.8\,MHz ($\lambda =23$\,cm; L-band) and 12 sub-bands. We also present results on several new objects found in these images.

The paper is organised as follows: Section~\ref{data} describes the MeerKAT observations, reduction, and data used to create the source lists, with data release details in Section~\ref{products}. In Section~\ref{sec:sdet} we describe the source detection methods used and Section~\ref{sec:cats} describes the new MeerKAT source catalogues including spectral indices (Section~\ref{sec:SI}), while in Section~\ref{scompare} we compare our work to previous catalogues of point sources towards the \ac{SMC}. In Section~\ref{sec:results} we present some of the early science results from this survey which includes the study of \acp{SNR} (Sections~\ref{SNR_samp},~\ref{sec:autosnr}), pulsars (Section~\ref{sec:psr}), and \ac{PNe} (Section~\ref{PN_samp}), radio stars (Section~\ref{sec:stars}), while in Section~\ref{Other_sources}, we briefly discuss other sources of interest, including those behind the \ac{SMC}.  The bulk of the discussion concerns the L-band mosaics.

\section{Data}
 \label{data}

\subsection{Observations}
\subsubsection{L-band}
Observations of the \ac{SMC} were made with the MeerKAT array primarily at L-band (856 to 1712~MHz) using 8-second integrations and 4096 spectral channels across the band. All four combinations of the linearly polarised feeds were recorded. The MeerKAT array is described in more detail in  \citet{Jonas2016,Camilo2018,DEEP2}. These observations used project code SSV-20190715-FC-02.

Since the \ac{SMC} is substantially larger than the field of view of the MeerKAT antennas, the observations consisted of a mosaic of pointings on a hexagonal pattern with an offset between centres of 29.6~arcmin which gives relatively uniform sensitivity. The mosaic was initially covered in seven sessions of eight or nine pointings each, cycling among 5~min scans on the pointings for roughly 10~hours in each session. This gives approximately 1~hour of integration on each pointing but is composed of many snapshots spread widely in hour angle to improve $uv$ coverage. The centre of the mosaic is at RA(J2000)=01:00:00.0, Dec(J2000)=--73:00:00. A total of 60 individual pointing centres were observed across the field of 7.08\D$\times$7.08\D.

PKS~B1934$-$638 and PKS~0408$-$65 were used for the delay, band-pass and flux density calibration; J0252--7104 was the gain (astrometric) calibrator and 3C138 was the polarised calibrator. PKS~B1934$-$638, PKS~0408$-$65, and 3C138 were observed for 10 min every few hours as available and J0252--7104 was observed for 2~minutes every half hour. The flux density scale is set to the spectrum of PKS~B1934$-$638 given in \cite{Reynolds94}: 
\begin{eqnarray}
  \log(S) = & -30.7667 + 26.4908 (\log\nu)
  - 7.0977 (\log \nu)^2 \nonumber \\
   & +0.605334 (\log\nu)^3\, , \qquad\qquad\qquad\qquad\qquad
\end{eqnarray}
\noindent where $S$ is the flux density in Jy and $\nu$ is the frequency in MHz.

The observations were obtained between 21~July~2019 and 29~September~2019 and are summarised in Table~\ref{Obs} where times include calibration overheads. Two of the sessions were divided into multiple blocks which were calibrated individually. After an initial imaging, several of the pointings had undiagnosed imaging artifacts and these pointings were reobserved. Furthermore, several pointings were dynamic range limited and additional observations were made. Later reduction managed to recover the bulk of the affected pointings' data which was then incorporated into the final imaging. The additional observations had the net effect of improving the dynamic range in the parts of the mosaic with large-scale, bright emission. 

\begin{table}
\caption{MeerKAT L-band observations of the \ac{SMC}.}
 \begin{center}
  \begin{tabular}{lcl}   
\hline
Date        & Total Time & Note\\
            & (hours)    & \\
\hline
21 Jul 2019 & 10.9 & Session 1 \\
24 Jul 2019 & 10.5 & Session 2 \\
31 Jul 2019 &  9.7 & Session 3 \\
04 Aug 2019 & 10.3 & Session 4\\
06 Aug 2019 &  5.3 & Session 5\\
06 Aug 2019 &  6.0 & Session 5 \\
11 Aug 2019 &  3.2 & Session 6 \\
12 Aug 2019 &  5.5 & Session 6 \\
13 Aug 2019 &  6.7 & Session 6 \\
13 Aug 2019 &  8.9 & Session 7 \\
11 Sep 2019 &  4.5 & Reobserve \\
12 Sep 2019 & 12.2 & Reobserve \\
29 Sep 2019 & 12.1 & Reobserve \\
\hline
  \end{tabular}
 \end{center}
 \label{Obs}
\end{table}

\subsubsection{UHF}
A single pointing observation of the \ac{SMC} was also made in the UHF band (544--1088\,MHz) on 6 and 7~Feb~2020. There were 13.5~hours of observations, including calibration, and 10.3~hours on the \ac{SMC}. The pointing centre was RA(J2000)=00:52:44.0 and Dec(J2000)=--72:49:42.  Fifty three of the 64 antennas were in the array on 6~Feb and 59 on 7~Feb.  Data were recorded with 8-second sampling, 4096 spectral channels and all four correlation products. These observations are under project code SSV-20200131-SP-01.
\subsection{Calibration}

Initial calibration and data editing were as described in \cite{DEEP2} and \cite{XGalaxy}, and we used the {\small OBIT} package described by \cite{2008PASP..120..439C}\footnote{{http://www.cv.nrao.edu/$\sim$bcotton/Obit.html}}. This processing trims the edge channels and divides the remainder into 8 `spectral windows' for calibration purposes. A standard list of channels always affected by strong \ac{RFI} was flagged. Calibration and editing steps are interleaved.

The polarisation calibration was based on the hardware `noise diode' calibration performed at the beginning of each observing session. This involves injecting a signal into each feed which allows for measuring the phase and delay differences between the orthogonal feeds after the injection point. Residual differences were estimated from the polarised calibrator and the instrumental polarisation (a.k.a. `leakage') was estimated from PKS~B1934$-$638 and PKS~0408$-$65 (assumed unpolarised) and J0252$-$7104. This method is described in more detail in \cite{XGalaxy}.

\subsection{Imaging\label{Imaging}}

The L-band data were imaged in Stokes~I, Q, U and V using the {\small OBIT} task {\sc MFImage} which is described more fully in \cite{Cotton2018}. UHF data were imaged only in Stokes I.  This imaging program uses multiple constant fractional bandwidth frequency sub-bands which are imaged independently and CLEANed jointly to accommodate the frequency dependence of the sky and the antenna gain pattern. Faceting \citep{Perley1999A} is used to correct for the non-coplanarity of the array. This allows for fully covering a given field of view while including outlying facets around stronger sources.

Imaging used 5~per~cent fractional bandwidth sub-bands, fully imaged at L-band to a radius of 1.2~degrees with outlying facets to 1.5~degrees centred on sources estimated from the 843~MHz \ac{SUMSS} catalogue \citep{mau03} to appear brighter than 1~mJy. At ultra high frequency (UHF) the field of view had a radius of 2.5~degrees and used outliers to 3~degrees. All pointings were subjected to two-phase self-calibrations and also amplitude and phase self-calibration if the peak brightness in the image exceeded 0.7~Jy~beam$^{-1}$. A Briggs Robust factor of --1.5 ({\sc AIPS}/{\small OBIT} usage where the scale goes from --5 (pure uniform) to 5 (natural) weights) was used to result in a typical resolution of 7.5~arcsec  at L-band. At UHF the restoring beam is 13.3 $\times$ 11.0~arcsec$^2$. CLEANing used a gain of 0.05 and proceeded to a maximum of 125\,000 components or a residual of 60~$\mu$Jy~beam$^{-1}$ in Stokes~I, 10\,000 components or a residual of 40~$\mu$Jy~beam$^{-1}$ in Stokes~Q and U and 500 components or a residual of 20~$\mu$Jy~beam$^{-1}$ in Stokes~V. 

The procedure described above makes only direction-independent gain corrections and is insufficient to produce noise-limited images in pointings with bright and extended emission. For a number of cases, `peeling' \citep{Noordam2004,Smirnov2015} was applied to allow direction-dependent gain correction for the brighter sources in the field of a given pointing. In a few cases, a multiresolution {\sc CLEAN} was also used to better image large-scale emission. For these, the second resolution of approximately 30~arcsec \ac{FWHM} was used. 

\subsection{Astrometric corrections}
 \label{astrometry}
A number of systematic astrometric errors affecting MeerKAT images are discussed in \cite{2022A&A...657A..56K}; the most important of these for the \ac{SMC} images is thought to be the low-accuracy geometric model used in the correlator.
The pointing images had residual astrometric variations on the sub-arcsecond level. These errors were estimated by comparing the positions of compact radio sources as extracted by \aegean\ (Section~\ref{sec:cat}) to quasar positions from MilliQuas\footnote{\url{https://heasarc.gsfc.nasa.gov/W3Browse/all/milliquas.html}} \citep{2021arXiv210512985F}. Sources were selected from the MeerKAT source catalogue with a ratio of integrated vs peak flux density $<$1.3, and measured uncertainties less than 2~arcsec in RA and Dec, i.e. compact and point sources with good positional precision.

MilliQuas sources in the imaged region were retrieved from the HEASARC interface to version 7.4 of the catalogue \cite[Dec 2021]{2021arXiv210512985F}, and matched to the nearest MeerKAT catalogue source within 2~arcsec. Of the 1489 MilliQuas sources, 745 are matched to MeerKAT sources. There was a clear systematic offset of 0.193$\pm$0.031~arcsec in RA and --0.734$\pm$0.020~arcsec in Dec; this correction was applied (added) to all L-band pointings. After removing that global mean offset, matched sources within 0.4~degrees of each pointing centre were considered. A pointing was determined to have a significant measurable residual offset if it had: at least five matched sources, a residual offset (after removing the full-mosaic offset) of more than 0.15~arcsec and a residual offset of greater than the dispersion of individual source offsets for that field. The fields with sufficiently significant offsets to be applied were corrected and are listed  in Table~\ref{offsets}. Based on the remaining dispersion in individual source offsets after these per-field corrections, we estimate that the astrometric uncertainty of a typical source in the mosaic is better than 0.25~arcsec. Additional comparisons with other catalogues are given in Section~\ref{scompareA}.

\begin{table}
  \caption{Pointings with individual astrometric corrections applied. Note that the pointings were arranged in a hexagonal pattern in a set of rows (`R') and columns (`C').}
    \begin{tabular}{lcccc}   
    \hline
      Name & Centre RA & Centre Dec & RA offset & Dec offset  \\
           &           &            & (arcsec)  & (arcsec) \\
    \hline
      SMC1R01C09 & 00:32:08.52 & --73:59:15.5 & --0.078 & +0.216 \\
      SMC1R01C11 & 00:32:08.52 & --74:58:31.1 & --0.110 & --0.239 \\
      SMC1R03C09 & 00:46:09.36 & --73:59:15.6 & --0.185 & --0.266 \\
      SMC1R03C11 & 00:46:09.36 & --74:58:31.1 & +0.130  & --0.250 \\
      SMC1R06C02 & 01:06:05.37 & --70:31:51.1 & --0.653 & +0.471 \\
      SMC1R06C04 & 01:06:05.37 & --71:31:06.7 & --0.309 & --0.050 \\
      SMC1R06C06 & 01:06:05.37 & --72:30:22.2 & +0.353  & +0.047 \\
      SMC1R07C01 & 01:13:00.74 & --70:02:13.3 & +0.476  & +0.336 \\
      SMC1R07C03 & 01:13:00.74 & --71:01:28.9 & +0.108  & --0.370 \\
      SMC1R10C04 & 01:33:06.86 & --71:31:06.7 & +0.318  & --0.199 \\
      \hline
    \label{offsets}  
    \end{tabular}
\end{table}

\subsection{Mosaicing}
\label{Mosaicing}

Individual L-band pointing images in Stokes~I, Q, U and V were combined into single mosaics covering the \ac{SMC} using a process similar to that described in \cite{Brunthaler2021}.
The optimal sensitivity combination used
\begin{equation}
 \label{mosaic_eq}
M(x,y)\ =\ {{\sum_{i=1}^n P_i(x,y) I'_i(x,y)}\over{\sum_{i=1}^n P_i^2 (x,y)}},
\end{equation}
\noindent where $M(x,y)$ is the combined image as a function of sky coordinates, $n$ is the number of pointing images contributing to the mosaic, $P_i$ is the antenna power gain pattern for pointing $i$ and $I'$ is the pointing image interpolated to the grid of the mosaic. This procedure will correct the mosaic image for variations in the antenna power gain across the field of view. Prior to the combination, the pointing images were convolved to a resolution of 8~arcsec and the coordinates were corrected as described in Section~\ref{astrometry}.

\begin{figure*}
		\centering
   		\includegraphics[width=\textwidth,trim= 0 0 0 0, clip]{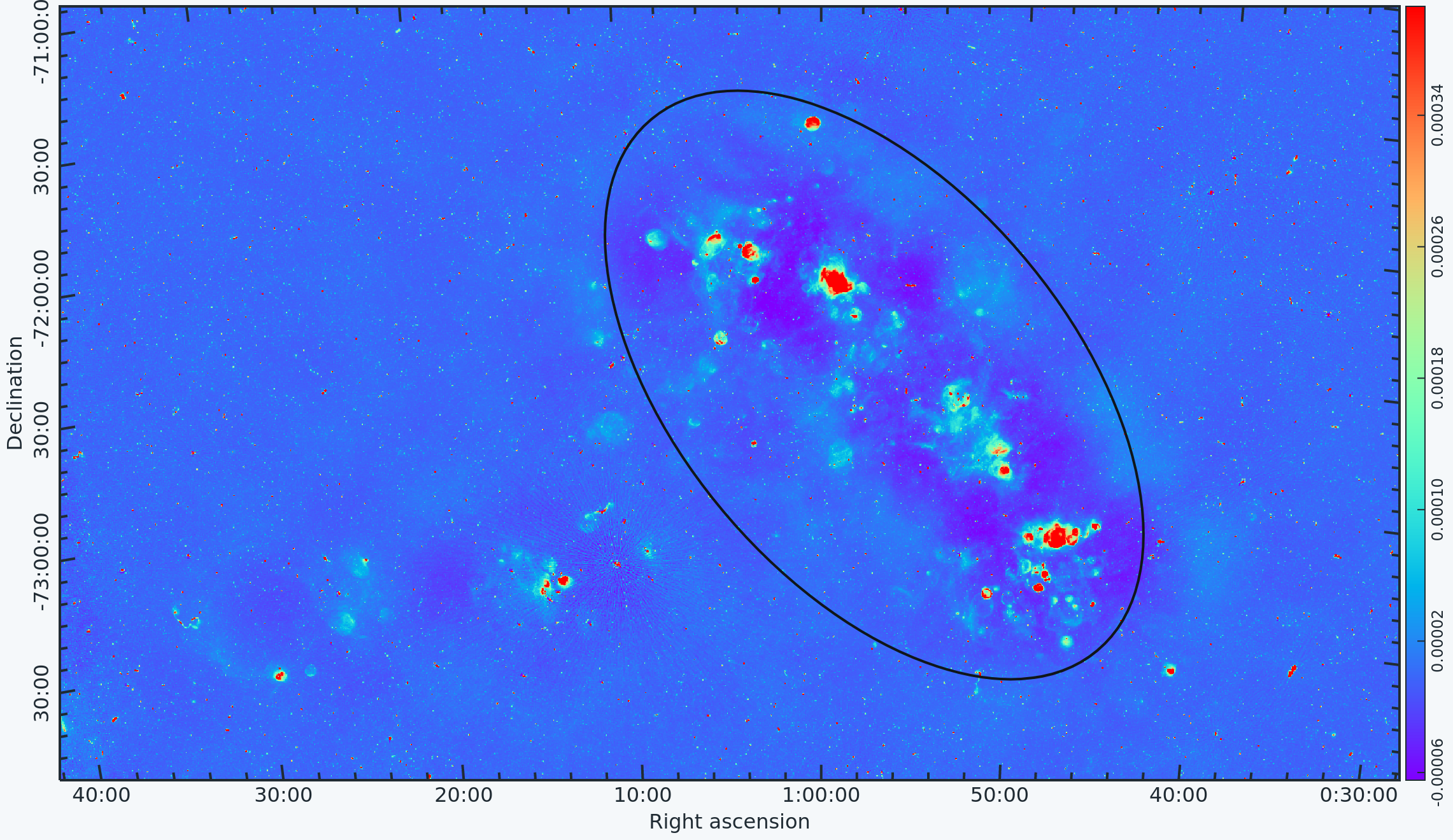}
			\caption{MeerKAT broad band Stokes~I image of the central portion of the \ac{SMC} field. The black ellipse indicates an area of the so-called \ac{SMC} bar region. The colour bar (linear scale) on the right-hand side is in units of Jy~beam$^{-1}$. 
   }
        \label{fig:1}
\end{figure*}

The sub-band mosaics were formed using Eq.~\ref{mosaic_eq} and used to derive a total flux density and spectral index (defined by $S \propto \nu^{\alpha}$, where $S$ is flux density, $\nu$ is frequency and $\alpha$ is the spectral index) as described in Section \ref{products}.
The off-source \ac{RMS} in quieter portions of the Stokes~I image is 11~$\mu$Jy~beam$^{-1}$, 7.1~$\mu$Jy~beam$^{-1}$ for Stokes~Q, 7.6~$\mu$Jy~beam$^{-1}$ for Stokes~U and 7.9~$\mu$Jy~beam$^{-1}$ for Stokes~V. In Fig.~\ref{fig:1} we show the central part of the Stokes~I image.

\subsection{UHF}
Since the UHF data were in a single pointing, the data were primary beam corrected rather than formed into a mosaic. The broadband image was constructed from a weighted average of the subband planes using weighting proportional to the inverse of the plane RMS.  This results in an effective frequency at the pointing center of 870.0 MHz.  The RMS in the image uncorrected for the primary beam is 8.6 $\mu$Jy beam$^{-1}$.

As the gain of the antennas away from the pointing center varies with frequency, the effective frequency declines with distance from the pointing center.  A frequency dependent correction to the subband pixel values following the technique of Cotton et al. 2024 (submitted) using an assumed spectral index of $\alpha$=-0.6 was used to adjust all parts of the image to the same effective reference frequency.

\subsection{Data products\label{products}}
There are a number of data products being made available with this paper from the observations of the \ac{SMC}.
\begin{itemize}
 \item {\bf Visibility data:}
Raw L-band visibility data is available from the \ac{SARAO} archive (\url{https://archive.sarao.ac.za/}) under project code SSV-20190715-FC-02.  The UHF data is under project code SSV-20200131-SP-01.
Catalogues and images are available from DOI \url{https://doi.org/10.48479/1fdb-r233}.
 \item {\bf L-band mosaic images:}
The mosaic FITS images are as described in \cite{MFImage}.
The mosaics in Stokes~I, Q, U and V are $7.08^\circ\ \times\ 7.08^\circ$ with 1.5~arcsec pixels centred at J2000 position RA=01:00:0.0 and Dec=--73:00:00.0.
The mosaic cubes contain 16 planes.
The broad-band flux densities are averages, weighted by Stokes~I plane RMS$^{-1}$, giving a reference frequency of 1310.89~MHz.
More details of the derivation of the reference frequency is given in Cotton et al.2024 (submitted).
For each pixel in the Stokes~I mosaic with an adequate signal-to-noise ratio and an average Stokes~I intensity in excess of 250 $\mu$Jy beam$^{-1}$, a least squares fit was performed to derive the spectral index. Other pixels are blanked.
The first plane is the band average flux density, the second plane is the spectral index, and subsequent planes are the frequency bins centred at the frequencies given in Table~\ref{tab:subchannumbers}.
Two planes are totally blanked due to \ac{RFI}.
Stokes~Q, U and V images have a second plane which is blanked. This plane is nominally the spectral index but since Q and U are also subjected to Faraday rotation, this is not meaningful. 
A second total intensity image containing the band average flux density, spectral index and least squares fitting error estimates is also provided.
\item {\bf UHF image:}
The primary beam and effective frequency corrected Stokes I is given in two forms similar to the L-band mosaics.  The minimum Stokes I for fitting spectral index was 50 $\mu$Jy beam$^{-1}$. The effective reference frequency is 870.0 MHz. These files are SMC\_UHF\_I\_Cube.fits.gz and SMC\_UHF\_I\_Fit.fits.gz.
 \item {\bf Catalogues:}
The catalogues underlying this article are available on the CDS/\textsc{VizieR} (\url{https://cds.u-strasbg.fr/}). The full point and compact extended source catalogues (Tables~\ref{tab:main} and \ref{tab:bcecat}) are also provided through the CDS/\textsc{VizieR} service.
DS9 region files for the Bright Compact Extended sample of sources are also provided (Section~\ref{sec:cats}).
\end{itemize}

\subsection{Data limitations}
There are a number of limitations of the MeerKAT observations that can affect the scientific interpretation of the results. Some of these have been mitigated in the data products distributed but will remain in the raw visibility data. These limitations are similar to those discussed in \cite{2022A&A...657A..56K} although more have been corrected in the \ac{SMC} data products. 

\subsubsection{Dynamic range}
While the excellent $uv$ coverage of MeerKAT generally gives a good dynamic range, Direction Dependent Effects (DDEs) can result in significant artifacts due to bright sources. This is especially true if there are multiple bright, widely separated sources that are not adequately corrected by direction-independent self-calibration. The expected DDEs include pointing errors, variable ionospheric refraction and asymmetries in the array antenna patterns. DDEs are explored in detail in \cite{smirnov11}.

As described in Section~\ref{Imaging}, the pointing fields most affected by dynamic range limiting DDEs have had `peeling' applied to reduce the Stokes~I artifacts due to the strongest sources. These improved pointing images were then incorporated into the mosaic data products. The raw visibility data still includes these effects.

\subsubsection{Largest angular size}
The largest structure that can be imaged using interferometer data is determined by the $uv$ coverage of the shortest spacings. As the $uv$ location of a given measurement depends on the observing frequency, the factor of two between the highest and lowest frequencies in MeerKAT L-band observations results in a factor of two difference in the largest structure that can be imaged at the top and bottom of the band. This effect can result in a severe spurious steepening of the apparent spectrum for regions that are increasingly resolved out with increasing frequency. For a 1 Jy 10~arcmin \ac{FWHM} Gaussian source, the visibility on typical short baselines is 0.32 at the bottom of the band and 0.02 at the top. This effect can severely limit the maximum flux density that can be recovered. Sources significantly smaller than 10~arcmin should not be seriously affected but larger ones can be.

A further complication of the limited short baseline coverage is the `missing zero spacing' problem. If the deconvolution of a bright, extended feature does not recover all of the flux density, the feature will be surrounded by a negative `bowl'. This bowl will be frequency dependent, more pronounced at higher frequencies, and complicates the estimation of flux density and especially spectral index. These bowls can be easily seen in Fig.~\ref{fig:1}.

\subsubsection{Flux density and spectral index}
The wide fractional bandpass of the MeerKAT L-band data allows the estimation of a wideband flux density and spectral index in pixels with adequate signal-to-noise. 
The result of the imaging process is a cube of images in 14 frequency bins (2 of which are totally blanked because of \ac{RFI}). A simple weighted average of these will produce a brightness image whose reference frequency depends on the weighting, derived from the RMS$^{-1}$ of fluctuations in spectral regions not affected by \ac{RFI}. For pixels in which there is an adequate signal, a least squares fit can produce a spectral index at the reference frequency derived from the weighting.
The spectral index image is the fitted value where available, else blanked.
The \ac{SMC} field contains emissions with a wide range of spectral indices, from $\sim$1.0 to --1.5 or less.

The effective array antenna pattern is also not well known; variable pointing errors cause it to be broader than that of a single antenna. Errors in the assumed pattern (that of an individual antenna) cause the derived spectral index to appear more negative than it actually is with increasing distance from the pointing centre. The `primary beam correction' is applied in the mosaic formation (Section~\ref{Mosaicing}) which reduces its errors in the mosaic products. This problem remains in the visibility data and in the UHF image.

\subsubsection{Astrometry}
The low-precision model used in the interferometer delay calculation can result in systematic position offsets in individual pointing images. These have been measured and corrected in the mosaic and catalogue data products as described in Section~\ref{astrometry} but are uncorrected in the visibility data.

\subsubsection{Polarisation}
The images in the first plane of the Stokes~Q, U and V cubes are meant to be suggestive of where there is polarised emission but should not be used for scientific interpretation. This plane is the band average of the Stokes~Q, U and V frequency channel images and for Q and U is only meaningful in the unlikely case that the rotation measure, depolarisation and polarised spectral index are all zero. The band average linearly polarised intensity is reduced by a factor of $\ge$2 when $|\mbox{RM}|$ exceeds about 25~rad\,m$^{-2}$.

Uncorrected instrumental polarisation is high towards the top end of the bandpass and away from the pointing centre \citep{2022AJ....163..135D}. This effect is reduced in the mosaic and catalogue products but is uncorrected in the visibility data. Residual uncorrected instrumental polarisation limits the polarisation dynamic range near sources bright in total intensity.

\section{MeerKAT SMC Source Catalogues}
 \label{sec:cat}

\subsection{Source detection}
\label{sec:sdet}

In a similar way to \citet{2021MNRAS.507.2885F} and \citet[][]{2021MNRAS.506.3540P}, we used \aegean, a compact source detection and flux extraction algorithm \citep{hancock2018}, to create point source catalogues from the broad-band image (channel~0) and 12 sub-band images as listed in Table~\ref{tab:subchannumbers}. 

\begin{table}
    \caption{Number of sources in each sub-band. The bandwidth of channel-0 is 811.455~MHz while other channels have $\sim$5~per~cent ($\sim$40-70~MHz) fractional bandwidth. Channels~8 and 9 are not used because of \ac{RFI}. Compared to other channels, we note a dip in the number of matched sources in channels~3 and 14, likely due to localised \ac{RFI}. 
    }
    \centering
    \begin{tabular}{ccccc}
    \hline
        Channel & Frequency (MHz) & Total matched\\
    \hline
        0 & 1283 & \TOTALNUMBERCHZEROPEAKFLUX  \\
        2 & 908.037 & \TOTALNUMBERCHTWOPEAKFLUX \\
        3 & 952.342 & \TOTALNUMBERCHTHREEPEAKFLUX\\
        4 & 996.646 & \TOTALNUMBERCHFOURPEAKFLUX\\
        5 & 1043.46 & \TOTALNUMBERCHFIVEPEAKFLUX\\
        6 & 1092.78 & \TOTALNUMBERCHSIXPEAKFLUX\\
        7 & 1144.61 & \TOTALNUMBERCHSEVENPEAKFLUX\\
        10 & 1317.23 & \TOTALNUMBERCHTENPEAKFLUX\\
        11 & 1381.18 & \TOTALNUMBERCHELEVENPEAKFLUX\\
        12 & 1448.05 & \TOTALNUMBERCHTWELVEPEAKFLUX\\
        13 & 1519.94 & \TOTALNUMBERCHTHIRTEENPEAKFLUX\\
        14 & 1593.92 & \TOTALNUMBERCHFOURTEENPEAKFLUX\\
        15 & 1656.2 & \TOTALNUMBERCHFIFTEENPEAKFLUX\\
    \hline
    \end{tabular}
    \label{tab:subchannumbers}
\end{table}

\subsubsection{L-band Broad-band image}
\label{sec:broadbandsdet}

The broad-band (channel~0) results are derived from the weighted average Stokes~I image with an effective frequency 1311~MHz rather than the band centre 1284~MHz. 

We created the initial \ac{RMS} and background map in the channel~0 using the Background and Noise Estimation tool (\textsc{BANE}; part of the \aegean\ suite), and the initial point source catalogue by applying a simple blind search with default parameters for both \textsc{BANE} and \aegean. This initial search yielded $>$122\,000 sources relatively evenly distributed across the field with a noticeable difference in the source density distribution in the central bar region structure (because of confusion and extended emission) and at the edge of the mosaic. The produced \ac{RMS} image was visually inspected and, as expected, because of the primary beam attenuation, the \ac{RMS} noise shows a rapid rise at the edge of the mosaic and consequently a fall-off in sensitivity.

\begin{figure*}
		\centering
     \includegraphics[]{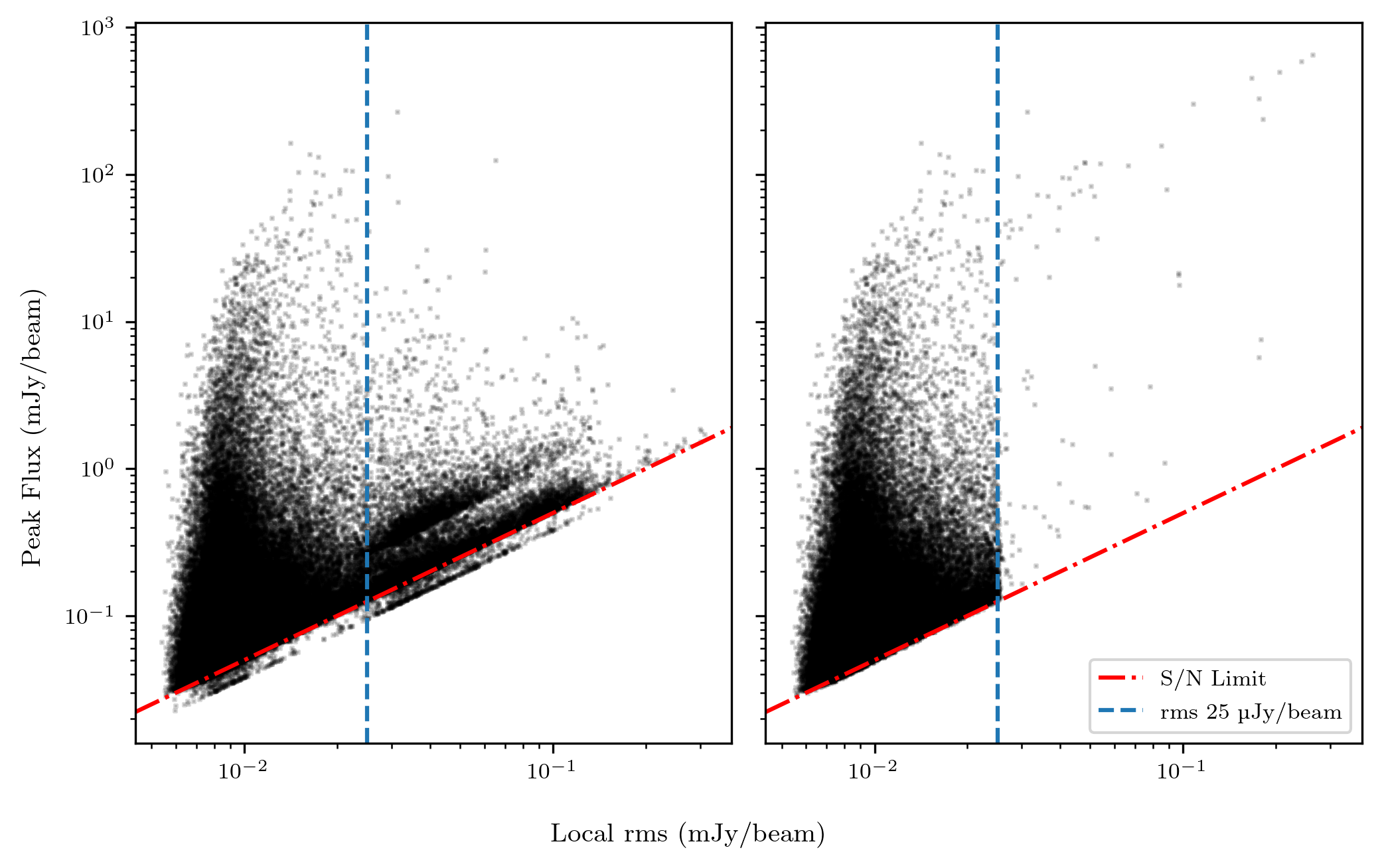}
			\caption{The plot of local \ac{RMS} values vs. peak flux density of the associated source from the initial catalogue of $>$122\,000 sources (left) and the final catalogue (right). 
			The distribution appears smooth except for a visible density excess in the region \ac{RMS}~$>$25~$\mu$Jy~beam$^{-1}$ (vertical blue dashed line).
			We point to a set of detections below the applied detection limit (S/N=5; red dash-dot line) which were deleted from the final catalogue. 
			}
        \label{fig:rms_peak}
\end{figure*}

In Fig.~\ref{fig:rms_peak} we show the plot of local \ac{RMS} values vs. peak flux density of the associated sources from the initial catalogue. As can be seen, the distribution appears smooth except for a visible density excess in the region with \ac{RMS}~$>$25~$\mu$Jy~beam$^{-1}$ (vertical blue dashed line). 
We visually inspected about 10~per~cent of sources from this excess. We noticed that all sources in the excess are located close to the edge of the mosaic (i.e. no sources are located in the high \ac{RMS} region of the central bar) and that more than 50~per~cent of these sources appear to be false detections. Also, we noted a set of detections below the applied detection limit (S/N=5; red dash-dot line in Fig.~\ref{fig:rms_peak}). These detections are relatively evenly distributed across the image mosaic. After visual inspection, we found that similar to the excess set, it consists of many false positive detections.

In the case of bright compact extended sources (BCE; sources with angular sizes larger than resolving beam $>\times$1.5~beam size or from $\sim$10~arcsec to 7.5~arcminutes; Fig.~\ref{fig:regions}; e.g. resolved background sources, blended objects), the \aegean\ algorithm will attempt to fit multiple Gaussian components to the detected `island' of emission (for more details see the \cite{hancock2018} or \aegean\ documentation\footnote{https://github.com/PaulHancock/Aegean/wiki}). This poses a concern that the complex emission regions would get broken into multiple sources which would create a problem when comparing flux densities from this new catalogue to values from previous or future lower-resolution catalogues. Another issue we noticed by visual inspection is that a significant number of faint point sources in the vicinity of BCE sources have not been picked up by the \aegean\ algorithm (i.e. false negative detections). The most likely reason is that the averaged \ac{RMS} noise, close to the bright and complex emission, is significantly higher than the real noise measured outside these complex regions, and therefore the real but faint sources fail to reach the S/N ratio needed for detection. We visually assessed such deconvolution errors and determined that the deconvolution is sufficiently good that the fainter sources near bright ones are real and we kept them in the catalogue.

To resolve the above-identified issues, we produced the final catalogue (see Table~\ref{tab:main}) using the following steps:
\begin{enumerate}
    \item Using the initial \ac{RMS} map we produced a masked channel~0 image where the region with averaged \ac{RMS}$>$25~$\mu$Jy~beam$^{-1}$ and at the edge of the field 
    was masked-off, i.e. it was ignored in the \ac{RMS} averaging and source detection. It is important to note that the inner mosaic regions with high \ac{RMS} were not excluded from the process;
    \item We used {\sc Blobcat}, the extended source ($>$10~arcsec) detection and extraction software \citep{2012MNRAS.425..979H} first to detect and then create masks for all BCE sources in the mosaic (more details and results are given at the end of Section~\ref{sec:cats}). Masks are applied to the channel~0 image in the same way as described in (i);
    \item We applied \textsc{BANE} and \aegean\ algorithms (using default parameters for both packages) to the masked channel~0 image;
    \item Only sources with flag 0 set by the \aegean\ finding algorithm (i.e. no fitting error reported) were used;
    \item Detected sources with S/N ratio $<$5 (\aegean\ applied detection limit) were excluded from the catalogue.
\end{enumerate}


\begin{table*}
\small
 \begin{center}
\caption{Example of the MeerKAT broad-band (1283.8\,MHz) point source catalogue of \TOTALNUMBERCHZEROPEAKFLUX\ objects in the directions of the \ac{SMC} with its positions, integrated flux densities with associated uncertainty, and spectral index that is calculated using other sub-band flux densities. The columns provided are as follows: (1) Source name; (2 and 3) source position; (4 and 5) source peak and integrated flux density; (6) spectral index ($\alpha$) fit to all available source components with the associated error; (7) reduced $\chi^2_\nu$ values obtained in the spectral index fit; (8) number of flux densities used in the spectral index fit. The full catalogue is provided through the \textsc{VizieR} service and as supplementary material. In the full catalogue, we also list peak and integrated flux densities for each sub-band.}
\label{tab:main}
\begin{tabular}{lllccccccccccc}
\hline
Name          & RA (J2000)      &  Dec (J2000)  & S$_{\rm peak\,1283.8\,MHz}$     & S$_{\rm int\,1283.8\,MHz}$       &   $\alpha\pm\Delta\alpha$ & $\chi^2_\nu$ & n\\
              &                 &               &(mJy~beam$^{-1}$)& (mJy)         &  &  \\
\hline
  J001553--743313 & 00:15:52.8 & --74:33:13 & 0.132$\pm$0.025 & 0.107$\pm$0.026 & ...             & ...  & ...\\
  J001614--743535 & 00:16:13.9 & --74:35:35 & 0.482$\pm$0.023 & 0.545$\pm$0.031 & ...             & ...  & 9\\
  J001625--742633 & 00:16:24.9 & --74:26:33 & 0.235$\pm$0.022 & 0.243$\pm$0.026 & ...             & ...  & 3\\
  J004115--733600 & 00:41:14.8 & --73:36:00 & 2.712$\pm$0.042 & 2.827$\pm$0.050 & --0.75$\pm$0.02 & 0.20 & 12 \\
  J004347--750932 & 00:43:46.9 & --75:09:32 & 1.488$\pm$0.007 & 1.533$\pm$0.009 &   0.27$\pm$0.01 & 0.28 & 12 \\
  J004516--730607 & 00:45:15.6 & --73:06:06 & 1.792$\pm$0.013 & 1.823$\pm$0.015 & --0.75$\pm$0.03 & 0.13 & 12 \\
  J004613--730704 & 00:46:13.2 & --73:07:04 & 3.485$\pm$0.062 & 4.137$\pm$0.086 & --0.82$\pm$0.02 & 0.09 & 12 \\
  J004852--731907 & 00:48:51.8 & --73:19:07 & 1.455$\pm$0.047 & 1.454$\pm$0.054 & --0.77$\pm$0.05 & 0.24 & 12 \\
  J004857--730954 & 00:48:57.2 & --73:09:53 & 1.369$\pm$0.027 & 2.395$\pm$0.055 & --0.02$\pm$0.04 & 0.33 & 12 \\
  J005049--724940 & 00:50:48.5 & --72:49:39 & 1.247$\pm$0.059 & 1.163$\pm$0.064 & --0.45$\pm$0.03 & 0.06 & 12 \\
  J005242--723621 & 00:52:42.2 & --72:36:20 & 3.516$\pm$0.035 & 3.494$\pm$0.040 & --0.70$\pm$0.02 & 0.20 & 12 \\
  J011104--731155 & 01:11:03.9 & --73:11:55 & 1.473$\pm$0.015 & 1.522$\pm$0.018 & --0.92$\pm$0.03 & 0.19 & 12 \\
  J012629--751001 & 01:26:29.0 & --75:10:00 & 1.527$\pm$0.015 & 1.563$\pm$0.017 & --0.78$\pm$0.03 & 0.27 & 12 \\
%
\hline
  \end{tabular}
 \end{center}
\end{table*}

\subsubsection{Sub-band images}
\label{sec:bandsdet}

Similarly, as in Section~\ref{sec:broadbandsdet}, the initial image catalogue in each band was generated by blind-fitting with default \textsc{BANE} and \aegean\ parameters. Prior to fitting, we masked each sub-band image using the same mask created in the channel~0 fitting procedure. The resulting catalogues were then cross-matched, using \textsc{Topcat} \citep{2005ASPC..347...29T}, against the catalogue produced in channel~0 and within an 8~arcsec radius. We examined by eye all unmatched sources (fewer than 100 per sub-band). Most of them appear to be artifacts at the edge of the masking regions or close to bright sources. We did not find any reliable detection in the set of unmatched sources and so they are removed (about 1100 sources) from the catalogue.

\subsection{Catalogues}
\label{sec:cats}

Our base catalogue consists of \TOTALNUMBERCHZEROPEAKFLUX\ point sources found in the broad-band 1283.8~MHz image (Table~\ref{tab:main}). The number of point sources found in each of the sub-band images is listed in Table~\ref{tab:subchannumbers}.

\citet{2012MNRAS.422.1812H} suggested that the \aegean\ detection reliability is $>$98~per~cent at S/N$>$5 and 100~per~cent at S/N$>$10. We visually inspected our catalogues and we estimate that $\approx$1~per~cent of the catalogued sources could be false positives (artifacts). 


Also, we present a catalogue of compact extended sources (BCE) detected with {\sc Blobcat} and used for masking. We used the following parameters in the {\sc Blobcat} detection process:
\begin{itemize}
        \item The \ac{RMS} was set to a constant value of15 $\mu$Jy~beam$^{-1}$. This value was intentionally chosen to be higher than the median \ac{RMS} value in the image (11$\mu$Jy~beam$^{-1}$) in order to avoid collecting very faint large extended sources.
        \item Detection threshold was 7$\times$RMS.
        \item For size limits we used $>$300 pixels (1~pixel = 1.5~arcsec; 7.5~arcmin) not `thinner' along RA/Dec than 20~arcsec. Extended objects outside these size limits were filtered out.
        \item The edge buffer was selected to be 10 pixels (15~arcsec; i.e. two blobs can not be closer than 20 pixels or 30~arcsec).
    \end{itemize}

\begin{figure}
		\centering
			\includegraphics[width=\columnwidth]{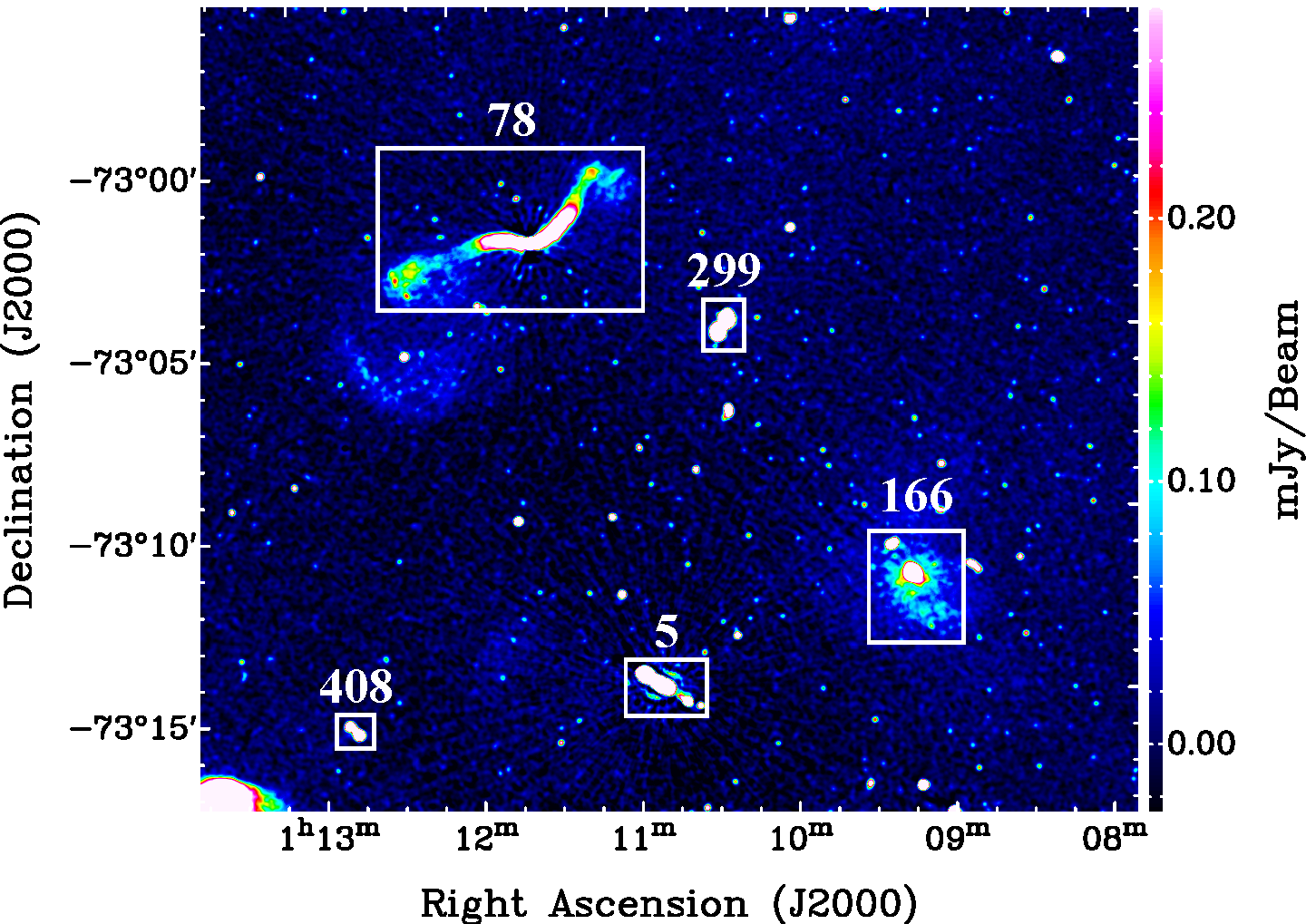}
			\caption{MeerKAT image of the \ac{SMC} field showing five BCE regions. The number over each region indicates the region's ID as tabulated in Table~\ref{tab:bcecat}.}
        \label{fig:regions}
\end{figure}

We visually inspected the catalogue of BCE sources as the image artifacts and large extended regions ($>$7.5~arcmin) in the \ac{SMC} bar were excluded from that catalogue. During the visual inspection, we found another \MANUALBCE\ extended objects which were manually added to the final BCE catalogue of sources. The final catalogue contains \TOTALBCE\ BCE objects (Table~\ref{tab:bcecat} and for example of BCE sources see Fig.~\ref{fig:regions}). The catalogue contains an ID (as produced by {\sc Blobcat}) and the position (RA and Dec) of the extended object's centroid. We also provide a DS9 region file (only in the electronic version of this paper) containing encapsulating regions for objects from this catalogue. Other observational parameters (e.g. classification, flux densities and angular sizes) will be published elsewhere. However, to measure flux densities of \acp{SNR} and \ac{SNR} candidates (Section~\ref{SNR_samp}) we used the method described in \citet[][Section 2.4]{2019PASA...36...45H,2019PASA...36...48H} which includes careful region selection that excludes all obvious point sources within each given \ac{SNR}.

\begin{table}
    \caption{Example of the catalogue of compact extended (BCE) sources. We show only the first 10 out of \TOTALBCE\ BCE detections. We provide an ID (as produced by {\sc Blobcat}) in Column~1 and the position of the centroid. We also provide a DS9 region file (only in the electronic version of this paper) containing encapsulating regions for objects from this catalogue.}
    \centering
    \begin{tabular}{ccccc}
    \hline
        ID   & RA (J2000) & Dec (J2000)\\
    \hline
          5  & 01:10:48.9 & --73:14:26\\
          9  & 01:02:13.2 & --75:46:57\\
          10 & 00:42:22.3 & --75:48:41\\
          19 & 00:43:29.4 & --70:41:48\\
          22 & 01:28:12.6 & --75:13:00\\
          24 & 01:14:29.1 & --73:21:46\\
          27 & 00:38:08.6 & --73:50:25\\
          28 & 00:31:21.0 & --70:36:48\\
          29 & 00:21:50.3 & --74:15:10\\
          30 & 00:24:09.6 & --73:57:07\\
    \hline
    \end{tabular}
    \label{tab:bcecat}
\end{table}

\begin{figure}
	\centering
		\includegraphics[width=\columnwidth]{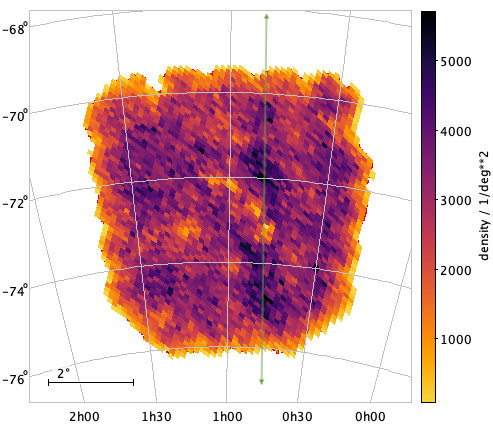}
		\includegraphics[width=\columnwidth]{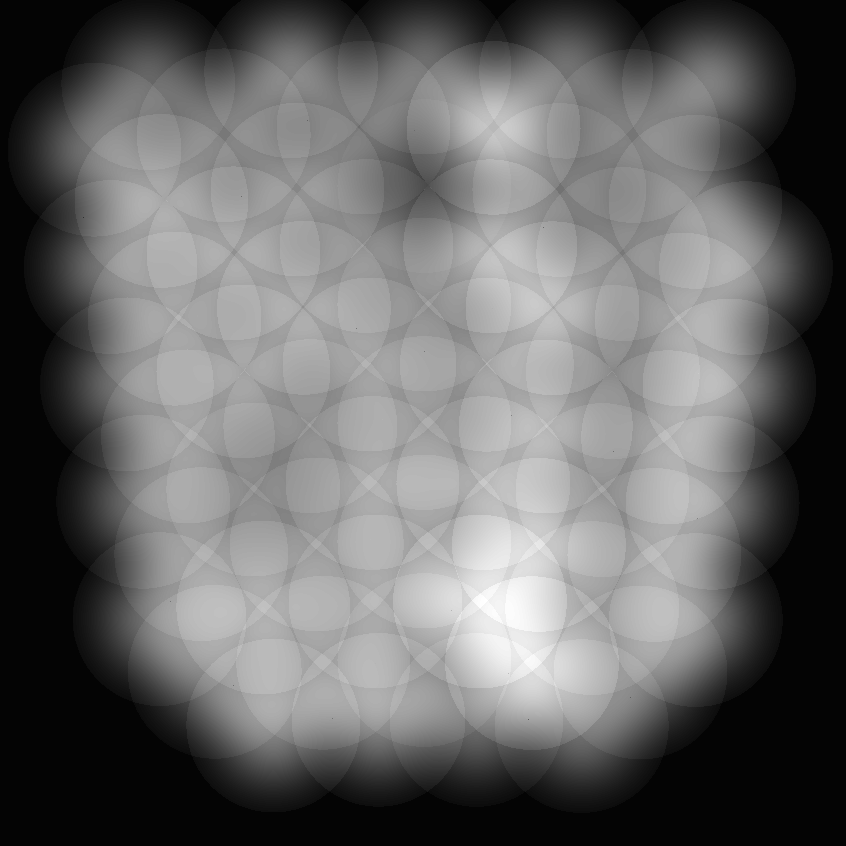}
        \caption{{\bf Top:} The MeerKAT \ac{SMC} field point source density image. A vertical green line indicates source overdensity. {\bf Bottom:} The summed weight ($P$ in Eq. \ref{mosaic_eq}) image, with circles indicating individual pointing positions.}
		\label{fig:source_dens}
\end{figure}

In Fig.~\ref{fig:source_dens} (top) we show the point source density image based on all sources catalogued in Table~\ref{tab:main}. We investigated a `vertical wall' of source over-density (marked with the vertical green line) and found that it corresponds to a region of excess in the weight image (Fig.~\ref{fig:source_dens} bottom). This is the area where we naturally pick up more sources as the exposure is deeper. This is a simple sensitivity issue, not an artifact. At the same time, this confirms that our source search software \citep[\aegean;][]{hancock2018} is working as  expected. Similarly, the lower source densities in the \ac{SMC} bar and the Eastern Wing \citep{1996ASPC..112...91F,2009ApJ...690L..76G} correspond to regions of source confusion where the software is appropriately more conservative. 


\subsection{Spectral indices}
 \label{sec:SI}

For all detected objects with at least two detections in sub-channel images, we can calculate a nominal spectral index $\alpha$ defined as \mbox{$S_{\nu}\propto\nu^{\alpha}$}. 
We use the \textsc{scipy.odr} \citep{2020SciPy-NMeth} code to fit the spectral index, and only accept fits that successfully converged. This produces nominal spectral index values for \TOTALSPINDEX\ objects. Following the results from recent deep radio surveys such as EMU-PS \citep{2021PASA...38...46N} and ATCA/ASKAP-LMC \citep{2021MNRAS.507.2885F,2021MNRAS.506.3540P}, the overall expectation is that the radio source spectral index will be in a range of --2.5$<\alpha<$+1.5 where point-like \ac{AGN} dominate the population. 



\begin{figure*}
		\centering
			\includegraphics[width=0.5\textwidth, trim=5 0 35 0,clip]{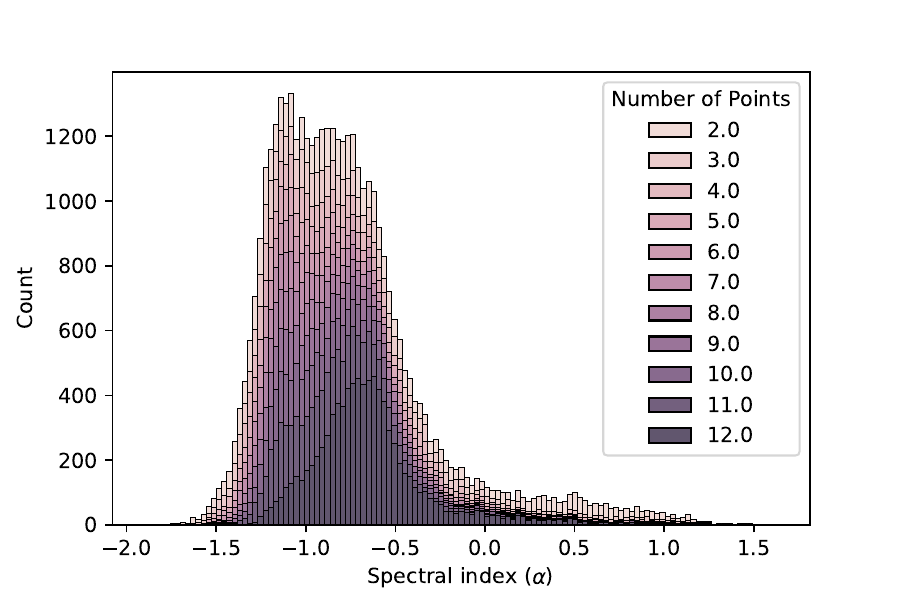}~
			\includegraphics[width=0.5\textwidth, trim=5 0 35 0,clip]{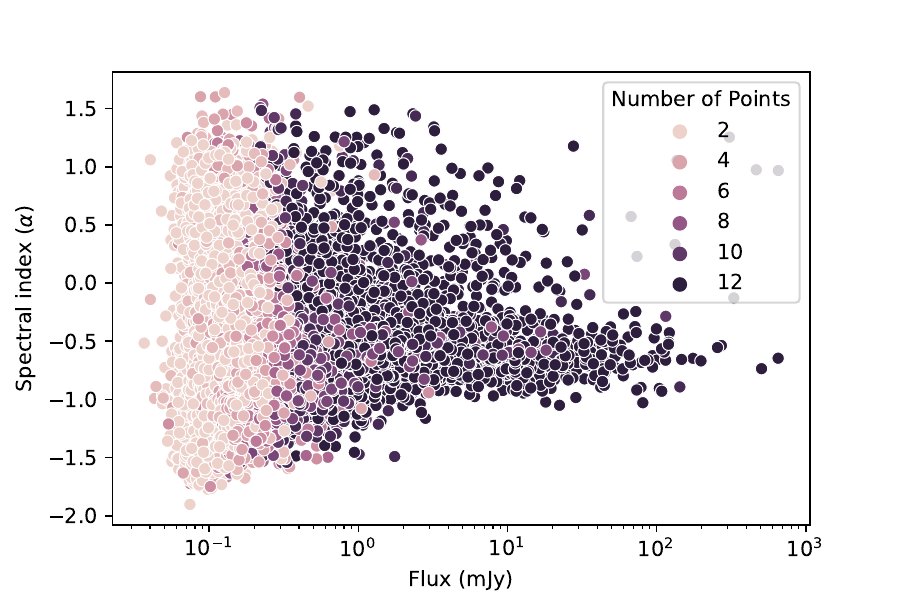}
        \caption{{\bf Left}: point source spectral index distribution. This is a stacked histogram, with colour representing the number of channels used in the spectral index fit. {\bf Right}: distribution of spectral indices per integrated flux density, with colour representing the number of channels used in the spectral index fit. In both figures, light colours represent small numbers, and dark colours represent high.}
        \label{fig:ch5_check_ch_missing2}
\end{figure*}

\begin{figure}
    \centering
    \includegraphics{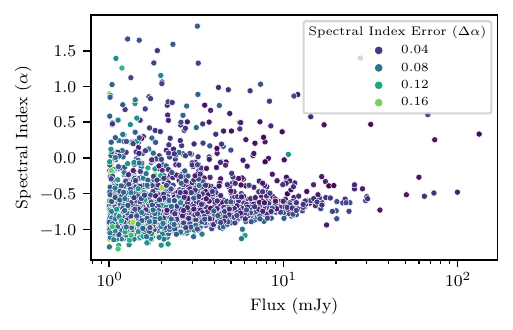}
    \caption{Distribution of spectral indices per integrated flux density for 2357 sources with 12 sub-band flux density measurements, broad-band flux density $>$1~mJy, $\chi_\nu^2$<10, and $\Delta \alpha <0.2$. The colours represent the spectral index fit error according to the inset box.}
    \label{fig:2357}
\end{figure}

We investigate the \ac{SMC} MeerKAT catalogue sources' spectral index distribution obtained from sub-band flux density measurements (see Fig.~\ref{fig:ch5_check_ch_missing2}).
Not surprisingly, we find that increasing the number of sub-bands included increases the quality of
the fit. Using only those sources with 12 sub-band measurements yields a mean spectral index of $\alpha=-0.65 $ (STDEV=0.37), with a median of $\alpha=-0.74$ (see Fig.~\ref{fig:2357}) . This is very similar to EMU-PS \citep{2021PASA...38...46N} and ATCA/ASKAP-LMC \citep{2021MNRAS.507.2885F,2021MNRAS.506.3540P} results. We also note that the spectral index distribution is similar to previous results with a `long tail' of sources with flat and inverted spectrum (0$<\alpha<$1.5). We suggest that the subset of these sources with reliable spectral index values are most likely to be \ac{GPS} and variable quasars \citep{2018MNRAS.477..578C}. 

However one should not forget that these flux density measurements are made within the relatively narrow bandwidth of 800~MHz, where a small change (or error) in sub-band flux density leads to large changes and unrealistic estimates in spectral index. 
There is a large range of unrealistic spectral index errors (0.2<$\Delta\alpha$<3) which indicates poor fitting. This is mainly for weaker sources ($<$1~mJy) and those with a small number of sub-band flux density measurements.

The highest-quality spectral index values arise from objects with flux density measurements in all 12 sub-bands, broad-band flux density $>$1~mJy, $\chi_\nu^2$<10, and $\Delta \alpha < 0.2$. There are 2357 such sources, included in the main catalogue (Table~\ref{tab:main}) and  displayed   in Fig.~\ref{fig:2357}. 
Some sources outside these constraints could have reliable spectral index values, but we are conservative in our presentation to limit contamination by unreliable values. For other point sources of interest, the individual sub-band values should be evaluated on a case-by-case basis.

\begin{figure}
    \centering
    \includegraphics[width=\columnwidth,trim=0 30 0 50,clip]{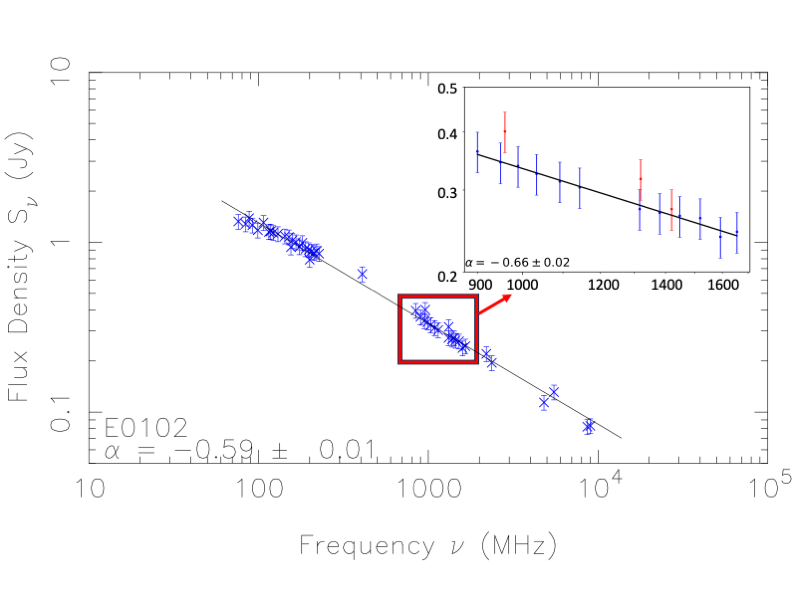}
\caption{Radio continuum spectrum of \ac{SMC} \ac{SNR} 1E0102.2--7219. The red box indicates measured MeerKAT flux densities with assumed 10\% uncertainty. Other flux density measurements are from Alsaberi et al. (in prep.). The best power-law weighted least-squares fit is shown (thick black line), with the spatially integrated spectral index $\alpha= -0.59 \pm 0.01$. The inset in the top right corner shows MeerKAT in-band measurements with $\alpha= -0.66 \pm 0.02$ which is in good agreement with the overall spectral index estimate. Three flux density measurements marked in red in this inset are from \ac{ASKAP} and \ac{ATCA} surveys. 
    }
    \label{fig:SI2}
\end{figure}

Finally, we present the spectrum for the well-known \ac{SMC} \ac{SNR} 1E0102--72 (MCSNR\,J0104--7201), an extended object with a diameter of 40~arcsec, finding consistency with previously measured flux densities (see Fig.~\ref{fig:SI2} and also Alsaberi et al., in prep.). This provides  some confidence in the reported flux density measurements for  extended BCE sources.

\subsection{Comparison with previous catalogues}
\label{scompare}

\subsubsection{Astrometry}
\label{scompareA}
\begin{figure}
	\centering
		\includegraphics[width=0.85\columnwidth]{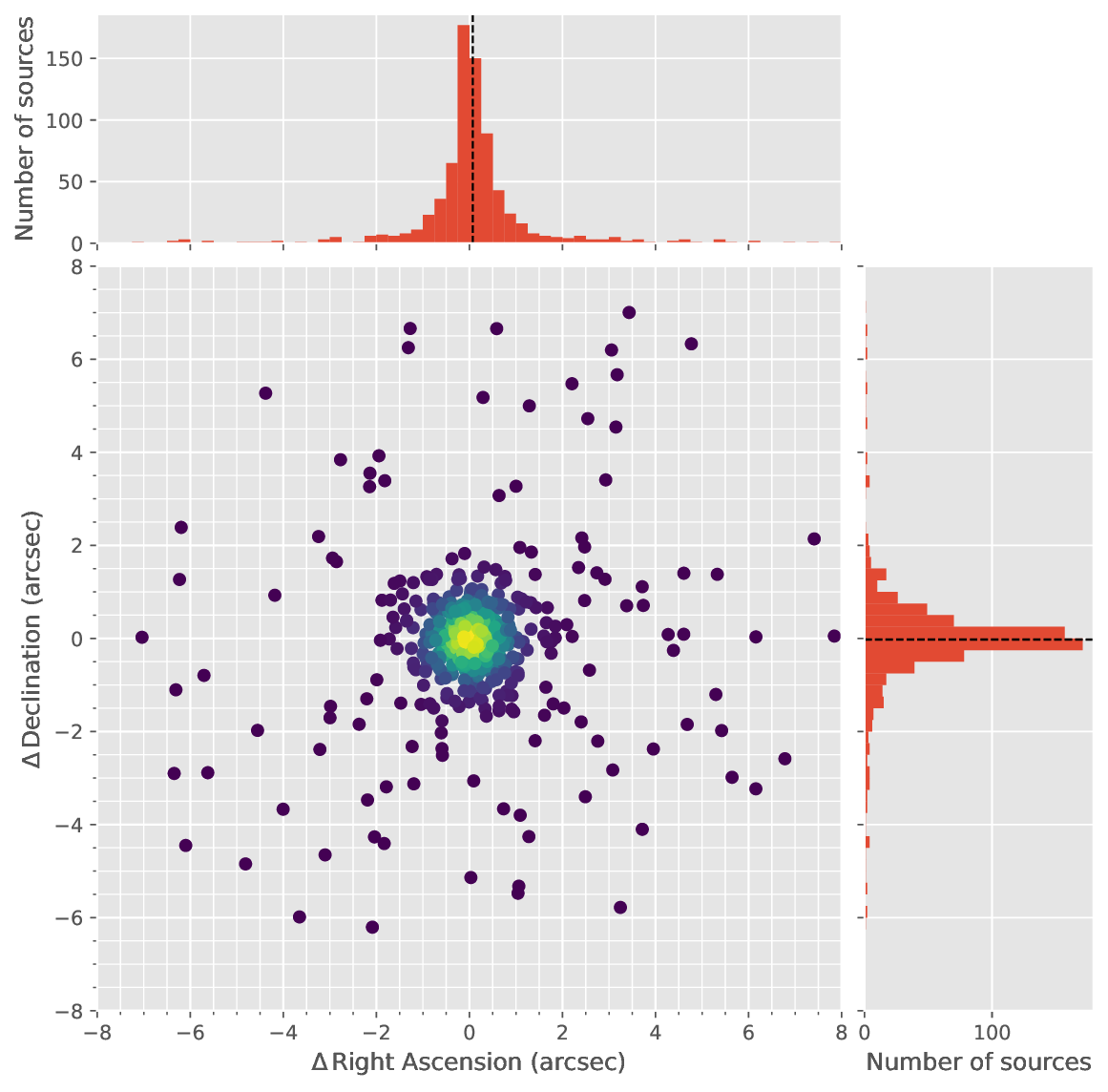}
        \includegraphics[width=0.85\columnwidth]{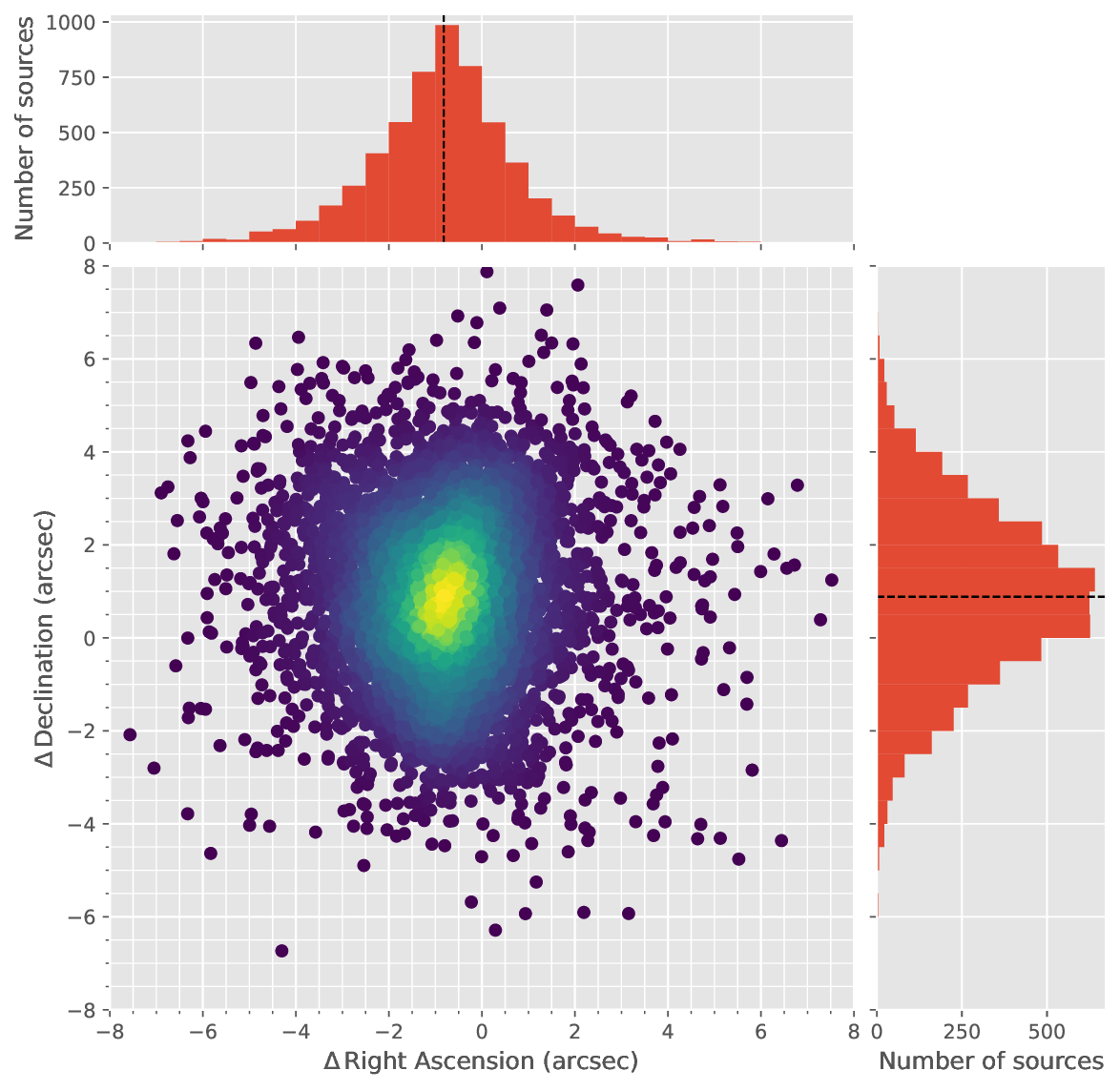}
        \includegraphics[width=0.85\columnwidth]{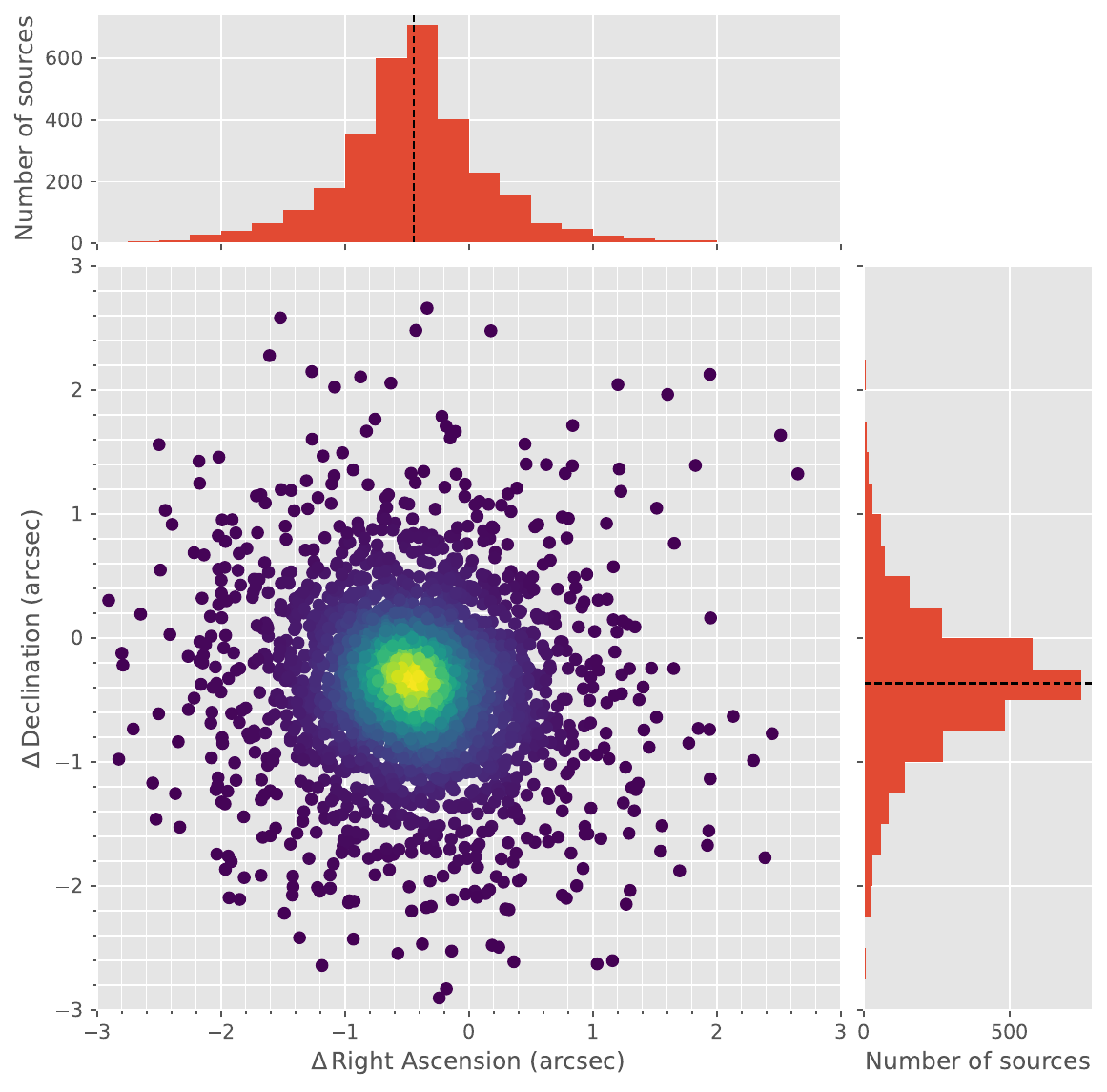}
        \caption{
        {\bf Top:} RA and DEC offsets between MilliQuas and MeerKAT catalogues (745 sources), with mean $\Delta$RA=+0.07~arcsec (STDEV=1.40) and $\Delta$DEC=$-0.02$~arcsec (STDEV=1.33). 
        {\bf Middle:} Positional difference (\ac{ASKAP} -- MeerKAT) of 5652 sources found in both catalogues. The mean offsets are
        $\Delta$RA=$-0.82$~arcsec (STDEV=1.59) and $\Delta$DEC=+0.89~arcsec (STDEV=1.84). 
        {\bf Bottom:} Positional difference (GASKAP -- MeerKAT) of 3066 sources found in both catalogues. The mean offsets are
        $\Delta$RA=$-0.45$~arcsec (STDEV=0.63) and $\Delta$DEC=--0.37~arcsec (STDEV=0.62).
        }
		\label{fig:position}
\end{figure}

We analyse our point source catalogue (Table~\ref{tab:main}) by examining the accuracy of source positions. We compared our catalogue source positions with the existing \ac{ASKAP} catalogue \citep{2019MNRAS.490.1202J} at 1320~MHz (5652 sources) as well as the MilliQuas catalogue (see Section~\ref{astrometry} and Fig.~\ref{fig:position}). We found an offset of $\sim$1~arcsec when compared to \ac{ASKAP} (Fig.~\ref{fig:position} middle) but excellent matching with the precise optical MilliQuas catalogue ($<$0.1~arcsec for 745 sources within 4~arcsec; Fig.~\ref{fig:position} top). The offset with the \ac{ASKAP} early science observations is not unexpected given that  \ac{ASKAP} (at the time) was still being commissioned \citep[for more details see][]{2019MNRAS.490.1202J}. 

For radio pulsars that can be identified (all point radio sources; see Section~\ref{sec:psr}), the MeerKAT position agrees with their timing position to the sub-arcsecond level (see Table~\ref{tab:psr}). 

Finally, we compare our  MeerKAT catalogue with the recent Galactic \ac{ASKAP} \citep[GASKAP;][]{2022PASA...39...34D} survey of the \ac{SMC} at 1220~MHz (bandwidth = 18.5~MHz, beam size = 8~arcsec) and its 4311 sources found using the SELAVY source finder \citep{2012PASA...29..371W}. This catalogue comprises point and complex and extended (BCE) sources\footnote{For images and tables see CASDA (\url{ https://data.csiro.au/domain/casdaObservation/}) scheduling block 10941.}. We found 3066 sources in common (within a search radius of 3~arcsec) and a small offset of $\Delta$RA=--0.45~arcsec (STDEV=0.63) and $\Delta$DEC=--0.37~arcsec (STDEV=0.62) (see Fig.~\ref{fig:position} bottom), indicating good agreement between the two survey source positions.

\subsubsection{Flux density}
\label{scompareF}

\begin{figure}
		\centering
			\includegraphics[width=\columnwidth,trim=0 150 0 0,clip]{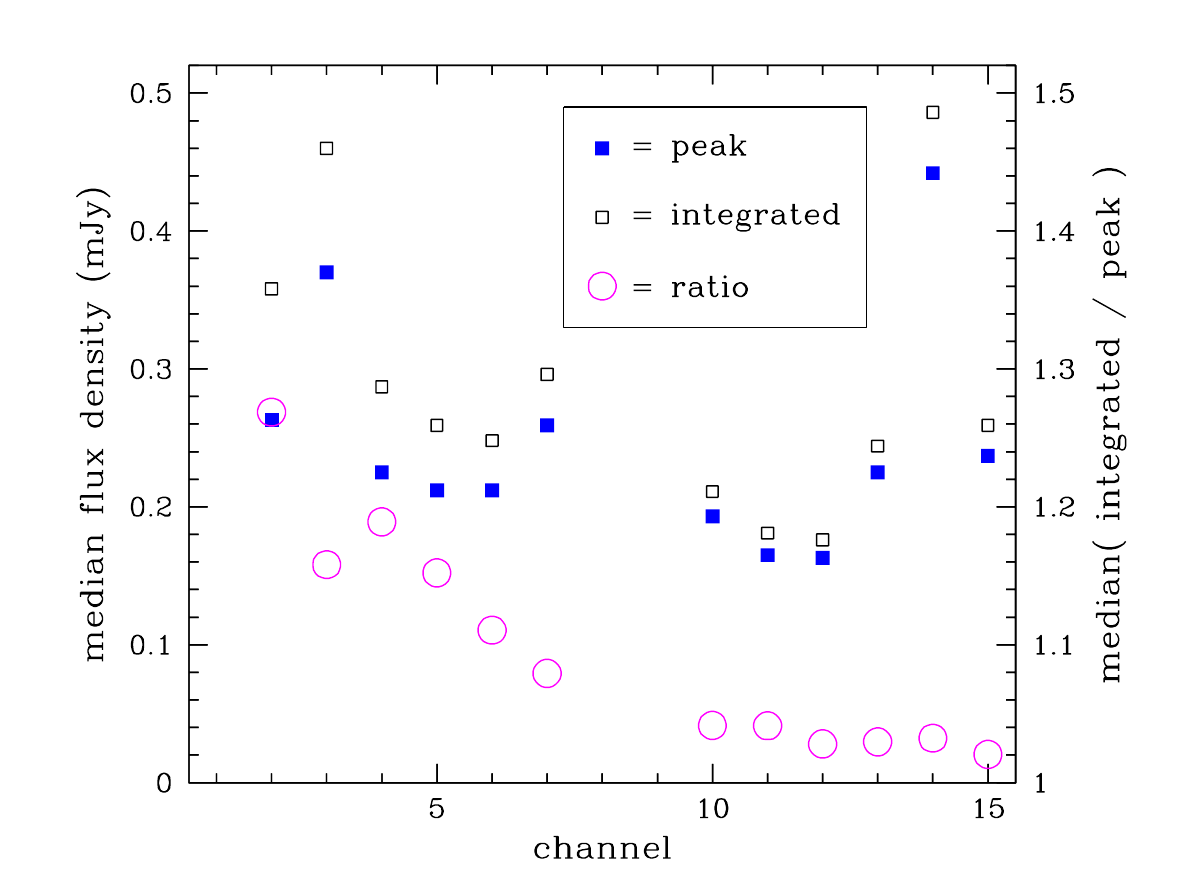}
        \caption{The median flux density in each channel, separately for peak (blue filled squares) and integrated (open squares) flux density for all sources found in this study. 
        Note also that the variation in both peak and integrated flux density is due to varying noise levels and hence sensitivity, where noise is larger the median flux density is higher. 
        }
		\label{fig:jacco}
\end{figure}

In Fig.~\ref{fig:jacco} we show the median flux density in each channel, separately for peak and integrated flux density for all sources found in this study. This shows the varying sensitivity (depth) in each of these channels, but it also shows a trend with the frequency of the ratio of integrated-to-peak flux. This ratio is observed to diminish with frequency, which is contrary to what one might expect if more sources become resolved at a higher frequency. We suggest that the opposite is the case and more extended radio emission is included at lower frequencies.

\begin{figure*}
		\centering
			\includegraphics[width=0.9\columnwidth]{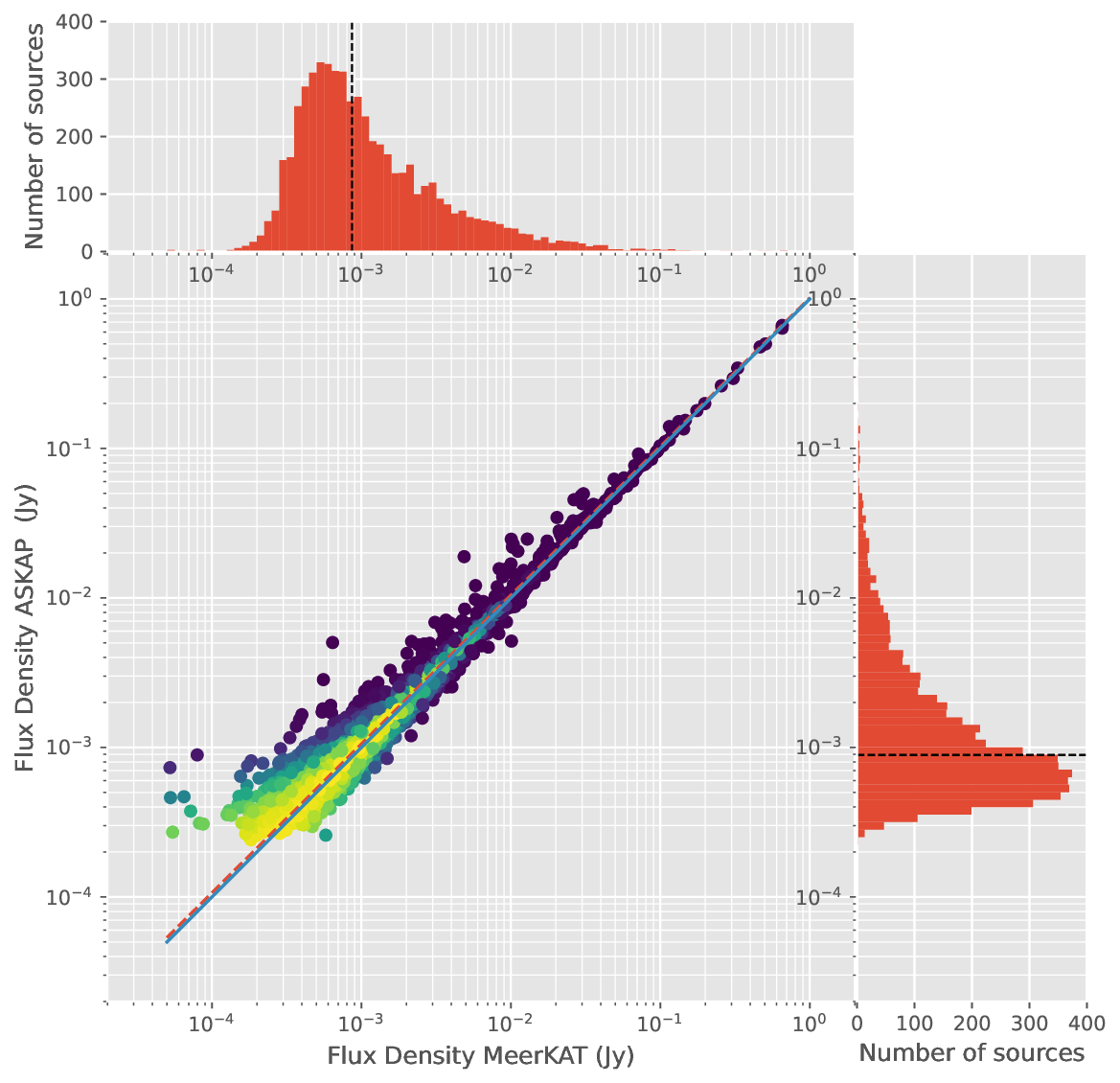}\\
			\includegraphics[width=0.9\columnwidth]{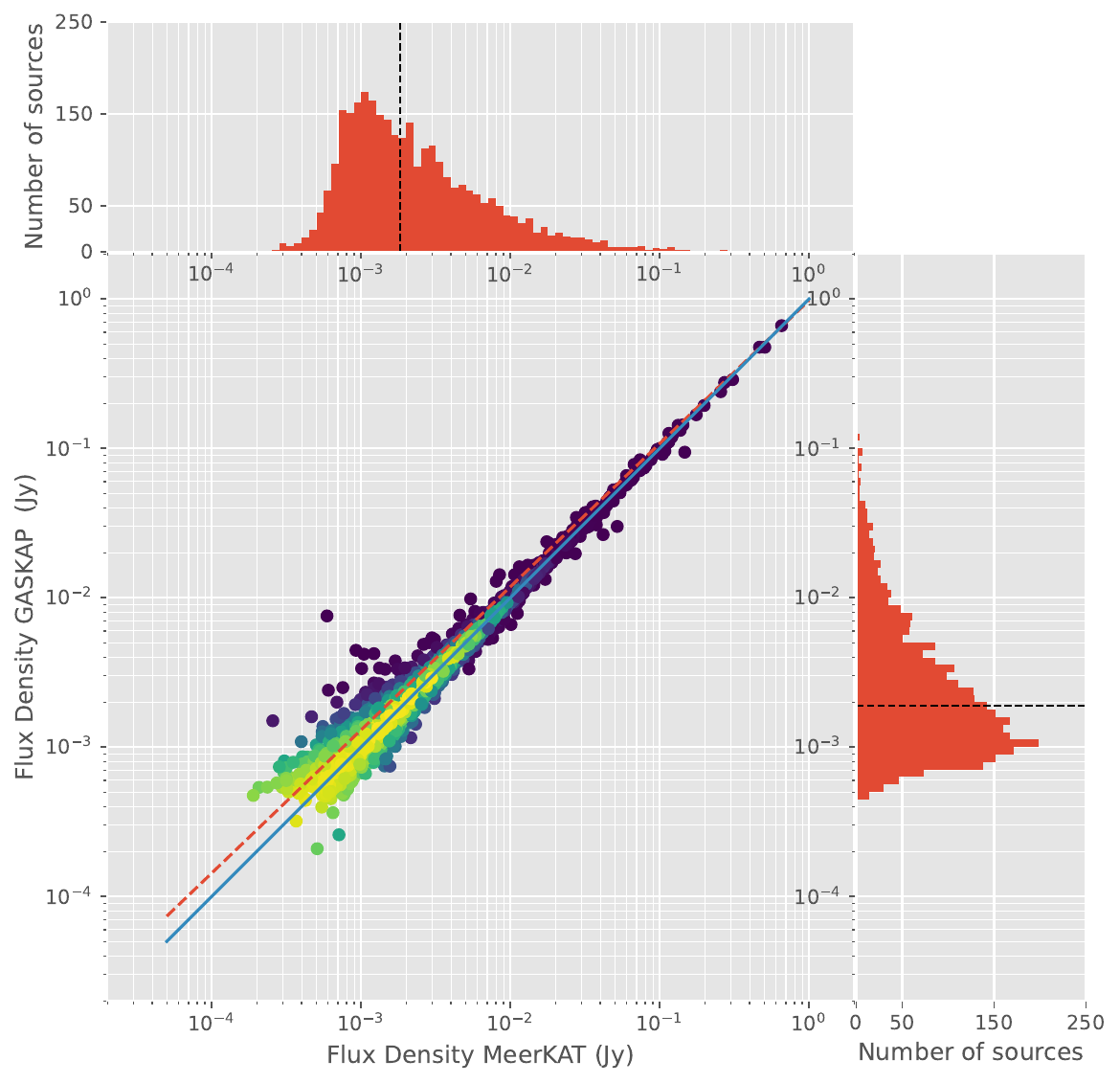}\\
        \caption{
        {\bf Top:} Integrated flux density comparison of 5652 sources found in both the 1283.8\,MHz (MeerKAT) and  1320\,MHz \ac{ASKAP} catalogues \citep{2019MNRAS.490.1202J}. The best-fit slope (linear) is 1.016$\pm$0.001 (dashed orange) while the blue line represents a 1-to-1 ratio (see Section~\ref{scompareF}). The points are colour coded to indicate local density, yellow for high density to purple for low density. The source integrated flux density distributions are shown in the side and top panels, with the black dashed line at the median integrated flux density.
        {\bf Bottom:} Integrated flux density comparison of 3066 sources found in both the 1283.8\,MHz (MeerKAT) and  1420\,MHz GASKAP catalogues \citep{2022PASA...39...34D}. The best-fit slope (linear) is 0.979$\pm$0.002 (dashed orange) while the blue line represents a 1-to-1 ratio (see Section~\ref{scompareF}). The points are colour coded to indicate local density, yellow for high density to purple for low density. The source integrated flux density distributions are shown in the side and top panels, with the black dashed line at the median integrated flux density. 
        }
		\label{fig:meerkat_vs_askap}
\end{figure*}

We also compared \ac{ASKAP} \citep{2019MNRAS.490.1202J} and MeerKAT integrated flux densities (see Fig.~\ref{fig:meerkat_vs_askap} top) for the same population of point sources in common to both surveys. We found no significant discrepancy for S/N above 50 (MeerKAT). 
Similarly, our comparison of flux densities for sources in common between the GASKAP and our MeerKAT catalogue indicates good matching as shown in Fig.~\ref{fig:meerkat_vs_askap} (bottom).

Our \ac{SMC} MeerKAT point source catalogue  flux density distributions including corresponding errors are also as expected i.e. following a power-law distribution
and very similar to other recent surveys such as EMU-PS \citep{2021PASA...38...46N}, ASKAP-LMC \citep{2021MNRAS.506.3540P} and ATCA-LMC \citep{2021MNRAS.507.2885F}. The flux density differences compared to those of the \ac{ASKAP} and GASKAP source distributions are at the $\sim\pm$2~per~cent level. 

\begin{figure*}
		\centering
			\includegraphics[width=\textwidth,angle=-90,trim= 0 120 0 0, clip]{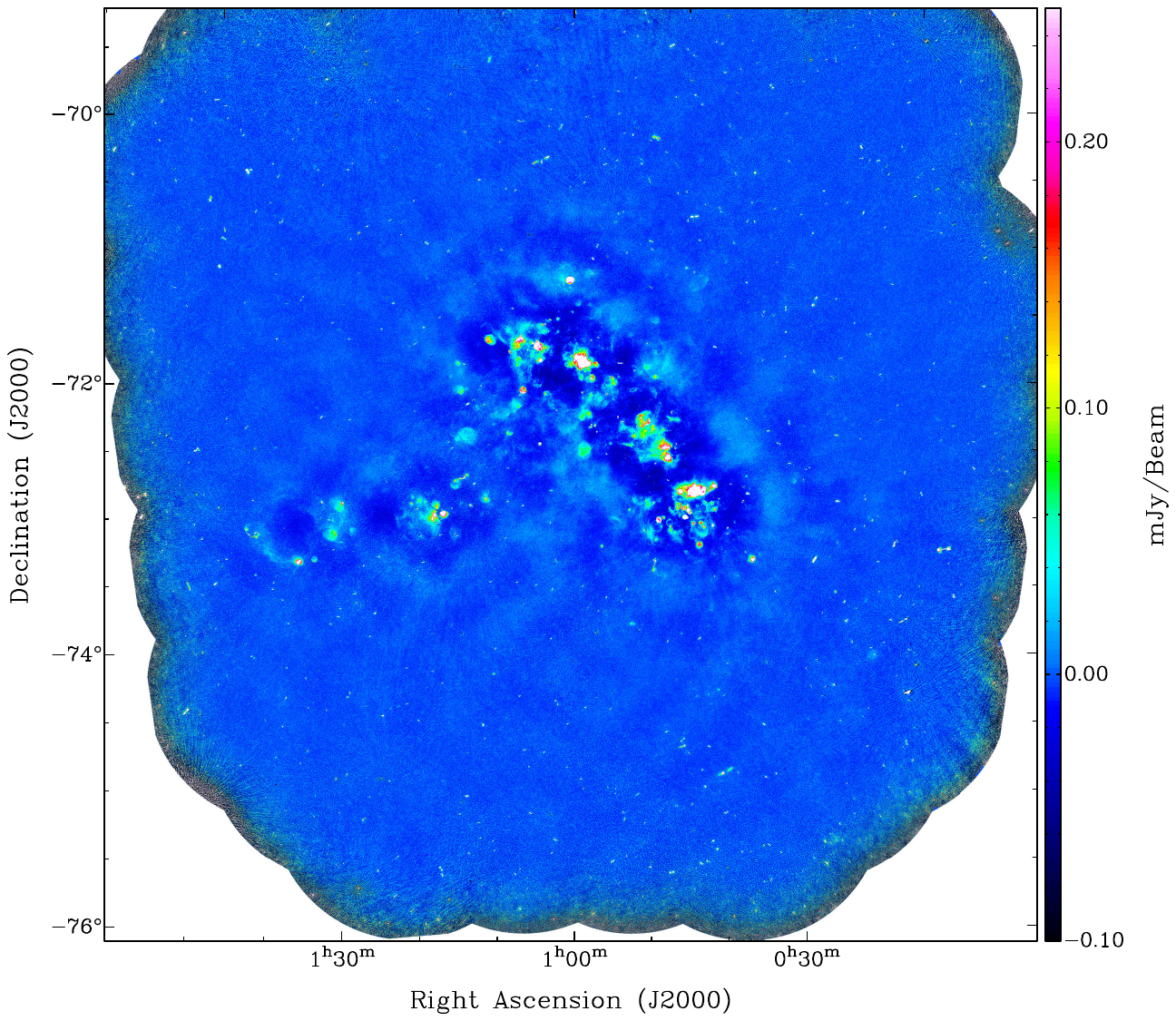}
        \caption{The MeerKAT residual image of the \ac{SMC} field after subtracting all catalogued point sources listed in Table~\ref{tab:main}. The colour bar (linear scale) on the right-hand side is in units of mJy~beam$^{-1}$.}
        \label{fig:residual}
\end{figure*}

We also show the residual image (Fig.~\ref{fig:residual}), where all catalogued points sources (Table~\ref{tab:main}) have been subtracted.

\section{Results}
 \label{sec:results}
 

Based on recent findings from  \ac{ATCA} and \ac{ASKAP} surveys of the \ac{LMC} \citep{2021MNRAS.507.2885F,2021MNRAS.506.3540P}, we expect that the vast majority of point radio sources catalogued here are unrelated to the \ac{SMC}, with only a small number of intrinsic \ac{PNe} ($\sim$40), compact \HII~regions and \ac{YSOs} \citep[$\sim$40; ][]{2013MNRAS.428.3001O} as well as foreground stars ($\sim$10). 
This is backed up by the paucity of sources with thermal spectral indices.
However, we expect that about 50\% of the \TOTALBCE\ compact extended sources are intrinsic \ac{SMC} objects such as \acp{SNR} and complex \HII~regions, and the other 50\% are  likely to be background radio \acp{AGN} with clear radio jets.

The \ac{SMC} is the prime target for  various multi-frequency surveys and therefore a vast amount of data is readily available for cross-comparison. Also, because  the \ac{SMC} is nearby and located well away from the Galactic Plane, it provides a good opportunity to study in great detail intrinsic  populations of sources such as \acp{SNR}, pulsars, and \ac{PNe}, alongside its ISM. The study of circularly polarised and other interesting (mainly background) sources also benefit from the breadth of available multi-wavelength data.

\subsection{MeerKAT \ac{SMC} SNR population}
 \label{SNR_samp}

\subsubsection{Previously known SMC SNRs and SNR candidates}
\label{sec:snrbonafide}

\begin{figure*}
    \centering
    \includegraphics[scale=0.265, trim=0 5 0 0,clip]{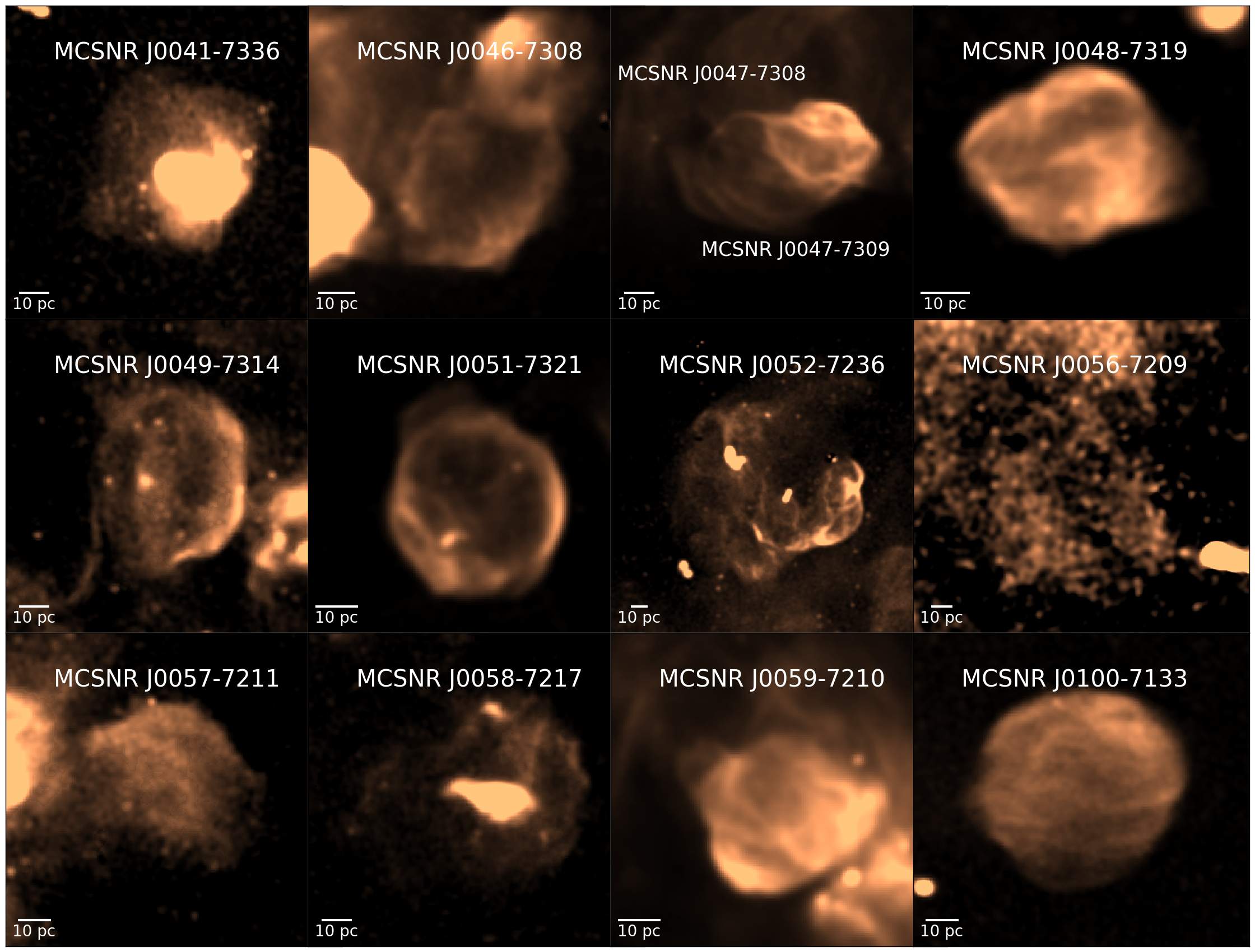}
    \includegraphics[scale=0.265, trim=0 0 0 10,clip]{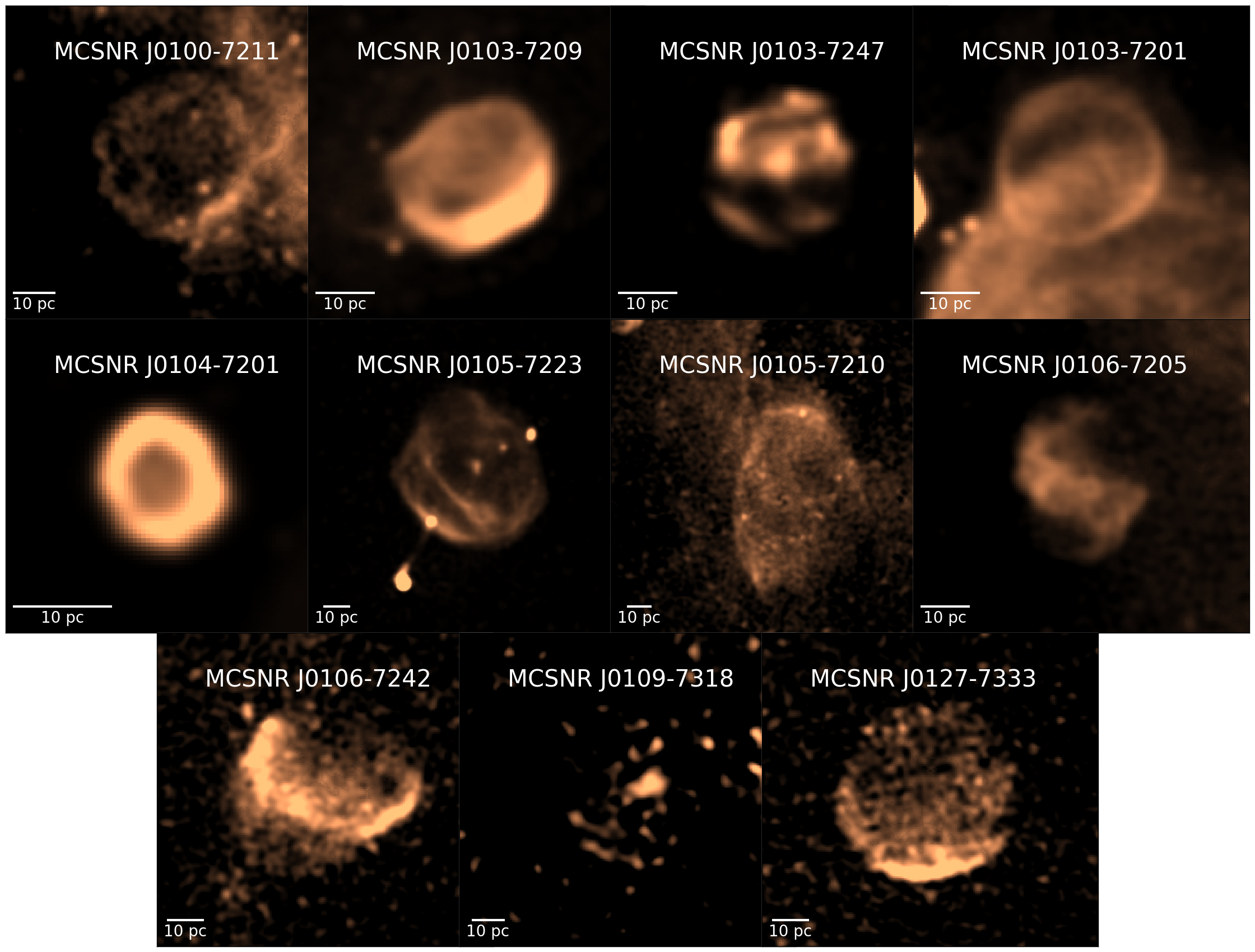}
    \caption{MeerKAT view of 24 bona fide \acp{SNR} in the \ac{SMC}. Nearby point sources are subtracted in order to better see the whole \ac{SNR} extent. The linear scale of 10~pc (34~arcsec at the \ac{SMC} distance of 60~kpc) is shown in the bottom left corner of each panel.
    We use a linear intensity scale for all images.
    \label{fig:snrMK1}}
\end{figure*}

\begin{figure*}
    \centering
    \includegraphics[scale=0.275, trim=0 0 0 0,clip]{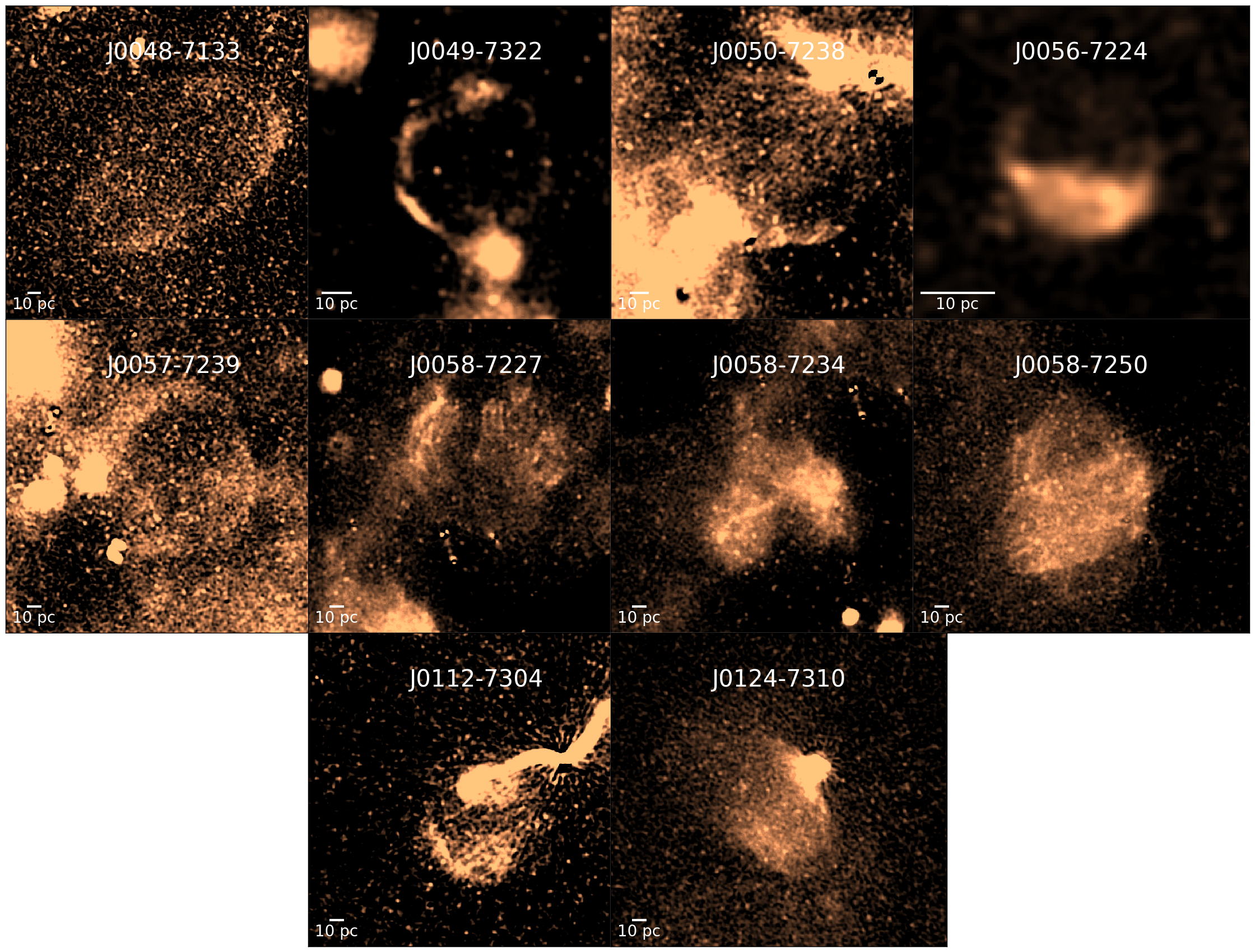}
    \caption{MeerKAT view of 10 new \ac{SNR} candidates in the \ac{SMC}. Nearby point sources are subtracted in order to see better the whole \ac{SNR} candidate extent. The linear scale of 10~pc (34~arcsec at the \ac{SMC} distance of 60~kpc) is shown in the bottom left corner of each panel. Note that the northern part of \ac{SNR} candidate J0112--7304 overlaps with  a  background radio galaxy. We use a linear intensity scale for all images.}
    \label{fig:snrcand}
\end{figure*}

\begin{figure*}
		\centering
			\includegraphics[width=\textwidth]{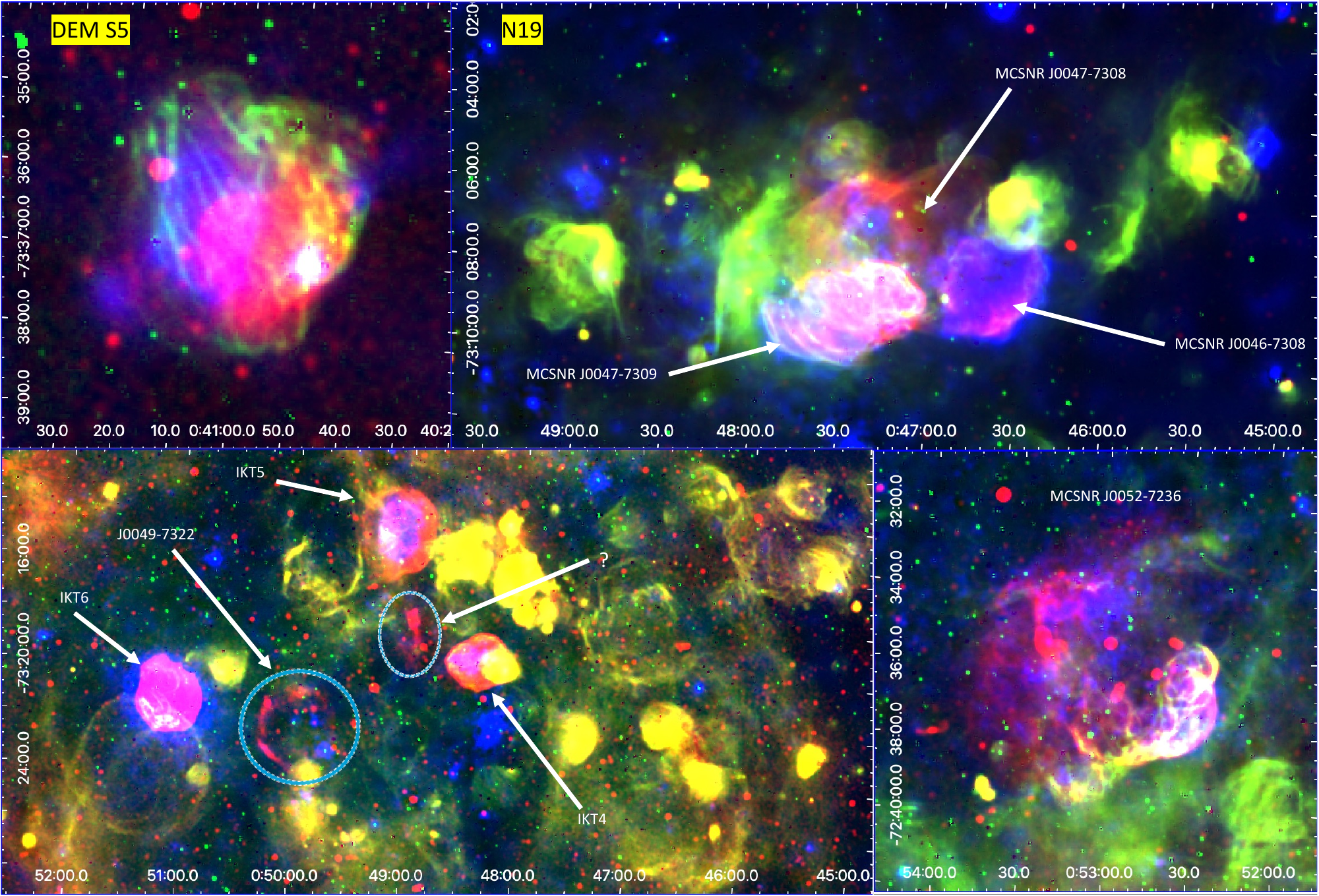}
        \caption{RGB images of selected \ac{SMC} regions. R corresponds to MeerKAT 1.3 GHz emission (plotted using a linear transfer function), G is H$\alpha$ (MCELS) and B is 0.3--1~keV X-rays from \xmm. This image helps to  discriminate between \acp{SNR} (dominantly in red colour) and other objects such as \HII~regions. Red dots are radio point sources of predominantly background nature but some \ac{PNe}, pulsars and foreground stars are seen as well. We show in these multi-colour/multi-band images well known \ac{SMC} \acp{SNR} (Table~\ref{tbl:snrbf}) DEM\,S5 (top left), N\,19 region \acp{SNR} (top-right; MCSNR~J0046--7308, MCSNR~J0047--7308 and MCSNR~J0047--7309), south of N\,19 region \acp{SNR} (bottom-left; IKT\,4, IKT\,5 and IKT\,6) and MCSNR~J0052--7236 (bottom right). In the bottom-left image, we show the new MeerKAT \ac{SNR} candidate J0049--7322 (marked with a cyan dashed circle) and also a newly discovered mysterious source (marked with `?' and cyan dashed ellipse).}
        \label{fig:snr1}
\end{figure*}

\begin{figure*}
		\centering
			\includegraphics[width=\textwidth]{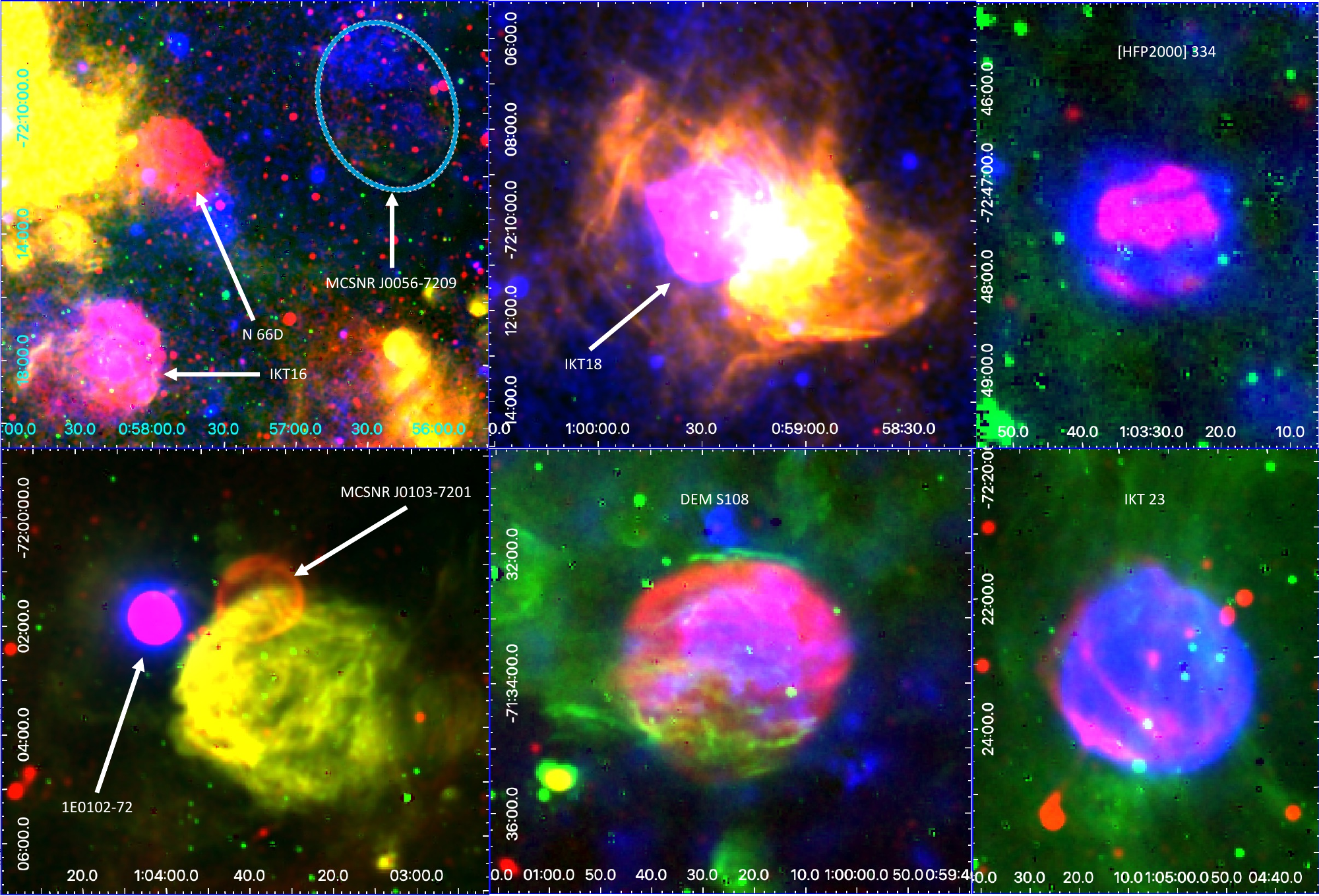}
        \caption{Same as Fig.~\ref{fig:snr1}. We show in these multi-colour/multi-band images well known \ac{SMC} \acp{SNR} (Table\ref{tbl:snrbf}): IKT\,16, N\,66D and MCSNR~J0056--7209 (top-left), IKT\,18 (top-middle), [HFP2000]\,334 (top-right), 1E0102--72 and MCSNR~J0103--7201 (bottom-left), DEM\,S108 (bottom-middle) and IKT\,23 (bottom-right).}
        \label{fig:snr2}
\end{figure*}

\begin{figure*}
		\centering
			\includegraphics[width=\textwidth]{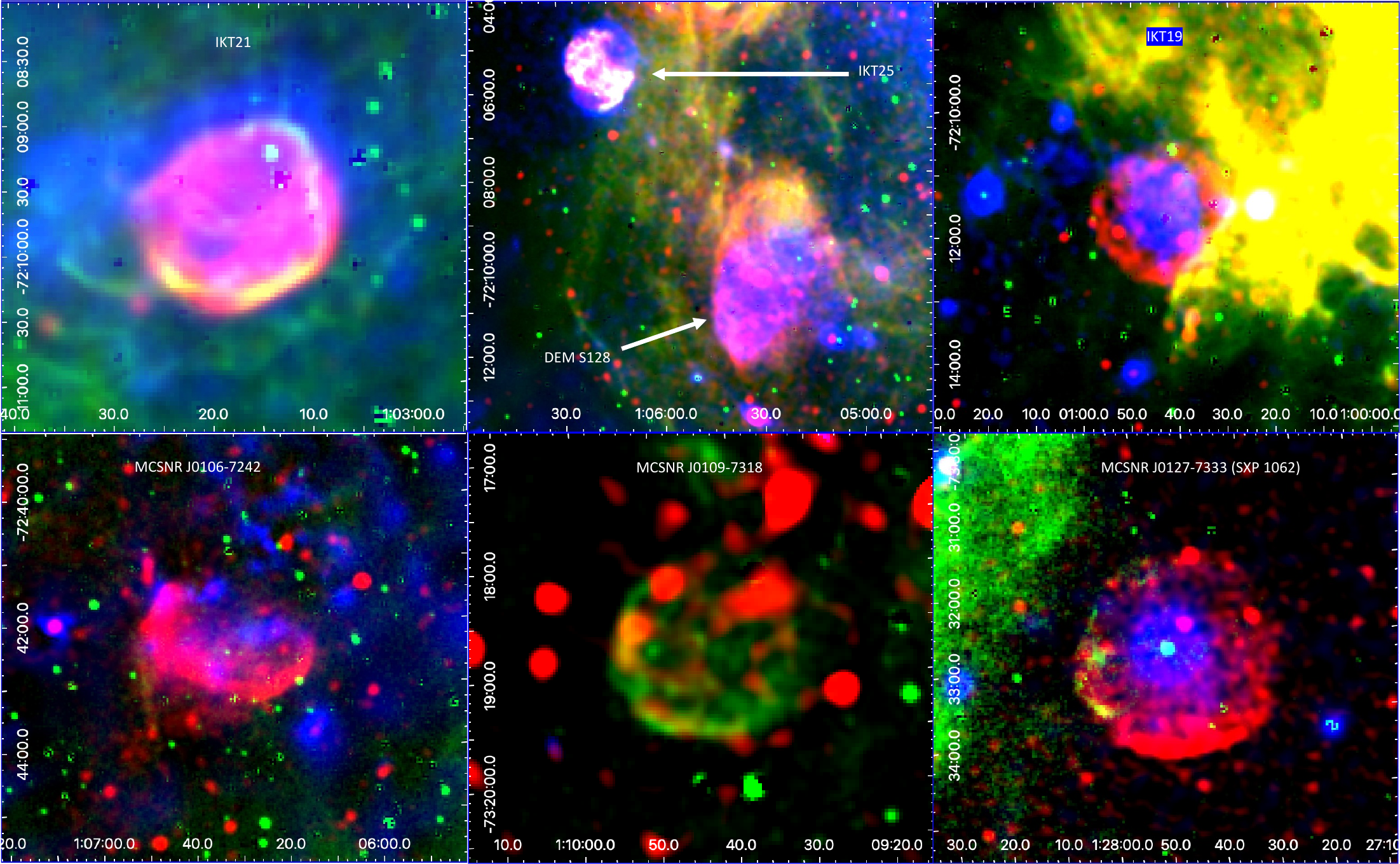}
        \caption{Same as Fig.~\ref{fig:snr1}. We show in these multi-colour/multi-band images well known \ac{SMC} \acp{SNR} (Table\ref{tbl:snrbf}): IKT\,21 (top-left), IKT\,25 and DEM\,S128 (top-middle), IKT\,19 (top-right), MCSNR~J0106--7242 (bottom-left), MCSNR~J0109--7318 (bottom-middle) and SXP\,1062 (bottom-right).}
        \label{fig:snr3}
\end{figure*}

Previous studies of \acp{SNR} in the \ac{SMC} \citep{2005MNRAS.364..217F,2007MNRAS.376.1793P,2008A&A...485...63F,2011A&A...530A.132O,2012A&A...545A.128H,2014AJ....148...99C,2015ApJ...803..106R,2019MNRAS.486.2507A,2019MNRAS.485L...6G,2019A&A...631A.127M,2019MNRAS.490.1202J} have established 21 objects as bona fide \acp{SNR} with two more (J0106--7242 and J0109--7318)\footnote{To distinguish between \ac{SNR} candidates and confirmed \acp{SNR}, \citet{2016A&A...585A.162M} established a nomenclature where bona fide \ac{MC} \acp{SNR} are named with the prefix `MCSNR~J' and candidates with only `J'.} considered as good candidates (see Table~\ref{tbl:snrbf} and Figs.~\ref{fig:snrMK1}--\ref{fig:snr3}). 

These two \ac{SNR} candidates (Fig.~\ref{fig:snr3}) are also identified in our new MeerKAT radio images as steep radio spectrum sources ($\alpha=-0.74\pm0.1$ for MCSNR~J0106--7242) which is indicative of non-thermal emission that is typical for \acp{SNR}. Also, we see an indication for X-ray emission from J0106--7242 in the \xmm\ soft band image and we detect J0109--7318 in optical wavebands (MCELS) with the typical \ac{SNR} enhanced \SII/\Ha>0.5 ratio. Therefore, we suggest that they are now confirmed \acp{SNR}, applying the principles established in \citet{2021MNRAS.500.2336Y} where bona fide \acp{SNR} must be identified (as possible \acp{SNR}) in at least two out of three (radio, optical, X-ray) wavebands.

\begin{figure*}
		\centering
			\includegraphics[width=\textwidth]{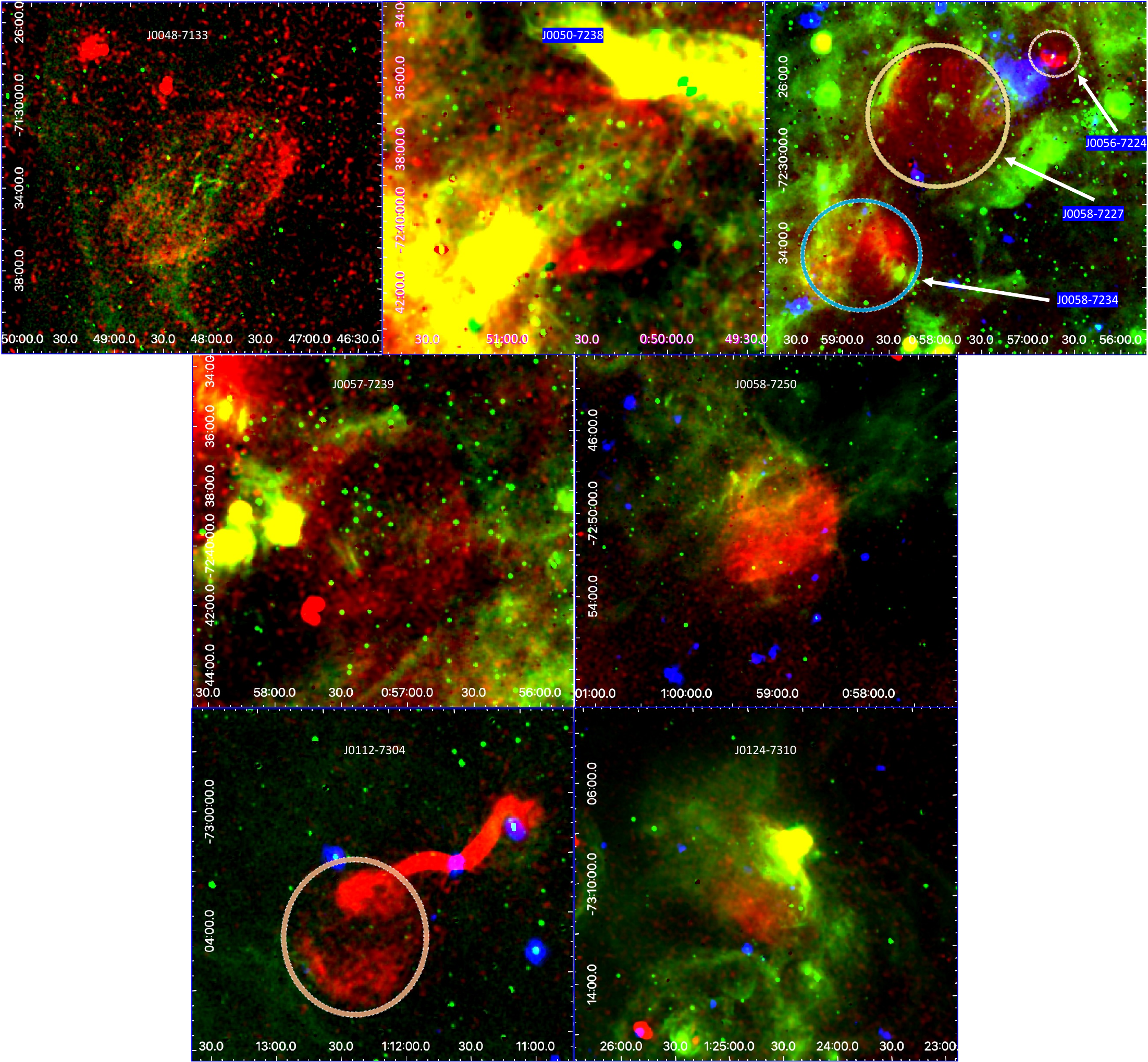}
        \caption{Same as Fig.~\ref{fig:snr1}. We show in these multi-colour/multi-band images new MeerKAT \ac{SMC} \acp{SNR} candidates (Table\ref{tbl:snrcand}): J0048--7133 (top-left), J0050--7238 (top-middle), J0056--7224, J0058--7227 and J0058--7234 (top-right), J0057--7239 (middle-left), J0058--7250 (middle-right), J0112--7304 (bottom-left) and J0124--7310 (bottom-right). The MeerKAT \ac{SNR} candidate J0049--7322 is shown in Fig.~\ref{fig:snr1}.}
        \label{fig:snr4}
\end{figure*}

In Tables~\ref{tbl:snrbf} and \ref{tbl:snrcand} (Section~\ref{sec:snrcand}) we show our complete radio continuum sample of the previously known \ac{SMC} \acp{SNR} and new \ac{SNR} candidates (see Section~\ref{sec:snrcand}). We measure extensions, integrated flux densities and spectral index (where possible) for all 34 objects. 

As our \ac{SMC} \ac{SNR} sample is somewhat diverse, various approaches (and initial parameters) were employed in order to measure the best \ac{SNR} flux densities. We used the \textsc{miriad} \citep{1995ASPC...77..433S} task \texttt{imfit} to extract integrated flux density, extensions (diameter/axes) and position angle for each radio detected \ac{SNR}. For cross-checking and consistency, we also used \aegean\ and found no significant difference in integrated flux density estimates (see Section~\ref{sec:cats}). Namely, we measured \ac{SNR} local background noise (1$\sigma$) and carefully selected the exact area of the \ac{SNR} which also excludes all obvious unrelated point sources. We then estimated the sum of all brightnesses above 5$\sigma$ of each individual pixel within that area and converted it to \ac{SNR} integrated flux density following \citet[][eq.~24]{1966ARA&A...4...77F}. We also made corrections for an extended background where applicable i.e. for sources where nearby extended objects such as \HII~region(s) are evident. These non-SNR sources are also morphologically complex radio objects of a distinctively different origin (thermal emission) with their colours in our multi-wavelength RGB images (e.g., Fig.~\ref{fig:snr1}) set to green-to-yellow. However, for most of our \ac{SMC} \acp{SNR}, this extended background flux density contribution is minimal ($<$5~per~cent). We also estimate that the corresponding radio flux density errors are below 15~per~cent as determined in our previous work \citep{2022MNRAS.512..265F,2023MNRAS.518.2574B}. In this estimate, various contributions to the flux density error are taken into account including source overlapping of extended structures that are not part of the \ac{SNR} (e.g., an \ac{AGN} in the case of J0112--7304) as well as missing short spacings. Certainly, for weaker sources, this uncertainty is more pronounced while for brighter objects such as \ac{SNR} 1E0102--72 the flux density error is much smaller.

For the spectral index estimate (Table~\ref{tbl:snrbf}; Column~10), we used all available flux density measurements including ones from \citet{2019A&A...631A.127M} and this work. While all our \ac{SMC} \ac{SNR} measurements shown in Table~\ref{tbl:snrbf} are significant improvements, they are within the errors of previous estimates \citep{2019A&A...631A.127M}.

\subsubsection{New MeerKAT SMC SNR candidates and SNRs}
 \label{sec:snrcand}
One of our goals with the MeerKAT \ac{SMC} survey is to discover new \acp{SNR}. With its unique coverage and depth, this MeerKAT survey allowed us to search for new and predominantly low surface brightness \ac{SNR} candidates. At the same time, we can measure the physical properties of the already established \ac{SMC} \acp{SNR}.

As MeerKAT images are sensitive to a low surface brightness and with good resolution, we can employ the same methodology  described in \citet{2023MNRAS.518.2574B} to search for the new \acp{SNR}. Apart from the typical morphological appearance (approximately circular shape), \acp{SNR} at radio frequencies show predominantly non-thermal emission. However, in H$\alpha$ and in a majority of X-ray detections thermal emission dominates. We used the Magellanic Clouds Emission Line Survey \citep[MCELS; ][]{2012ApJ...755...40P} optical images as well as the new X-ray images from \xmm\ \citep{2012A&A...545A.128H} to search for new \ac{SMC} \acp{SNR}. In our multi-colour/multi-waveband RGB images (Figs.~\ref{fig:snr1}--\ref{fig:snr4}) \acp{SNR} show distinctively red colour as predominantly radio emitters. 

Using this method, we identified 11 new MeerKAT \ac{SMC} \ac{SNR} candidates (Table~\ref{tbl:snrcand}) from which one (J0100--7211; Fig.~\ref{fig:snr3} top-right) we confirm as a true \ac{SNR} and include in Table~\ref{tbl:snrbf}. Apart from the typical shell-like \ac{SNR} structure, MCSNR~J0100--7211 also harbours the well-known X-ray magnetar \citep[CXOU\,J0110043.1$-$721134;][]{2002ApJ...574L..29L} at its centre. Several magnetars in the Milky Way are found in their parent \ac{SNR} \citep[e.g.][]{2021ASSL..461...97E}, consistent with an \ac{SNR} nature for this object. 
Multi-waveband images of the remaining 10 newly identified  \ac{SMC} \ac{SNR} candidates are shown in Figs.~\ref{fig:snr1} (one object) and \ref{fig:snr4} (nine objects).

\subsubsection{SMC SNRs and SNR candidate properties}
 \label{snr:stat}

Our final sample of this MeerKAT study of \ac{SMC} \acp{SNR} consists of 24 confirmed and 10 candidate remnants. Most of the new \acp{SNR} and \ac{SNR} candidates display approximately semi-circular structures which together with their multi-waveband appearance are consistent with a typical \ac{SNR} spherical morphology. As expected, the group of new MeerKAT \ac{SNR} candidates is of low radio surface brightness (see below), which largely explains their previous non-detections. 

As none of the 10 \ac{SMC} \ac{SNR} candidates identified here have been detected at other radio frequencies, we cannot estimate their spectral index. To compare the surface brightness of our sample with established \ac{SMC} \acp{SNR} (Section~\ref{sec:snrbonafide}), we assumed a typical \ac{SNR} spectral index of $\alpha=-0.5$ \citep{2012SSRv..166..231R,2014SerAJ.189...15G,2017ApJS..230....2B,2019A&A...631A.127M,book2}. Using this assumed spectral index allowed us to estimate the flux density and surface brightness ($\Sigma_{\rm 1\,GHz}$) of these sources at 1~GHz.

\begin{figure*}
    \centering\includegraphics[scale=0.69,trim=0 45 70 90,clip]{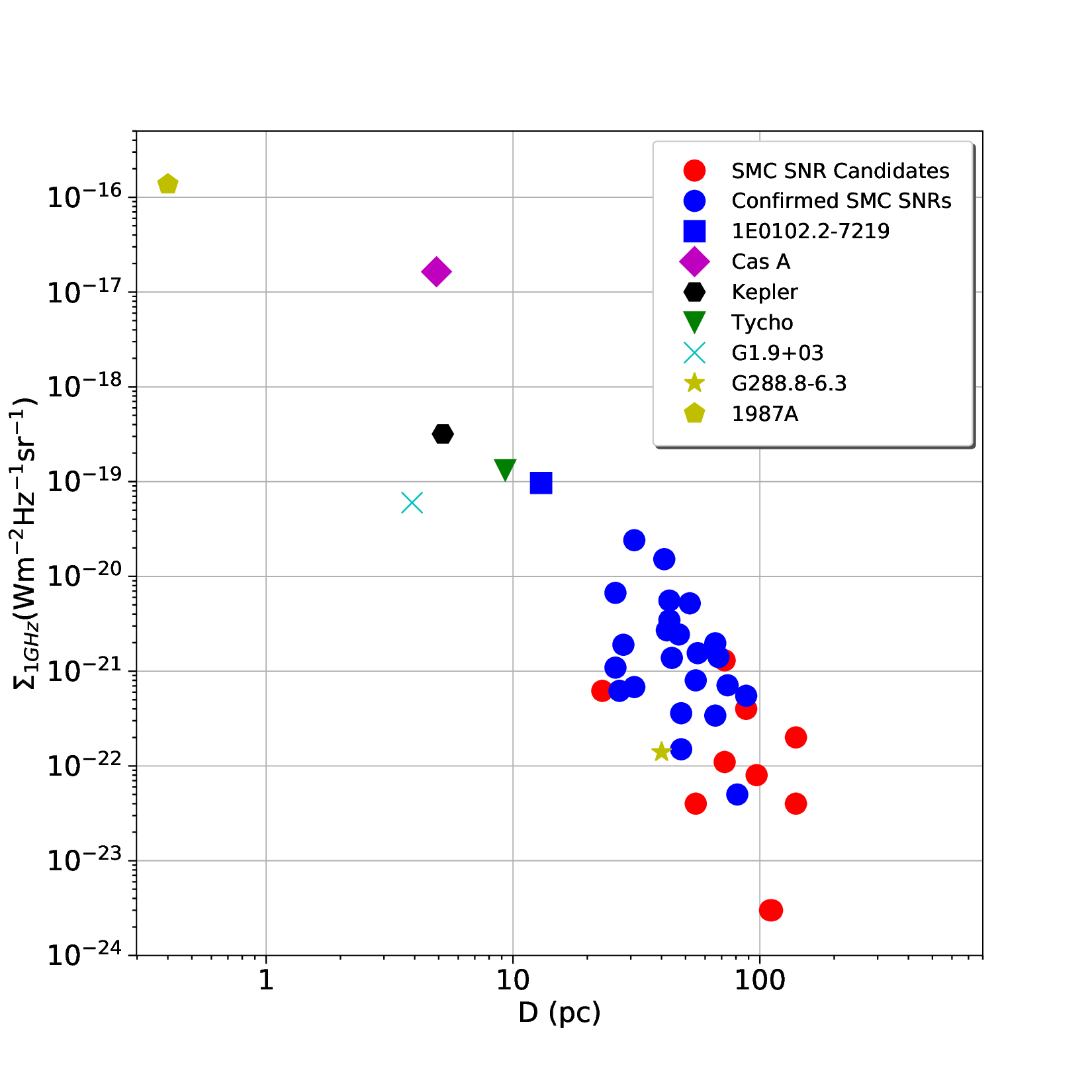}
    \caption{Radio surface brightness–to–diameter diagram for the 24 \ac{SMC} \acp{SNR} (blue circles) and 10 candidates (red circles) at a frequency of 1\,GHz (see Tables~\ref{tbl:snrbf} and \ref{tbl:snrcand}. We also use several estimates of known Galactic \acp{SNR} from \citet{2018ApJ...852...84P} as well as G288.8--6.3 \citep[][marked with a yellow star in the bottom right corner]{2023AJ....166..149F} which is one of the lowest Galactic radio surface brightness \ac{SNR}. Cassiopeia~A is shown with a magenta-filled diamond while the cyan cross represents the youngest Galactic \ac{SNR}, G1.9+0.3 \citep{2020MNRAS.492.2606L}. We also included Kepler \ac{SNR} (black hexagon), Tycho (green triangle) and SN\,1987A (yellow pentagon).}
    \label{fig:evol}
\end{figure*}

These 10 new \ac{SMC} \ac{SNR} candidates escaped  previous detection because of their low surface brightness, which typically indicates an advanced evolutionary stage. Some of these objects are most likely evolved and expanding in a rarefied environment, and we note that they occupy the bottom right portion of the \ac{SNR} $\Sigma$--D diagram \citep{2018ApJ...852...84P} as can be seen in Fig.~\ref{fig:evol}. 

The arithmetic average of surface brightness from the sample of 24 confirmed \ac{SMC} \acp{SNR} (Table~\ref{tbl:snrbf}) is 7.1$\times10^{-21}$~W~m$^{-2}$~Hz$^{-1}$sr$^{-1}$ (median 1.4$\times10^{-21}$~W~m$^{-2}$~Hz$^{-1}$sr$^{-1}$) while for the sample of 10 new candidates (Table~\ref{tbl:snrcand}) it is 1.6$\times10^{-22}$~W~m$^{-2}$~Hz$^{-1}$sr$^{-1}$ (median is 9.5$\times10^{-23}$~W~m$^{-2}$~Hz$^{-1}$sr$^{-1}$). This order of magnitude difference suggests that we are discovering low surface brightness \acp{SNR} in the \ac{SMC}. At the same time, our sample \ac{SNR} candidate diameters are also significantly larger compared to established \ac{SMC} \acp{SNR} ($D_{\rm av, 10\,SNRcand}$=98~pc; STDEV=30~pc vs. $D_{\rm av, 24\,SNR}$=48~pc; STDEV=19~pc). In turn, this immediately implies that our sample of 10 new objects belongs to a more evolved and mid-to-older ($>5$~kyrs) \ac{SNR} population \citep{2017ApJS..230....2B}.

\begin{table*}
\centering
\scriptsize
\caption{24 confirmed \acp{SNR} in the \ac{SMC}. The errors on the integrated flux densities (Column~9) are estimated to be $<$15\%. Column~3 abbreviations are: DEM\,S: \citet{1976MmRAS..81...89D}, [HFP2000]: \citet{haberl2000}, IKT: \citet{1983IAUS..101..535I}, SXP: \citet{2012A&A...537L...1H}. Position angle in Column~8 is measured from north to east. In Column~11 we list \ac{SNR} surface brightness at 1~GHz where results in brackets assumed a typical \ac{SNR} spectral index of --0.5. In Column~12 we indicate radio, X-ray and/or optical (\SII\ or \OIII) detection. Along with the MCSNR Name (Column~2) we mark with * newly promoted bona fide \acp{SNR} in this study. Names in column (3) marked with $^{\dagger}$ are names of associated central X-ray point sources. The assumed distance to the \ac{SMC} is 60~kpc.} 
 \begin{tabular}{lllcccccccccccccccc}
 \hline
No.  & MCSNR		&	Other 		&	RA			&	DEC				& D$_{\rm maj} \times$ D$_{\rm min}$    & D$_{\rm av}$  & PA    &  $S_{1283.8\,\mathrm{MHz}}$	&	$\alpha \pm \Delta\alpha$   &$\Sigma_{1 \rm{GHz}} (\times 10^{-20}$)& Notes & Fig.	\\
     & Name			&	Name		&	(J2000)		&	(J2000)			& (arcsec)                              & (pc)          & (\D)  & 	(Jy)					&	    	                    & (W\,m$^{-2}$Hz$^{-1}$sr$^{-1}$) &  & \\
(1) & (2) & (3) & (4) & (5) & (6) & (7) & (8) & (9) & (10) & (11) & (12) & (13) \\
\hline
1    & J0041--7336	&	DEM\,S5    	&	00 41 01.7	&	$-$73 36 30.4	&	246$\times$220  & 68          & 105   &   0.1560			&	--0.28$\pm$0.01	& 0.141 & R, X, O  & \ref{fig:snr1} \\
2    & J0046--7308	&[HFP2000]\,414	&	00 46 40.6	&	$-$73 08 14.9	&	185$\times$140  & 47          & 35    &   0.1012			&	--0.37$\pm$0.04 & 0.244 & R, X, O  & \ref{fig:snr1} \\
3    & J0047--7308	&	IKT\,2     	&	00 47 16.6	&	$-$73 08 36.5	&	110$\times$100  & 31          & 45    &   0.3537			&	--0.54$\pm$0.03 & 2.410 & R, X, O  & \ref{fig:snr1} \\
4    & J0047--7309	&	DEM\,S32	&	00 47 36.5	&	$-$73 09 20.0	&	180$\times$120  & 43          & 75    &   0.2594			&	--0.58$\pm$0.11	& 0.556 & R, X, O  & \ref{fig:snr1} \\
5    & J0048--7319	&	IKT\,4     	&	00 48 19.6	&	$-$73 19 39.6	&	165$\times$130  & 43          & 90    &   0.1071			&	--0.59$\pm$0.03	& 0.345 & R, X, O  & \ref{fig:snr1} \\
6    & J0049--7314	&	IKT\,5     	&	00 49 07.7	&	$-$73 14 45.0	&	190$\times$190  & 55          & 0     &   0.0416			&	--0.66$\pm$0.05	& 0.080 & R, X, O  & \ref{fig:snr1} \\
7    & J0051--7321	&	IKT\,6     	&	00 51 06.7 	&	$-$73 21 26.4	&	145$\times$145  & 42          & 0     &   0.1008			&	--0.55$\pm$0.03	& 0.276 & R, X, O  & \ref{fig:snr1} \\
8    & J0052--7236	&	DEM\,S68 	&	00 52 59.9	&	$-$72 36 47.0	&	340$\times$270  & 88          & 135   &   0.0939			&	--0.52$\pm$0.03	& 0.055 & R, X, O  & \ref{fig:snr1} \\
9    & J0056--7209  &               &   00 56 28.1  &   $-$72 09 42.2   &	340$\times$225  & 81          & 30    &   0.0065			& (--0.5)     &(0.005)& R, X, O  & \ref{fig:snr2} \\
10   & J0057--7211  &   N\,66D      &   00 57 52.4  &   $-$72 11 49.7   &	199$\times$161  & 52          & 40    &   0.0290			&	--0.72$\pm$0.03	& 0.052 & R, O     & \ref{fig:snr2} \\
11   & J0058--7217	&	IKT\,16    	&	00 58 22.4 	&	$-$72 17 52.0	&	267$\times$241  & 74          & 0     &   0.0707			&	--0.51$\pm$0.03	& 0.071 & R, X, O  & \ref{fig:snr2} \\
12   & J0059--7210	&	IKT\,18    	&	00 59 27.7 	&	$-$72 10 09.8	&	140$\times$140  & 41          & 0     &   0.4810			&	--0.47$\pm$0.02	& 1.520 & R, X, O  & \ref{fig:snr2} \\
13   & J0100--7133	&	DEM\,S108	&	01 00 23.9 	&	$-$71 33 41.1	&	240$\times$216  & 66          & 0     &   0.1741			&	--0.50$\pm$0.02	& 0.197 & R, X, O  & \ref{fig:snr2} \\
14   & J0100--7211* & CXOU\,J0110043.1$-$721134$^{\dagger}$ &	01 00 44.3	&	$-$72 11 40.9	&	121$\times$124  & 66          & 0     &   0.0070  		    & (--0.5)           & (0.034)& R     & \ref{fig:snr3}\\
15   & J0103--7209	&	IKT\,21    	&	01 03 17.0 	&	$-$72 09 42.5	&	95$\times$83    & 26          & 135   &   0.0806			&	--0.67$\pm$0.03	& 0.670 & R, X, O  & \ref{fig:snr3} \\
16   & J0103--7247	&[HFP2000]\,334 &	01 03 29.1	&	$-$72 47 32.6	&	105$\times$85   & 28          & 0     &   0.0288	        &	--0.58$\pm$0.05	& 0.190 & R, X     & \ref{fig:snr2} \\
17   & J0103--7201	& SXP\,1323$^{\dagger}$    &	01 03 36.6	&	$-$72 01 35.1	&	98$\times$83    & 26          & 90    &   0.0410			& (--0.5)       & (0.109) & R, X, O  & \ref{fig:snr2} \\
18   & J0104--7201	&1E\,0102.2--7219&	01 04 01.2 	&	$-$72 01 52.3	&	43$\times$45    & 13          & 0     &   0.2970			&	--0.67$\pm$0.01	& 9.670 & R, X, O  & \ref{fig:snr2} \\
19   & J0105--7223	&	IKT\,23    	&	01 05 04.2 	&	$-$72 23 10.5	&	192$\times$192  & 56          & 0     &   0.0778			&	--0.61$\pm$0.03	& 0.155 & R, X     & \ref{fig:snr2} \\
20   & J0105--7210	&	DEM\,S128  	&	01 05 30.5	&	$-$72 10 40.4	&	184$\times$126  & 44          & 0     &   0.0499			&	--0.55$\pm$0.03	& 0.138 & R, X     & \ref{fig:snr3} \\
21   & J0106--7205	&	IKT\,25    	&	01 06 17.5 	&	$-$72 05 34.5	&	110$\times$80   & 27          & 25    &   0.0101			&	--0.41$\pm$0.04	& 0.062 & R, X, O  & \ref{fig:snr3} \\
22   & J0106--7242* &	        	&	01 06 33.7 	&	$-$72 42 29.5	&	175$\times$153  & 48          & 10    &   0.0126			&	--0.74$\pm$0.10	& 0.036 & R, X     & \ref{fig:snr3} \\
23   & J0109--7318* &	        	&	01 09 43.6	&	$-$73 18 46.0	&	105$\times$105  & 31          & 0     &   0.0154	        & (--0.5)      & (0.086) & R, O     & \ref{fig:snr3} \\
24   & J0127--7333	&	SXP\,1062$^{\dagger}$ &	01 27 44.1	&	$-$73 33 01.6	&	166$\times$166  & 48          & 0     &   0.0100			&	--0.42$\pm$0.12	& 0.015 & R, X, O  & \ref{fig:snr3} \\
\hline
 \end{tabular}
 \smallskip
 \flushleft
\label{tbl:snrbf}
\end{table*}

\begin{table*}
\centering
\caption{Ten new MeerKAT \ac{SNR} candidates in the \ac{SMC}. 
The errors on the integrated flux densities (Column~8) are estimated to be $<$15\%. In Column~10 we list \ac{SNR} candidate surface brightness at 1~GHz where we assumed a typical \ac{SNR} spectral index of --0.5. In Column~11 we indicate radio, X-ray and/or optical (\SII\ or \OIII) detection. The assumed distance to the \ac{SMC} is 60~kpc.} 
 \begin{tabular}{lllcccccccccccccccc}
 \hline
No.  & Candidate		&		RA			&	DEC				& D$_{\rm maj} \times$ D$_{\rm min}$    & D$_{\rm av}$  & PA    &  $S_{1283.8\,\mathrm{MHz}}$	&	$\alpha$   &$\Sigma_{1 \rm{GHz}} (\times 10^{-20}$)& Notes & Fig.	\\
     & Name			&		(J2000)		&	(J2000)			& (arcsec)                              & (pc)          & (\D)  & 	(Jy)					&	    	                    & (W\,m$^{-2}$Hz$^{-1}$sr$^{-1}$) &  & \\
(1) & (2) & (3) & (4) & (5) & (6) & (7) & (8) & (9) & (10) & (11) & (12)  \\
\hline
1    & J0048--7133  &	00 48 14.4	&	$-$71 33 28.7	&	625$\times$370  & 140          & 120   &   0.0174			&	    (--0.5)         & 0.004   & R, O?  & \ref{fig:snr4} \\
2    & J0049--7322	&	00 49 57.1	&	$-$73 22 23.7	&	188$\times$188  & 55           & 0     &   0.0026			&	    (--0.5)         & 0.004   & R, X?  & \ref{fig:snr1} \\
3    & J0050--7238	&	00 50 35.3	&	$-$72 38 38.3	&	356$\times$314  & 97           & 0     &   0.0158			&       (--0.5)         & 0.008   & R  & \ref{fig:snr4} \\
4    & J0056--7224	&	00 56 46.1	&	$-$72 24 49.8	&	80$\times$75    & 23           & 0     &   0.0060			&	    (--0.5) 	    & 0.062   & R  & \ref{fig:snr4} \\
5    & J0057--7239	&	00 57 14.5	&	$-$72 39 13.7	&	378$\times$378  & 110        & 0     &   0.0007 		&	   (--0.5)      	& 0.0003   & R  & \ref{fig:snr4} \\
6    & J0058--7227  &   00 58 00.1  &   $-$72 27 56.2   &	387$\times$382  & 112        & 0     &   0.0008			&	   (--0.5)          & 0.0003   & R  & \ref{fig:snr4} \\
7    & J0058--7234	&	00 58 53.7	&	$-$72 34 40.4	&	348$\times$262  & 88          & 100   &   0.0605	        &      (--0.5) 	        & 0.040   & R  & \ref{fig:snr4} \\
8    & J0058--7250	&	00 58 57.0	&	$-$72 50 46.2	&	512$\times$452  & 140         & 100   &   0.0760		    &      (--0.5)          & 0.020   & R  & \ref{fig:snr4} \\
9    & J0112--7304	&	01 12 18.8	&	$-$73 04 23.8	&	296$\times$206  & 72          & 0     &   0.0133			&     (--0.5)           & 0.013   & R, O?   & \ref{fig:snr4} \\
10   & J0124--7310	&	01 24 30.7	&	$-$73 10 33.0	&	292$\times$208  & 72          & 0     &   0.0107			&     (--0.5)           & 0.011   & R, X?  & \ref{fig:snr4} \\
\hline
 \end{tabular}
 \smallskip
 \flushleft
\label{tbl:snrcand}
\end{table*}

%

\subsubsection{Notes on individual SMC SNRs and SNR candidates}

Here, we describe properties of the \ac{SMC} \acp{SNR} and \ac{SNR} candidates where in our Figs.~\ref{fig:snr1}, \ref{fig:snr2}, \ref{fig:snr3} and \ref{fig:snr4} they exhibit mainly red colour.

\paragraph{SMC SNRs:}
DEM\,S5 \citep[Fig.~\ref{fig:snr1};][also see in Carli et al. (in prep.)]{2019MNRAS.486.2507A} and IKT\,16 \citep[Fig.~\ref{fig:snr2}; ][]{2021MNRAS.507L...1M,2015A&A...584A..41M,2011A&A...530A.132O,2022MNRAS.517.5406C} contain the only two confirmed \ac{PWN} in the \ac{SMC}. [HFP2000]\,334 \citep[Fig.~\ref{fig:snr2}; ][]{2014AJ....148...99C} and recently 1E\,0102.2--7219 (Alsaberi et al.~2022, submitted) were also examined for the existence of a \ac{PWN} but without a firm confirmation. Finally, \acp{SNR} MCSNR~J0127--7333 \citep[Fig.~\ref{fig:snr3};][]{2012A&A...537L...1H}, MCSNR~J0103--7201 (Fig.~\ref{fig:snr2}) and the here discovered MCSNR~J0100--7211 (Fig.~\ref{fig:snr3}) were found around well-known X-ray pulsars (the Be/X-ray binaries SXP\,1062 and SXP\,1323, and the magnetar CXOU\,J0110043.1$-$721134). The existence of a neutron star at the centre of an SNR with a circular ring-like shell establishes its core-collapse origin. The distinctive circular radio appearance of MCSNR~J0100--7211 and its central magnetar makes this object a new bona fide \ac{SNR}, despite the fact that no X-ray emission from the SNR itself could be detected yet.

Three \acp{SNR} in the N\,19 region (Fig.~\ref{fig:snr1} top-right; [HFP2000]\,414, IKT\,2 and DEM\,S32) are clearly overlapping. Defining their borders (and properties) is quite difficult, but we manage to measure the properties consistent with their \ac{SNR} nature.

The region south from N\,19 (Fig.~\ref{fig:snr1} bottom-left) consists of three previously confirmed \acp{SNR} (IKT\,4, IKT\,5 and IKT\,6). Here we propose a new \ac{SNR} candidate J0049--7322 which shows a filamentary ring structure. 
Interestingly, there is some indication for a similar structure seen on the \xmm\ mosaic image, which needs confirmation by a dedicated future observation.

The previously confirmed MCSNR~J0052--7236 (DEM\,S68; Fig.~\ref{fig:snr1} bottom-right) looks much more extensive in our new radio images (and also in the \xmm\ soft X-ray images) than previously defined in \citet{2019A&A...627A.142V}. Especially its northern end makes this \ac{SNR} among the largest D$_{\rm av}$=88.2~pc. 

Well-established \acp{SNR} N\,66D, IKT\,18, DEM\,S108, IKT\,23 (Fig.~\ref{fig:snr2}) and IKT\,21, DEM\,S128 \citep{2000A&A...353..129F} and IKT\,25 (Fig.~\ref{fig:snr3}) show typical \acp{SNR} characteristics. Three low surface brightness confirmed \acp{SNR} include MCSNR~J0056--7209 and here confirmed MCSNR~J0106--7242 and MCSNR~J0109--7318. 

\paragraph{SMC SNR candidates:}
Our 10 new \ac{SMC} \ac{SNR} candidates (Fig.~\ref{fig:snr4}) are exclusively seen at radio continuum frequencies. However, in a few candidates such as for J0048--7133 (\SII) and J0112--7304 (\OIII) we can see some possible traces of corresponding optical emission in the MCELS images. The shape of \ac{SNR} candidate J0048--7133 is elongated and with a major axis of $\sim$180~pc might represent one of the largest \acp{SNR} that expands in the rarefied environment as it is located some 0.75~degree from the main body of the \ac{SMC}. This might be a similar type of \ac{SNR} to our recent discovery of the intergalactic \ac{SNR} MCSNR~J0624--6948 \citep{2022MNRAS.512..265F} or MCSNR~J0509--6402 \citep{2021MNRAS.500.2336Y}. 
Besides J0049--7322 (see above) there is also some faint diffuse X-ray emission indicated in \xmm\ images at the location of J0124--7310. Again, a dedicated observation is required for confirmation.
Among these 10 new \acp{SNR} candidates, we detect three large-scale objects (J0050--7238, J0057--7239 and J0058--7250) with diameters exceeding 100~pc. While somewhat unlikely, they also might represent (super)bubbles \citep[][]{2021ApJ...918...36Y,2019A&A...621A.138K,2017ApJ...843...61S}. \ac{SNR} candidate J0112--7304 is `hidden' around the southeast jet of the nearby large (apparent) \ac{AGN} and therefore its true extent might be somewhat different than measured here. 

\subsection{Automating the SNR search}
 \label{sec:autosnr}
Scientific interpretation of diffuse emission from the \ac{ISM} generally requires constraining the emission mechanism. A common calculation in the centimetre continuum regime is the spectral index ($\alpha$), which can discriminate between optically thin thermal free-free ($\alpha\sim$--0.1) and non-thermal ($\alpha<$--0.5), or even contributions from optically thick free-free, or dust emission with larger positive spectral indices. Spectral indices calculated over modest frequency ranges have large uncertainties, and any differences in spatial filtering or primary beam response as a function of frequency only increase the uncertainties. The thermal/non-thermal fraction can also be fitted with spectral data, with similarly large uncertainties. Even with MeerKAT's large fractional bandwidth, these calculations are challenging. 

\acp{SNR} can be identified visually as described in Section~\ref{SNR_samp}.
Although such identification is relatively unambiguous for an experienced investigator, it is desirable to explore more automated techniques, especially in the era of large-scale surveys. 
Thermal and non-thermal sources have different correlations between infrared and centimetre emission, so the ratio has been used to distinguish between emission mechanisms \citep{broadbent89,cohen07}.
We extend the use of a single infrared map to use infrared maps at several frequencies and a map of H$\alpha$, which is also expected to correlate strongly with thermal centimetre emission, and rather less strongly with non-thermal emission.  We first preprocess the maps to have the same spatial filtering characteristics, then demonstrate a supervised and unsupervised classification in the \ac{SMC}.

For the analysis of diffuse emission, we use the residual MeerKAT image produced by \aegean\ by removing all of the compact and point sources (see Section~\ref{scompareF}). This is not essential, and similar results can be achieved with the original image, but does make the preprocessing (see next section) easier, and helps highlight the extended sources.
We complement MeerKAT with images that trace the physically most relevant emission mechanisms: emission from warm dust is traced by {\it Spitzer} MIPS 24~$\mu$m at native 6~arcsec resolution from the SAGE survey \citep{meixner_sage}; and somewhat cooler dust by {\it Herschel} PACS 160~$\mu$m at 13~arcsec resolution from the HERITAGE survey \citep{meixner_heritage}. Although longer wavelength {\it Herschel} data are available, it is desirable to perform this analysis at as close to the MeerKAT resolution as possible.  Free-free emission from ionised gas, albeit subject to dust attenuation, is traced by MCELS continuum-subtracted H$\alpha$ \citep{mcels}.

\subsubsection{Image preprocessing}

\begin{figure*}
    \centering
    \includegraphics[width=\textwidth]{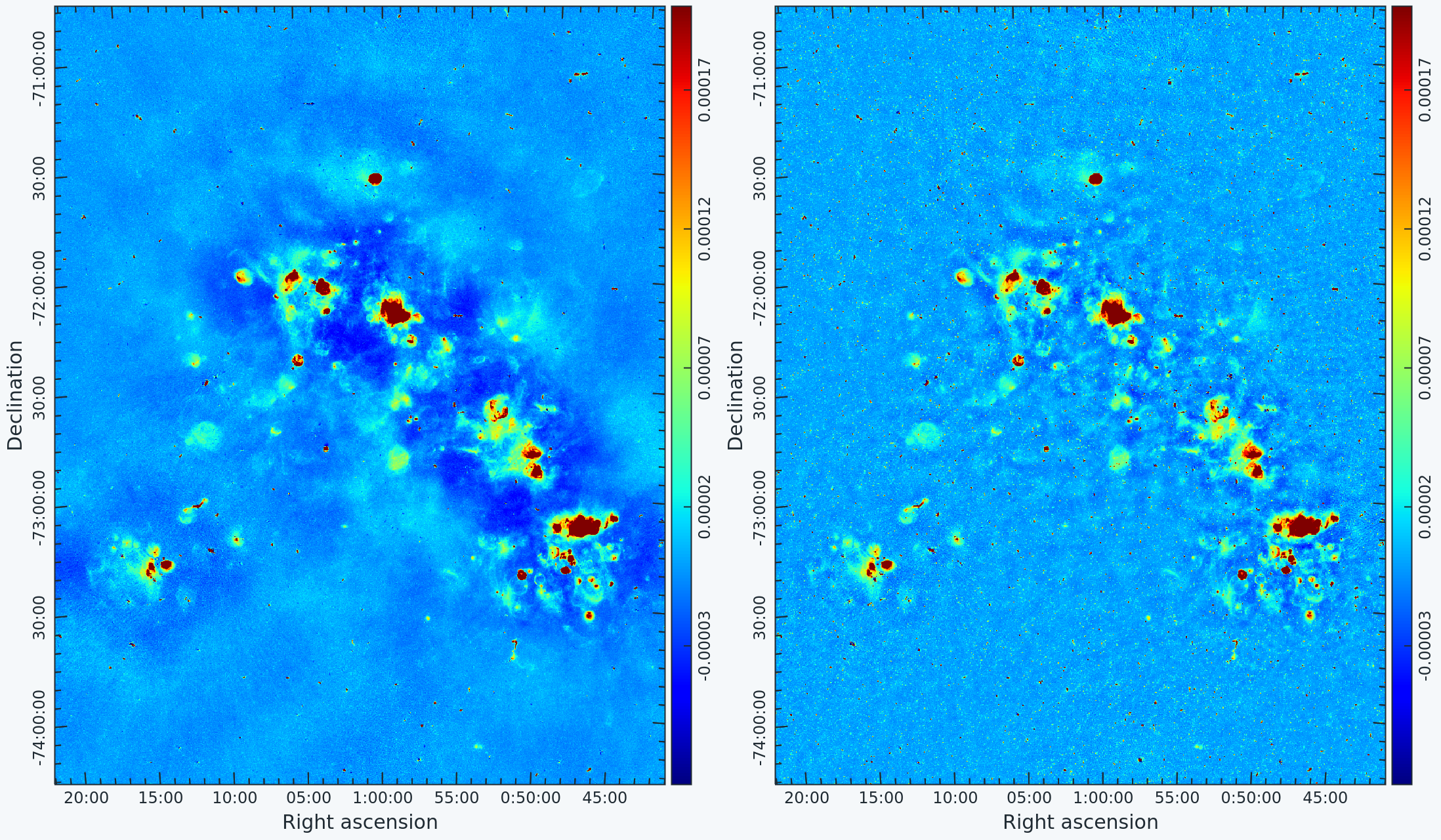}
    \caption{(Left) original (with point sources removed) and (right) background-subtracted or flattened MeerKAT mosaics of the \ac{SMC} (only showing the central part with diffuse emission on large angular scales). The linear colour scale spans --1$\times$10$^{-5}$ to 2.5$\times$10$^{-4}$~Jy~beam$^{-1}$.} 
    \label{fig:mk-bgsub}
\end{figure*}

At the time of publication, appropriately deep wide-band single-dish observations of the \ac{SMC} did not exist, so as with all interferometer data, the MeerKAT image lacks short spacing information, evident in negative bowls around bright emission at large angular scales. As is well known, there is no way to reliably extrapolate the interferometer measurements to zero $uv$ distance and recover the unmeasured information on those large angular scales. However, any correlation calculation with multiwavelength data would be stymied by these negative bowls, so we remove them cosmetically, using a method similar to the negative of the `feather' technique long used for the combination of interferometer and single dish data \citep{1999MNRAS.302..417S,2007MNRAS.382..543H,Cotton2017}. 
We performed the calculation on 2200$\times$2200 pixel overlapping tiles separated by 2000 pixels, but varied both the tile and overlap size and found minimal differences in the output (the pixel scale is 1.5~arcsec, so the tiles are 0.92~degree on a side). A robust \ac{RMS} is calculated in two iterations: a first $\sigma$ is calculated from the MAD (Median Absolute Deviation), then again from all pixels less than 15 times the first estimate. Then all pixels greater than 2.5$\times\sigma$ are set to zero. 
Each tile image is padded by another 200 pixels on all sides, extending the edge values outwards, for numerical stability, Fourier transformed, and all but the $\pm$7 pixels corresponding to angular scales with wavelengths from DC to (2400/7)$\times$0.92~$\deg$ = 5.3~arcmin are zeroed. We keep a sharp cutoff in Fourier space because the goal is to remove the existing ringing/bowls. That low-pass-filtered image is inverse Fourier transformed and subtracted from the original mosaic. The resulting `background-subtracted' tiles are mosaiced back together with a linear weighting of the overlap regions. The original and `cosmetically flattened' MeerKAT images of the \ac{SMC} are shown in Figure~\ref{fig:mk-bgsub}. As with the tile size, we varied the number of pixels to keep in the background image, and whether the cutoff is sharp or slightly tapered and chose values to minimise large-scale bowling without affecting the surface brightness of real sources.

In order to compare multiwavelength images and analyse the same range of angular scales, we must impose the same filtering on those images as well.

\subsubsection{Colour index}
\begin{figure}
    \centering
    \resizebox{3.3in}{!}{\includegraphics{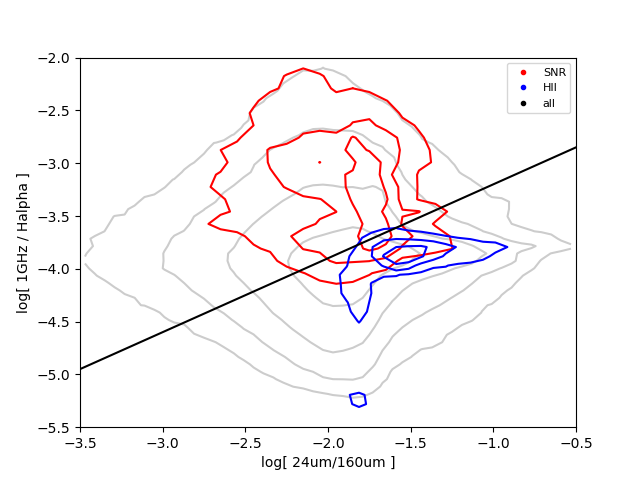}}
    \caption{Colour distributions of pixels corresponding to known \ac{SMC} \acp{SNR} (red), known \HII~regions (blue), and the entire \ac{SMC} (grey). The best-discriminating line between the known-object training sets, calculated with a support vector machine, is shown. The color index extends from --1.4 to 3.6 for points within the lowest grey contour.}
    \label{fig:diffuse_cmd}
\end{figure}

Next, we analyse the pixel-by-pixel colours (flux ratios) involving centimetre, H$\alpha$, 24~$\mu$m and 160~$\mu$m wavelengths.
Figure~\ref{fig:diffuse_cmd} shows the pixel colour distributions of the previously known \acp{SNR} (Table~\ref{tbl:snrbf}) and known \ac{SMC} \HII~regions \citep{1976MmRAS..81...89D}. The colour separation is very robust to different choices of large-spatial-scale filtering (including not doing any filtering at all on images other than the centimetre interferometric one), to modest changes in the training sets of objects, to different signal-to-noise thresholds, and even to smoothing the images up to 1~arcmin resolution. We calculate the best-discriminating line between the populations using a linear kernel support vector machine, specifically LinearSVM in scikit-learn \citep{scikit-learn}. Then for each pixel in the map, the perpendicular distance from its colours to that line defines a colour index (see~Fig.~\ref{fig:diffuse_cmd}). Quantitatively, 
\begin{eqnarray*}
    x &=& \log(F_{24\mu m}/|F_{160\mu m}|); \\
    y &=& \log(F_{cm}/|F_{H\alpha|}); \\
    {\rm dividing\ line\ } y &=& 0.7x-2.6; \\
    {\rm color\ index\ } ind &=& {{-0.7x +y +2.5}\over{\sqrt{0.7^2+1}}}.
\end{eqnarray*}

\begin{figure}
    \centering
    \resizebox{3.3in}{!}{\includegraphics{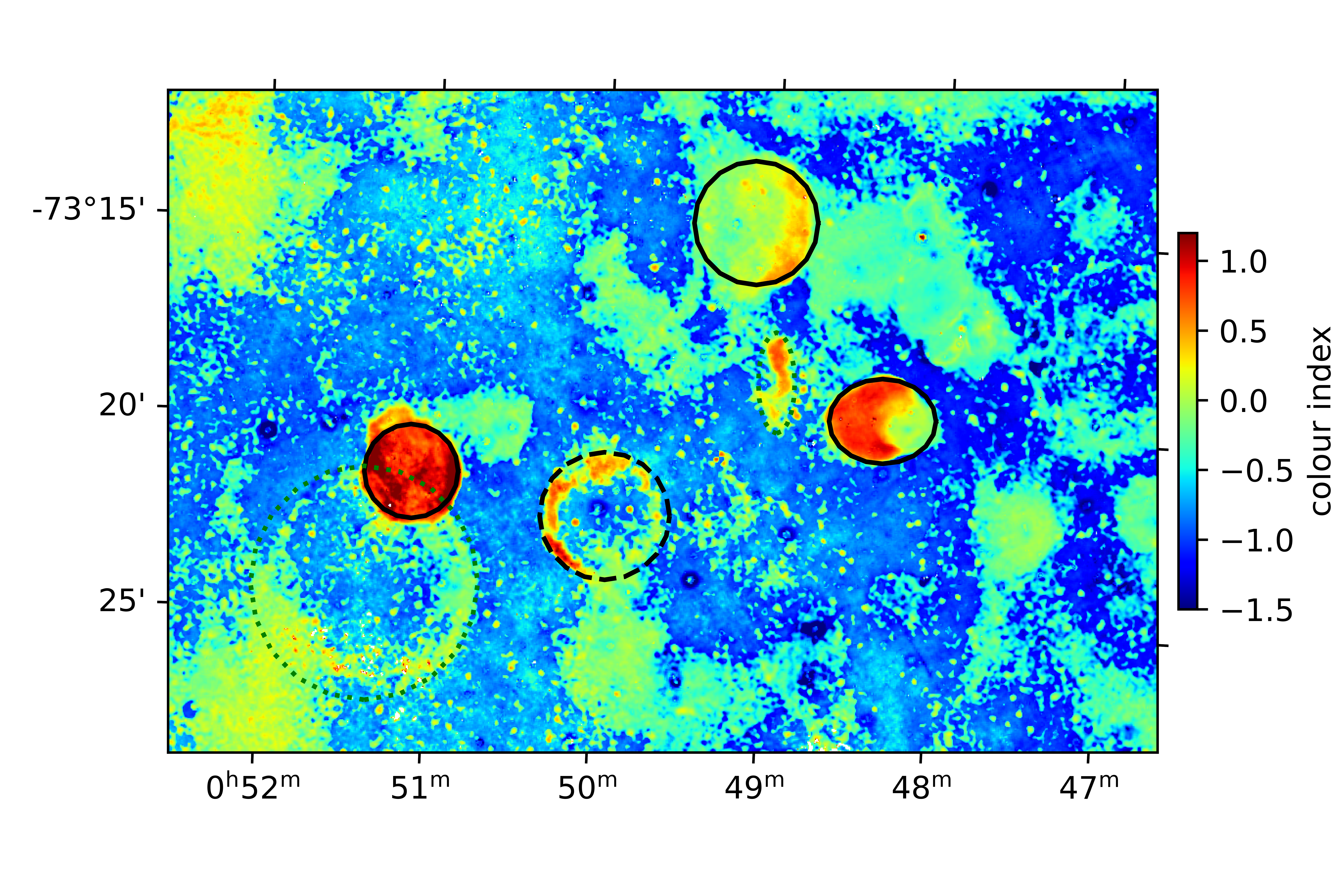}}
    \caption{Colour index calculated from centimetre (MeerKAT), H$\alpha$, 24~$\mu$m and 160~$\mu$m images as described in the text. The discriminant was trained to have redder index in known \acp{SNR}. Solid ellipses mark three such previously known \acp{SNR}. The dashed line marks a new MeerKAT \ac{SNR} candidate J0049--7322, and dotted lines mark the circularly shaped (super)bubble J005119--732408. 
    The colour index clearly reveals the dashed and dotted sources although it was not trained using those sources.}
    \label{fig:colorindex_zoom}
\end{figure}

Of the 24 confirmed \acp{SNR}, 19 have sufficient signal-to-noise  in all 4 images to calculate colour index maps; all of those have large portions of their area with colour index $>$0.3, and the median colour indices over their full extents are all positive. The five fainter \acp{SNR} only have high enough signal-to-noise to calculate indices for a modest number of pixels in their extent, but the median values of those pixels are $>$0.4 in all cases. 
Figure~\ref{fig:colorindex_zoom} shows the colour index map in the south-east \ac{SMC}, the same region as the 3$^{\rm rd}$ pane of Fig.~\ref{fig:snr1}.

\begin{figure}
    \centering
    \resizebox{3in}{!}{\includegraphics{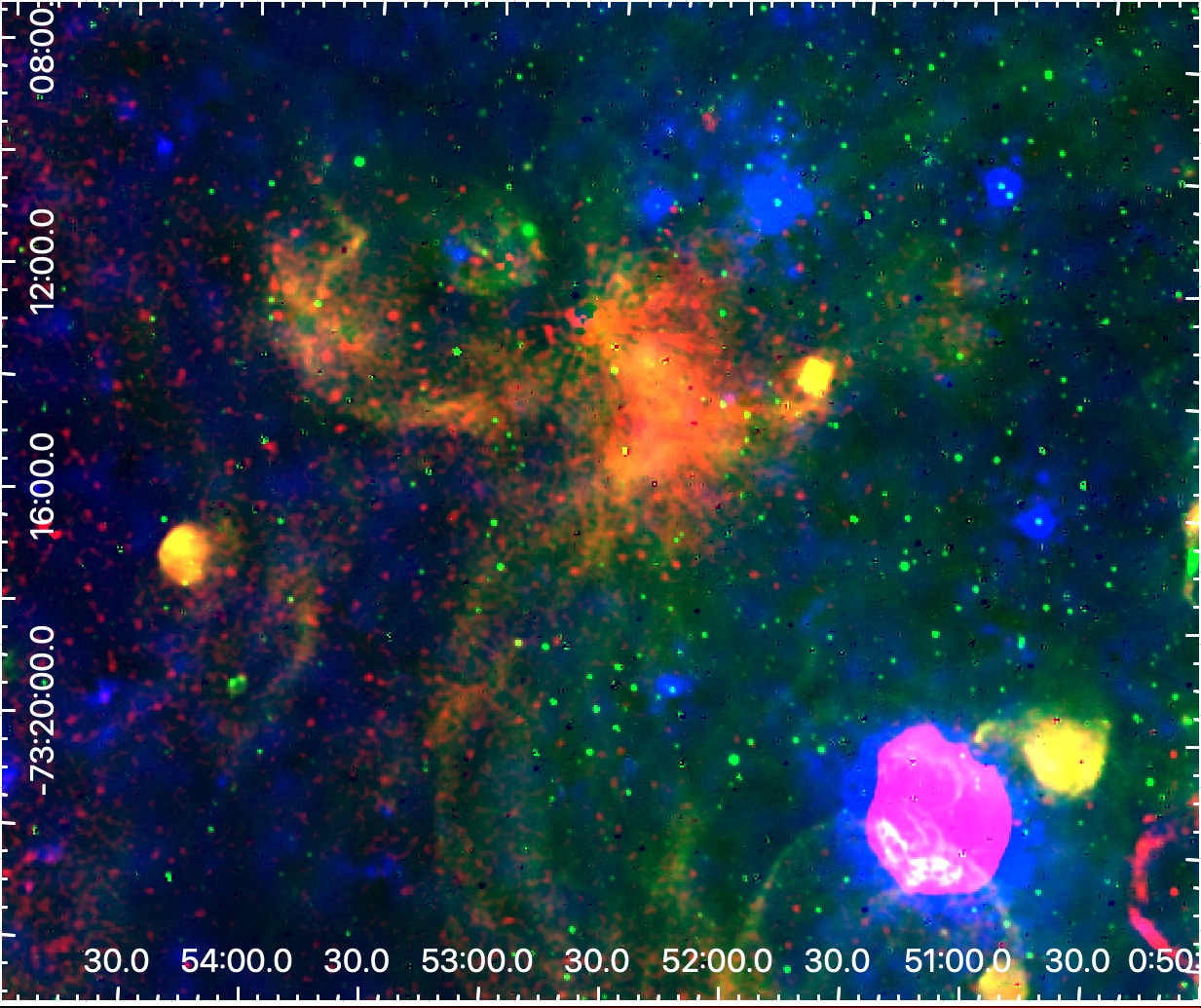}}
    \resizebox{3in}{!}{\includegraphics{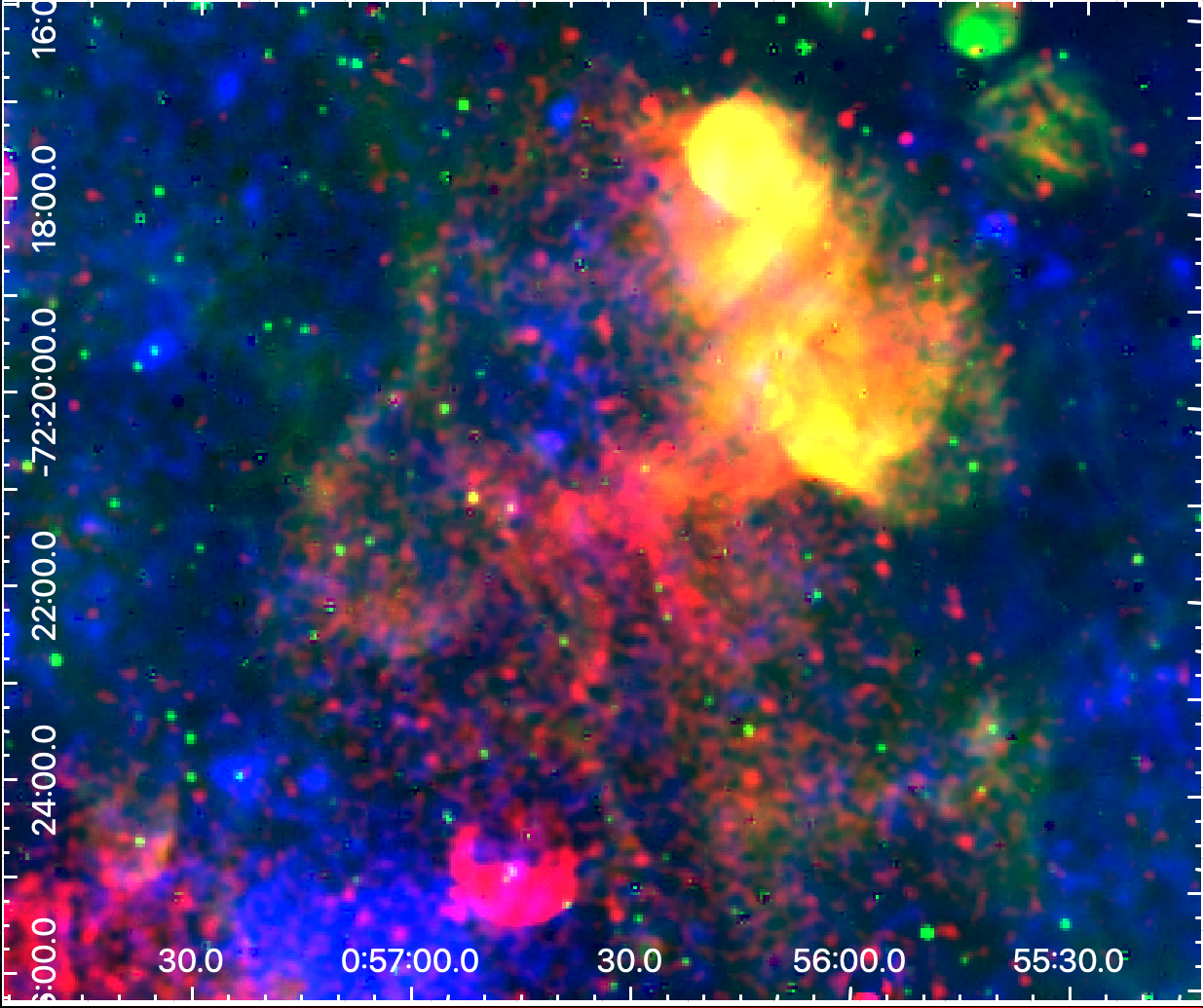}}
    \caption{Diffuse regions with high colour index suggestive of potential non-thermal origin. Red, green and blue are the MeerKAT centimetre, MCELS H$\alpha$ and soft X-ray images.
    In the top panel, the bright previously known \ac{SNR} IKT\,6 appears towards the bottom right of the field, while in the bottom panel,  the \ac{SNR} candidate J0056--7224 is visible near the bottom, left of center.}
    \label{fig:newdiffuse}
\end{figure}

To assess the efficacy of the colour index, we examined the 37 regions in the bar and wing of the \ac{SMC} (the central region shown in Figure~\ref{fig:1}) with high colour indices that were not already previously identified \acp{SNR}, all with median colour indices $>$0.25. Of these, 26 are compact, clearly background radio galaxies in the MeerKAT image. The remaining 11 include MeerKAT \ac{SNR} candidates J0049--7322, J0058--7234, J0058--7250, J0112--7304, J0124--7310, the `mystery object' (see Section~4.6.2.1) and (super)bubble J005119--732408 seen in Fig.~\ref{fig:colorindex_zoom}, the eastern lobe of the nearby (D=293$\pm$20~Mpc) radio galaxy MM~J01115--7302, QSO PKS~0128--738 ($z$=2.438), coincidentally in the middle of the S\,N90 \HII~region \citep{1956ApJS....2..315H}, and two diffuse regions shown in Fig.~\ref{fig:newdiffuse}. We note that the (super)bubble J005119--732408 is 331$\times$291~arcsec$^2$ (96$\times$86~pc$^2$) in diameter and shows a low \SII/\Ha\ ratio of $<$0.25 indicating that there are no shock excitations as one would find in \acp{SNR}. We also checked these two diffuse regions (Fig.~\ref{fig:newdiffuse}) and found that they are the most likely part of the wider \ac{SMC} \ac{ISM}.
Following this blind search, we checked the colour indices of the other five MeerKAT \ac{SNR} candidates. We found that four were too faint to have calculated the colour index map over their extent, but that the few pixels with calculated indices were all high, so consistent with the index measuring non-thermal emission. The final candidate, J0050--7238, happens to have relatively bright diffuse 24~$\mu$m and H$\alpha$ emission coincident with the centre of the \ac{SNR}, causing a lower colour index, but that emission is morphologically wispier than the rounder \ac{SNR} candidate, so likely unrelated and merely unfortunately confused along the line of sight.

\subsection{MeerKAT \ac{SMC} pulsar sample}
\label{sec:psr}

There are 10 known pulsars within the MeerKAT \ac{SMC} field. Eight of them are \ac{SMC} pulsars \citep{1991MNRAS.249..654M,2001ApJ...553..367C,2006ApJ...649..235M,2019MNRAS.487.4332T}, including an X-ray magnetar~\citep[CXOU\,J0110043.1$-$721134,][]{2002ApJ...574L..29L}. The other two are foreground Galactic pulsars: J0057$-$7201~\citep{2001ApJ...553..367C} and J0133$-$6957~\citep{1998MNRAS.295..743L}. In Table~\ref{tab:psr}, we list the parameters of these 10 pulsars obtained from the \ac{ATNF} pulsar catalogue~\citep{2005AJ....129.1993M}\footnote{http://www.atnf.csiro.au/research/pulsar/psrcat}.

\begin{table*}
\small
\begin{center}
\caption{MeerKAT pulsars in the \ac{SMC} field. The entries for the first four columns are from the \ac{ATNF} pulsar catalogue. The next four columns are based on MeerKAT data. $\dagger$ is from \citet{1994ApJ...423L..43K,1995ApJ...447L.117B} and $\ddagger$ is from \citet{2002ApJ...574L..29L}. The flux limits correspond to a 5$\sigma$ ($\sim$55\,$\mu$Jy\,beam$^{-1}$) no-detection.}
\label{tab:psr}
\begin{tabular}{lllcllcccl}
\hline
\hline
Name & RA (J2000) &  Dec (J2000) &S$_{1400}$ & RA (J2000) &  Dec (J2000) & Angular offset & S$_{1283.8}$ & $\alpha$ & Notes\\
     &     ATNF       &    ATNF          &(mJy)      & MeerKAT    &  MeerKAT     & (arcsec)         &  (mJy)     &          &      \\

\hline
J0043$-$73   & 00:43:25 (95)    & --73:11(7)          & 0.047 &            &               &      &$<0.05$ &         & \\ 
J0045$-$7042 & 00:45:25.69 (17) & --70:42:07.1 (13)   & 0.11  &            &               &      &$<0.1$ &         & \\ 
J0045$-$7319 & 00:45:35.16 (7)  & --73:19:03.0 (2)    & 0.3   & 00:45:35.2 & --73:19:03.3  & 0.5 &0.220  & --0.7(2)& Binary$^{\dagger}$ \\ 
J0052$-$72   & 00:52:28 (95)    & --72:05 (7)         & 0.039 &            &               &      &$<0.05$ &         & \\ 
J0057$-$7201 & 00:57:44.0 (4)   & --72:01:19 (2)      &       & 00:57:44.3 & --72:01:18.8  & 1.3 &0.083  &         & Galactic \\ 
J0100$-$7211 & 01:00:43.03 (11) & --72:11:33.6 (5)    &       &            &               &      &$<0.05$ &         & X-ray magnetar$^{\ddagger}$ \\ 
J0111$-$7131 & 01:11:28.77 (9)  & --71:31:46.8 (6)    & 0.06  &            &               &      &$<0.05$ &         & \\ 
J0113$-$7220 & 01:13:11.09  (3) & --72:20:32.20 (15)  & 0.4   & 01:13:11.2 & --72:20:31.8  & 0.6 &0.300  & --1.6(2)& \\ 
J0131$-$7310 & 01:31:28.51 (3)  & --73:10:09.34 (13)  & 0.15  & 01:31:28.5 & --73:10:09.9  & 0.5 &0.067  &         & \\ 
J0133$-$6957 & 01:33:32.46 (1)  & --69:57:29.68 (3)   & 0.361 & 01:33:32.6 & --69:57:29.1  & 0.8 &0.530  & --2.2(5)& Galactic \\
\hline
\end{tabular}
\end{center}
\end{table*}

In total, five pulsars were confidently ($>$5$\sigma$) detected by \aegean\ and we measured their positions and averaged flux densities at 1283.8\,MHz (see Table~\ref{tab:psr}). MeerKAT positions and timing positions agree with each other within the uncertainties. For PSR~J0057$-$7201, ours is the first published flux density value at 1.3\,GHz. For three relatively bright pulsars, J0045$-$7319, J0113$-$7220 and J0133$-$6957, we were able to measure their flux densities in multiple sub-bands and present the first measurements of their spectral indices. 
While the spectra of PSRs J0113$-$7220 and J0133$-$6957 are consistent with the steep spectrum nature of radio pulsars~\citep[e.g.,][]{2015MNRAS.449.3223D,2017PASA...34...20M,2018MNRAS.473.4436J}, PSR~J0045$-$7319 shows a relatively flat spectrum. Previous MeerKAT pulsar timing-mode observations showed that it is significantly brighter at lower frequencies~\citep{2020MNRAS.493.3608J}, suggesting a steep spectrum, and our measurement might be biased by emission related to its binary companion~\citep{1994ApJ...423L..43K,1995ApJ...447L.117B}.

The X-ray magnetar CXOU\,J0110043.1$-$721134 in MCSNR~J0100--7211 is not detected and we place a 5$\sigma$ upper limit of $55$\,$\mu$Jy on its radio flux density. However, the \ac{SNR} associated with this magnetar can be clearly identified (see Fig.~\ref{fig:snr3} top-right) as described in Section~\ref{SNR_samp}. At the position of two faint pulsars, J0052$-$72 and J0111$-$7131, we also identified potential counterparts, but with low significance. 
The position of PSR~J0052$-$72 is poorly known~\citep{2019MNRAS.487.4332T} and there are multiple radio sources within its error box of $\sim7$\,arcmin.

\subsection{MeerKAT \ac{SMC} \ac{PNe} sample}
\label{PN_samp}

The proximity of the \ac{SMC} also allows us to create a sample of radio continuum detected \ac{PNe}. At the distance of the \ac{SMC}, we expect that the largest \ac{PNe} would be a few arcsecs in diameter and therefore a point radio source in our survey. \ac{PNe} are essential to undertake a variety of studies, e.g. related to the chemical, atomic, molecular and solid-state galactic \ac{ISM} enrichment \citep{2005JKAS...38..271K,2015HiA....16..623K,book1,book2}. The MeerKAT survey can detect fainter \ac{PNe} than previous surveys, helping to build a more complete \ac{SMC} \ac{PNe} radio sample. 

Previous searches for radio continuum \ac{PNe} in the \ac{SMC} \citep{2008SerAJ.176...65P,2009MNRAS.399..769F,2010SerAJ.181...63B,2016Ap&SS.361..108L,2019A&A...621A.138K} and recently Asher~et al., in prep., yielded 24 \ac{PNe} detections. Our MeerKAT survey has revealed an additional 19 new PN radio detections (see Table~\ref{tbl:MeasurePNe}), bringing the total number of known \ac{SMC} \ac{PNe} detected in various radio surveys to 43 from the sample of 102 optically confirmed \citep[][]{2016JPhCS.728g2008D,2016Ap&SS.361..108L}. We did not detect five previously known radio \ac{PNe} --- they are below the survey's  threshold level due to an inverted (optically thick) radio spectrum. The remaining 38 (19 new and 19 previously known) MeerKAT \ac{SMC} \ac{PNe} radio detections presented here (Table~\ref{tbl:MeasurePNe}) have been visually inspected and confirmed as true point-source \ac{PNe} detections.

\begin{table*}
\small
\begin{center}
\caption{The MeerKAT radio continuum population of \ac{PNe} in the \ac{SMC}. In Column~1, with $^*$ we indicate 19 new \ac{SMC} \ac{PNe} radio detections while in Column~5 we list the separation between the optical position and our radio position determined here. 
Column~2 abbreviations are from: LHA -- \citet{1956ApJS....2..315H}, MGPN -- \citet{1985MNRAS.213..491M}, LIN -- \citet{1961AJ.....66..169L}, [MA93] -- \citet{1993A&AS..102..451M}, [MA95] -- \citet{1995A&AS..112..445M}, Jacoby -- \citet{1980ApJS...42....1J} and SMP -- \citet{1978PASP...90..621S}.
}
\label{tbl:MeasurePNe}
\begin{tabular}{llllccc}
\hline

Name & PN Name & RA & DEC & Separation & $S_{1283.8\,\mathrm{MHz}}$ & $\alpha\pm\Delta\alpha$ \\
     &         & (J2000) & (J2000) & (arcsec) & (mJy) & \\

\hline
  J002359--733804$^*$ & LHA 115-N1   & 00:23:59 & --73:38:03.9 & 0.1 & 0.084 $\pm$ 0.011 & --- \\
  J002921--721407$^*$ & MGPN SMC3    & 00:29:21 & --72:14:07.4 & 0.8 & 0.040 $\pm$ 0.009 & ---\\
  J003239--714158     & LIN14        & 00:32:39 & --71:41:57.6 & 2.4 & 1.513 $\pm$ 0.012 & --0.12$\pm$0.05\\
  J003422--731321     & LHA 115-N 4  & 00:34:22 & --73:13:21.5 & 0.3 & 0.302 $\pm$ 0.009 & --0.38$\pm$0.20\\
  J004037--732546$^*$ & [MA93] 14    & 00:40:37 & --73:25:46.3 & 0.9 & 0.092 $\pm$ 0.009 & ---\\
  
  J004046--751621$^*$ & SMP SMC 4    & 00:40:46 & --75:16:21.0 & 0.2 & 0.463 $\pm$ 0.010 & 0.04$\pm$0.12\\
  J004122--724517     & LHA 115-N 5  & 00:41:22 & --72:45:17.0 & 0.2 & 0.713 $\pm$ 0.009 & 0.21$\pm$0.08\\
  J004127--750251$^*$ & MGPN SMC 5   & 00:41:27 & --75:02:51.2 & 0.5 & 0.067 $\pm$ 0.008 & ---\\
  J004228--732055$^*$ & Jacoby SMC 1 & 00:42:28 & --73:20:55.2 & 0.2 & 0.160 $\pm$ 0.010 & --0.22$\pm$0.28\\
  J004309--730804$^*$ & [MA93] 44    & 00:43:09 & --73:08:04.5 & 0.9 & 0.065 $\pm$ 0.011 & ---\\
  
  J004325--723819     & LIN 43       & 00:43:25 & --72:38:19.0 & 0.2 & 0.542 $\pm$ 0.009 & 0.18$\pm$0.07\\
  J004426--735139$^*$ & MGPN SMC 6   & 00:44:26 & --73:51:39.4 & 0.2 & 0.070 $\pm$ 0.007 & ---\\
  J004521--732411     & LIN 66       & 00:45:21 & --73:24:10.9 & 0.9 & 0.221 $\pm$ 0.012 & --1.00$\pm$0.16\\
  J004528--734215$^*$ & LIN71        & 00:45:28 & --73:42:15.1 & 0.7 & 0.097 $\pm$ 0.007 & ---\\
  J004700--724917     & LHA 115-N 18 & 00:47:00 & --72:49:16.6 & 0.2 & 0.369 $\pm$ 0.007 & --0.20$\pm$0.09\\
  
  J004921--735258$^*$ & [MA93] 291   & 00:49:21 & --73:52:58.4 & 0.2 & 0.117 $\pm$ 0.007 & ---\\
  J004952--734421     & LIN 144      & 00:49:52 & --73:44:21.4 & 0.3 & 0.600 $\pm$ 0.008 & 1.03$\pm$0.06\\
  J005035--734258     & LIN 134      & 00:50:35 & --73:42:58.1 & 0.2 & 0.452 $\pm$ 0.007 & --0.08$\pm$0.09\\
  J005107--735738     & LIN 174      & 00:51:07 & --73:57:37.7 & 0.9 & 0.298 $\pm$ 0.008 & 1.01$\pm$0.17\\
  J005156--712444     & LIN 191      & 00:51:56 & --71:24:44.4 & 0.2 & 0.917 $\pm$ 0.011 & 0.57$\pm$0.05\\
  
  J005311--724507     & Jacoby SMC 20& 00:53:11 & --72:45:07.7 & 0.2 & 0.509 $\pm$ 0.014 & 0.19$\pm$0.05\\
  J005343--733707$^*$ & [MA93]700    & 00:53:43 & --73:37:06.9 & 0.2 & 0.074 $\pm$ 0.009 & ---\\
  J005605--701926$^*$ & LHA 115-N 54 & 00:56:05 & --70:19:25.6 & 1.2 & 0.088 $\pm$ 0.008 & ---\\
  J005620--720658     & LIN 302      & 00:56:19 & --72:06:58.4 & 1.0 & 0.211 $\pm$ 0.010 & --0.53$\pm$0.31\\
  J005631--722702     & [MA93]943    & 00:56:31 & --72:27:02.3 & 0.4 & 0.334 $\pm$ 0.010 & 1.00$\pm$0.23\\
  
  J005723--724817$^*$ & [MA93]999    & 00:57:23 & --72:48:17.6 & 1.2 & 0.065 $\pm$ 0.012 & ---\\
  J005837--713549     & LIN 333      & 00:58:37 & --71:35:49.1 & 0.4 & 0.541 $\pm$ 0.009 & --0.15$\pm$0.10\\
  J005842--725700     & Jacoby SMC 26& 00:58:42 & --72:57:00.0 & 0.5 & 0.278 $\pm$ 0.010 & --0.47$\pm$0.16\\
  J005916--720200     & LIN 347      & 00:59:16 & --72:02:00.0 & 0.3 & 0.573 $\pm$ 0.012 & 0.49$\pm$0.09\\
  J005940--713815$^*$ & LIN 357      & 00:59:40 & --71:38:15.4 & 1.0 & 0.066 $\pm$ 0.018 & ---\\
  
  J010418--732151     & LIN 430      & 01:04:18 & --73:21:51.3 & 0.5 & 0.165 $\pm$ 0.008 & --0.48$\pm$0.22\\
  J010451--730522$^*$ & MGPN SMC 10  & 01:04:51 & --73:05:21.6 & 1.2 & 0.098 $\pm$ 0.011 & ---\\
  J011211--712651$^*$ & [MA93] 1757  & 01:12:11 & --71:26:51.1 & 1.1 & 0.121 $\pm$ 0.012 & ---\\
  J011240--725347$^*$ & [MA93] 1762  & 01:12:40 & --72:53:47.0 & 0.4 & 0.074 $\pm$ 0.010 & ---\\
  J011424--710521$^*$ & MGPN SMC 12  & 01:14:24 & --71:05:20.9 & 0.3 & 0.138 $\pm$ 0.011 & ---\\
  
  J012111--731435     & LHA 115-N 87 & 01:21:11 & --73:14:34.8 & 0.1 & 0.836 $\pm$ 0.010 & 0.42$\pm$0.09\\
  J012412--740232     & LIN536       & 01:24:12 & --74:02:32.0 & 0.4 & 0.283 $\pm$ 0.010 & 0.67$\pm$0.23\\
  J012440--713727$^*$ & [M95]8       & 01:24:40 & --71:37:27.0 & 0.2 & 0.067 $\pm$ 0.010 & ---\\
\hline\end{tabular}
\end{center}
\end{table*}

The results for these 38 \ac{SMC} \ac{PNe} are presented in Table~\ref{tbl:MeasurePNe} which includes \ac{PNe} other names, position, separation between optical and radio position, flux density, and spectral index (where possible) based on the MeerKAT in-band measured flux densities. The average separation between our radio sources and corresponding previously established optical detections is 0.55~arcsec with SD of 0.13~arcsec, indicating that the majority of these correspond to true \ac{PNe} detections. However, \ac{PNe} LIN14 has a significantly larger separation (2.4~arcsec), and being the strongest radio source in this sample (1.5\,mJy) might indicate a possible different nature as examined in \citet{2009MNRAS.399..769F}. 
Most of the \ac{SMC} \ac{PNe} have typical radio continuum spectra \citep[--0.5$<\alpha<$+1.0;][]{2009MNRAS.399..769F,2021MNRAS.503.2887BF} and their flux densities are in agreement with previous measurements at similar frequencies where those exist. 
A more in-depth study of the entire \ac{SMC} \ac{PNe} sample will be presented elsewhere (Asher~et al., in prep.).

\subsection{Radio stars in the SMC field}
\label{sec:stars}

We have searched for radio stars in the field of view of the MeerKAT image, comprising both thin and thick disk populations from our Galaxy and stars in the \ac{SMC}. As noted before, our  point source catalogue contains \TOTALNUMBERCHZEROPEAKFLUX\ sources, and we cross-matched this with  SIMBAD \citep{2000A&AS..143....9W} and GAIA DR3 \citep{2016A&A...595A...2G,2022arXiv220800211G}. Due to the nature of the field, the GAIA catalogue is extremely crowded, making the cross-matching process difficult, with a high number of spurious matchings, calculated by shifting the MeerKAT source catalogue by 0.1~degree and matching it with the GAIA one. To find a compromise between having a good number of matches and not having too many spurious matches, we analysed the spurious match percentage $p$ and the number of new sources $N_\mathrm{new}$ at different matching radii. In the following, $N_{r_{i}}$ and $S_{r_{i}}$ are the number of matches and spurious matches, respectively, that we obtain using $r_{i}$ as our matching radius, where we step through radii in bins of 0.05 arcsec:
\begin{eqnarray}
N_\mathrm{new}=(N_{r_{i}}-S_{r_{i}})-(N_{r_{i-1}}-S_{r_{i-1}}); \\
p=100*S_{r_{i}}/N_{r_{i}}.
\end{eqnarray}

\begin{figure}
    \centering
    \includegraphics[width=\columnwidth]{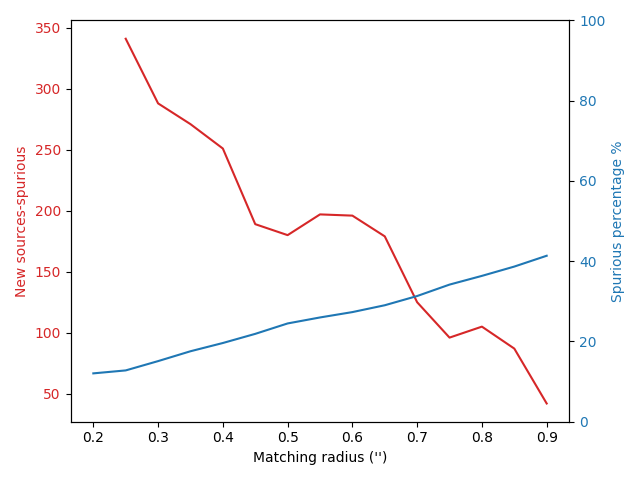}
    \caption{Plot of $N_\mathrm{new}$ (in red) and the spurious percentage $p$ (in blue) as a function of the matching radius. $p$ increases with matching radius, while $N_\mathrm{new}$ decreases, with a possible plateau around 0.6~arcsec. }
    \label{fig:stars}
\end{figure}

As can be seen in Fig.~\ref{fig:stars}, $p$ increases with  matching radius while $N_\mathrm{new}$ tends to decrease. A good compromise can be found at $r$=0.6~arcsec, where $N_\mathrm{new}$ has a local maximum of $N_\mathrm{new}\approx200$ and $p\lesssim$30~per~cent. At $r$=0.6~arcsec we found 3777 matches, with $\sim1000$ expected to be spurious. GAIA DR3 supplies several parameters for each source, in particular three of great relevance here: $p_\mathrm{star}$, $p_\mathrm{galaxy}$ and $p_\mathrm{quasar}$, the probability, knowing all the other parameters about that particular source, of the source being a star, a galaxy or a quasar, respectively \citep{2022arXiv220605864C}. It is important to note that the compact MeerKAT SMC catalogue sources suffer from the strong bias of being $\sim1\,$GHz radio sources, meaning that their probability of being a galaxy is higher than the typical GAIA source. To check the reliability of these parameters we matched the MeerKAT sources with SIMBAD, finding 415 matches within a 1~arcsec radius that also have a GAIA match. Among them, 101 sources have $p_\mathrm{star}>0.99$ in GAIA but 29  are classified as being galaxies in SIMBAD. It is therefore clear that this parameter alone cannot be used to discriminate between radio stars and galaxies.

To create  catalogues of radio star candidates we proceeded on two parallel tracks: exploiting the proper motion parameter and using the SIMBAD catalogue to isolate  known stars. Proper motion is a constraining parameter, whereby nearly all  sources with a measured proper motion should  be stars. 
Among all the sources in the MeerKAT catalogue, there are just 141  with a measured proper motion $\mu>5\,\mathrm{mas~year^{-1}}$, 85~per~cent of which are classified with the $p_\mathrm{star}$ parameter; 91 of them  have $p_\mathrm{star}>0.99$. Using a more stringent cut of $\mu>10\,\mathrm{mas~year^{-1}}$, we find 32 sources, 26 with the $p_\mathrm{star}$ parameter, and 24  with $p_\mathrm{star}>0.99$. 

We assemble a ``golden sample'' of radio stars from these 24 sources plus another 5 sources with a measured proper motion
$\mu > 5\,\mathrm{mas~year^{-1}}$ and classified as stars by SIMBAD. These 29 sources are listed in Table~\ref{tab:gsa}. With fewer restrictions (e.g., on proper motion cuts) one can assemble different samples that capture a larger numbers of stellar objects, including \ac{YSOs}, red giant branch candidate, and RR~Lyrae, even if at the cost of larger cross-contamination. That work will be presented elsewhere.

To compare the 29 stars in our golden sample with expectations from the literature, we considered the work of \citet{Yu2021}. Using their scaling relations for a detection threshold of $\sim\,20\mu\mathrm{Jy}$ (the faintest detection in our MeerKAT catalogue), and considering only M stars (our SMC field is away from the Galactic Plane, and at this sensitivity M stars should be closer than $\sim200\,\mathrm{pc}$ and distributed approximately isotropically, unlike some other stellar populations), we expect that $\sim80$ M stars might be detected across our 7\D$\times$7\D field. Given this very rough extrapolation, and considering that there are more stars detected in our field of view once the stringent constraints of the golden sample are relaxed, the detection rate is compatible with expectations. Considering that only $\sim 1000$ radio stars have been identified to date (see \citealt{Gudel}), the likely identification of $\sim 100$ new ones in the small field of view of our SMC survey points to the ability of high sensitivity and resolution surveys such as this one to significantly advance the study of radio stars.

\begin{table*}
\begin{center}
\caption{Golden sample of candidate radio stars, selected as described in Section~\ref{sec:stars}.}
\label{tab:gsa}
\begin{tabular}{lllllll}
\hline
Name & RA (J2000) & DEC (J2000) & GAIA designation & pm & $p_\mathrm{star}$& SIMBAD \\
 &  &  &  & (mas~yr$^{-1}$)  & & main\_type \\
\hline
J004231--741155 & 00:42:31.07 & --74:11:55.0 & Gaia DR3 4685749263432487040 & 5.96 & 0.99992 & Star \\
J004403--743734 & 00:44:03.49 & --74:37:34.0 & Gaia DR3 4685490259707560832 & 31.4 & 0.9998 & RotV* \\
J005205--744649 & 00:52:04.85 & --74:46:49.3 & Gaia DR3 4684816740118487936 & 32.8 & 0.9988 & Star \\
J005325--710243 & 00:53:25.46 & --71:02:42.7 & Gaia DR3 4689370676764568576 & 37.4 & 0.9998 & Star \\
J005611--724749 & 00:56:10.82 & --72:47:49.2 & Gaia DR3 4685980435808569344 & 5.89 & 0.000004 & Star \\
J010314--705059 & 01:03:14.46 & --70:50:59.0 & Gaia DR3 4690839693019667840 & 6.83 & - & Star \\
J011322--724520 & 01:13:21.92 & --72:45:20.4 & Gaia DR3 4687241365113885440 & 20.5 & 0.9997 & Star \\
J011517--731143 & 01:15:16.56 & --73:11:43.2 & Gaia DR3 4687166877510636032 & 31.1 & - & Star \\
J011705--722844 & 01:17:04.72 & --72:28:44.2 & Gaia DR3 4687350732164591104 & 6.54 & 0.99991 & Star \\
J011710--702150 & 01:17:09.72 & --70:21:49.7 & Gaia DR3 4690959294964458880 & 33.2 & 0.9997 & Star \\
J012711--701022 & 01:27:10.58 & --70:10:22.5 & Gaia DR3 4691120446431145088 & 10.3 & 0.999991 & Cepheid \\
J002741--744049 & 00:27:41.09 & --74:40:48.9 & Gaia DR3 4685345467763983104 & 16.5 & 0.99998 & - \\
J003329--713246 & 00:33:29.25 & --71:32:46.1 & Gaia DR3 4690033750987052800 & 52.0 & 0.99994 & - \\
J004627--734621 & 00:46:27.07 & --73:46:20.9 & Gaia DR3 4685776265879398656 & 10.9 & 0.9998 & - \\
J004733--731701 & 00:47:33.22 & --73:17:01.0 & Gaia DR3 4685942708820468224 & 12.8 & 0.997 & - \\
J004752--732455 & 00:47:52.47 & --73:24:54.9 & Gaia DR3 4685926422305317504 & 10.1 & 0.9997 & - \\
J005445--745554 & 00:54:45.49 & --74:55:53.9 & Gaia DR3 4684753312040026112 & 38.8 & 0.996 & - \\
J005621--730647 & 00:56:20.66 & --73:06:46.6 & Gaia DR3 4685912815770474368 & 13.0 & 0.995 & - \\
J005759--723741 & 00:57:59.21 & --72:37:40.8 & Gaia DR3 4685988613398832000 & 21.2 & 0.99997 & - \\
J005815--744629 & 00:58:14.85 & --74:46:29.0 & Gaia DR3 4684760561944723456 & 29.4 & 0.99991 & - \\
J010338--732726 & 01:03:38.34 & --73:27:25.8 & Gaia DR3 4685691573434514176 & 10.2 & 0.994 & - \\
J010631--714838 & 01:06:31.12 & --71:48:38.5 & Gaia DR3 4690540282257021440 & 10.9 & 0.99992 & -\\
J010711--753224 & 01:07:11.07 & --75:32:24.1 & Gaia DR3 4684485198699244672 & 30.0 & 0.99988 & - \\
J010746--704641 & 01:07:45.88 & --70:46:41.0 & Gaia DR3 4690832683625105152 & 12.4 & 0.99997 & - \\
J011249--701317 & 01:12:48.53 & --70:13:17.0 & Gaia DR3 4690956138162512256 & 20.5 & 0.99995 & - \\
J012221--714122 & 01:22:20.77 & --71:41:22.2 & Gaia DR3 4687587337618874240 & 18.4 & 0.99997 & - \\
J012740--734651 & 01:27:39.77 & --73:46:50.6 & Gaia DR3 4686223290360321664 & 36.2 & 0.99998 & - \\
J013527--725507 & 01:35:27.13 & --72:55:07.5 & Gaia DR3 4686843586718702464 & 24.4 & 0.999992 & - \\
J014028--702341 & 01:40:27.60 & --70:23:41.0 & Gaia DR3 4688123860632580352 & 10.1 & 0.9998 & - \\
\hline
\end{tabular}
\end{center}
\end{table*}

\subsection{Individual sources of interest} 
 \label{Other_sources}
 
We also show other interesting sources found in these MeerKAT \ac{SMC} images. They encompass various foreground and background sources such as stars and \ac{AGN}.

\subsubsection{Circularly polarised point sources}

\begin{figure*}
		\centering
			\includegraphics[width=\textwidth]{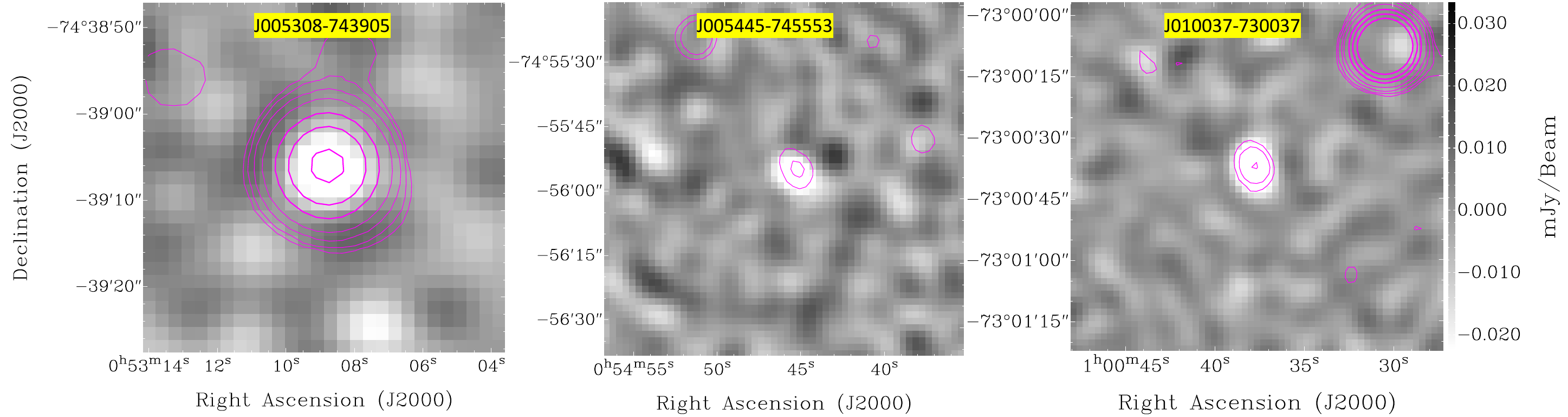}
        \caption{Three circularly polarised nearby stars in the field of the \ac{SMC}: V$^*$~CF Tuc (left), [HFP2000]\,665 (middle) and ISO-MCMS~J010037.4--730036 (right). Greyscale is from Stokes~V image while contours (3$\sigma$, 5$\sigma$, 10$\sigma$, 20$\sigma$, 50$\sigma$, 100$\sigma$ and 200$\sigma$) are from Stokes~I image.}
        \label{fig:vi}
\end{figure*}

We use our new MeerKAT Stokes~V image of the \ac{SMC} field to search for circularly polarised sources. Applying a cut-off |V/I| intensity of $>$1~per~cent we found 11 sources (Table~\ref{tab:v}) from which the brightest (|V/I|) two are nearby stars \citep[ISO-MCMS~J010037.4--730036 and HFP2000\,665;][]{haberl2000,sturm2013} at a distance of D=81~pc and 86~pc (Fig.~\ref{fig:vi}). Another source found in this study is the well-known nearby radio star V*~CF~Tuc \citep[][]{1999MNRAS.305..966B} at D=87~pc,  an RS~CVn type binary (Fig.~\ref{fig:vi}, left). 
These three and J013117$-$730356 (4XMM\,J013117.0$-$730356) are detected in X-rays (see below).
Using \textsc{VizieR} we found no counterpart at other wavelengths within 3~arcsec for three sources (J002208--712355, J005244--755103, and J011549--755652), while all others are detected at infrared wavebands by \Spitzer\ and \WISE\ \citep[CatWISE2020 catalogue;][]{2021ApJS..253....8M}.

The peak flux density for these 11 sources ranges over 2.5 orders of magnitude: 0.06--14~mJy~beam$^{-1}$. Similarly, the spectral index is also widely distributed. Source J005243$-$755102 displays a nominally very steep spectrum with $\alpha =-1.47\pm0.73$, although the uncertainty is very large despite 12 sub-band flux density measurements. At face value it is challenging to physically interpret this in terms of an incoherent emission mechanism.

In addition to being a highly circularly polarised source, ISO-MCMS J010037.4--730036 is a variable object. There were seven images from our MeerKAT survey including its position; three observations done on the same day all showed bright I and V emission but with none of the artefacts seen that are characteristic of rapidly variable sources. None of the other four observations (two on the same day but different from those in which signals were detected) showed detectable emission in I or V. Hence, the time scale on which substantial variability occurs is of order a day. It is unlikely that J010037.4--730036 is a flare star -- the variability time scale is too long. It may be an RS~CVn-type binary, which are known to produce highly circularly polarised emission at decimetre wavelengths that are polarised in the opposite sense of higher-frequency emissions. This has been interpreted in terms of plasma radiation \citep[e.g.,][and references therein]{1995ApJ...444..342W}. However, the degree of polarisation is typically less ($<50$~per~cent) than that observed on J010037.4--73003. A more likely possibility is that this source is a late-type dwarf and auroral emitter by means of the cyclotron maser instability \citep[e.g.,][]{2006A&ARv..13..229T}. A recent example of this class of radio emitter is the dM5.5 star UV~Cet discussed by \citet{2022ApJ...935...99B} which produces flux density similar in I and V when scaled to the appropriate distance.


\begin{table*}
\begin{center}
\caption{MeerKAT Stokes~V point sources in the \ac{SMC} field. The source positions (RA and Dec) are from the V image. In Column~5 we list the source peak flux density at MeerKAT broad-band frequency of 1283.8~MHz.}
\label{tab:v}
\begin{tabular}{llllcccllcccc}
\hline
No. & Name            & RA          &  Dec          & $S_{\rm p}$       & |V/I|   & $\alpha\pm\Delta\alpha$ & Notes \\
    &                 & (J2000)     &  (J2000)      & (mJy~beam$^{-1}$) & (\%)  &                         &  \\
\hline
1   & J002208--712355 & 00:22:08.5  & --71:23:55.5  & 0.169              & 24.5 & --1.15$\pm$0.21 & \\ 
2   & J005244--755103 & 00:52:43.7  & --75:51:02.7  & 1.021              & 7.1  & --1.47$\pm$0.72 & \\ 
3   & J005309--743905 & 00:53:08.9  & --74:39:05.9  & 2.333              & 1.7  & +0.04$\pm$0.02  & V$^*$~CF Tuc; \citet{1999MNRAS.305..966B} \\ 
4   & J005445--745554 & 00:54:45.5  & --74:55:53.9  & 0.059              & 80.8 & ---             & X-rays: HFP2000~665; \citet{haberl2000} \\ 
5   & J005507--701749 & 00:55:07.3  & --70:17:48.5  & 2.984              & 1.3  & --0.12$\pm$0.03 & \\ 
6   & J010038--730037 & 01:00:37.7  & --73:00:37.0  & 0.103              & 70.3 & --- & X-rays: ISO-MCMS~J010037.4--730036; \citet{sturm2013} \\ 
7   & J011549--755652 & 01:15:49.2  & --75:56:57.1  & 14.193             & 1.03 & +0.56$\pm$0.03  & \\ 
8   & J012921--724944 & 01:29:21.0  & --72:49:44.1  & 0.064              & 49.7 & ---             & \\ 
9   & J013117--730356 & 01:31:17.1  & --73:03:55.9  & 4.125              & 1.0  & --0.13$\pm$0.01 & X-rays: 4XMM\,J013117.0$-$730356; \citet{webb2020} \\ 
10  & J013601--725059 & 01:36:01.2  & --72:50:59.0  & 7.024              & 1.1  & --0.32$\pm$0.07 & \\ 
11  & J013635--744331 & 01:36:35.2  & --74:43:30.7  & 4.030              & 1.3  & --0.66$\pm$0.09 & \\ 
\hline
\end{tabular}
\end{center}
\end{table*}



Three of the circularly polarised objects in Table~\ref{tab:v} have been observed in X-rays. ISO-MCMS~J010037.4--730036 was serendipitously observed since 1979 more than 35 times with various X-ray observatories and shows long-term X-ray variability of at least a factor of five \citep{sturm2013}. Source J005445$-$745554 (a.k.a. [HFP2000]\,665) was detected in a deep (50\,ks) \ROSAT\ pointed observation in October 1992, but no variability information is available. Finally, for source J013117$-$730356 we identify 4XMM\,J013117.0$-$730356 as its X-ray counterpart \citep[4XMM-DR11 catalogue;][]{webb2020}. \xmm\ observed the source twice in 2016 with one detection at (7.9$\pm$0.9) $\times$ 10$^{-15}$ erg s$^{-1}$ cm$^{-2}$. During an \Einstein\ observation in 1979 the source was a factor of $\sim$80 brighter than during the \xmm\ observation. The X-ray detections and in particular the variability support the picture of stars with a highly active corona.

\subsubsection{Extended sources}

There are a large number of extended sources with various morphologies that we catalogued as BCE (Section~\ref{sec:broadbandsdet} and Table~\ref{tab:bcecat}). Many, but not all of these, are background objects such as \ac{AGN} and star-forming galaxies.  Here we introduce a small sample of interesting objects selected from the BCE list.


\paragraph{Non-thermal filaments in the southern part of the SMC}

\begin{figure}
  \includegraphics[width=\columnwidth, trim= 0 0 0 0, clip, angle=0]{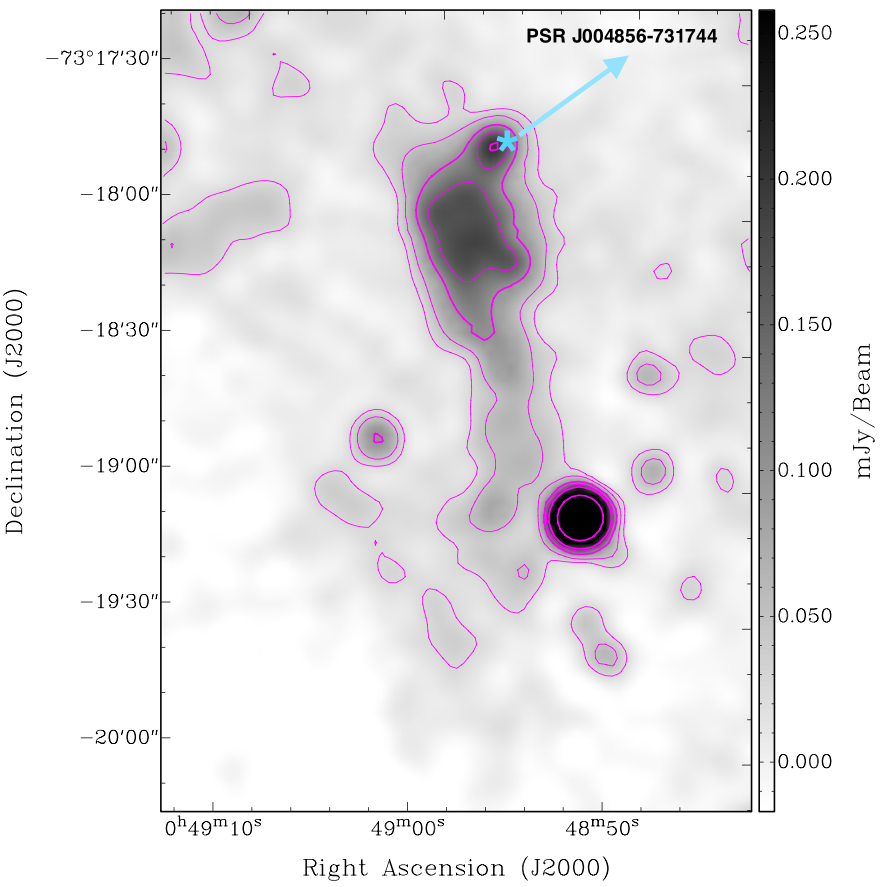}
  \caption{MeerKAT image of the unknown non-thermal feature (also marked with `?' in Fig.~\ref{fig:snr1}). The contours are 0.025, 0.05, 0.1, 0.15, 0.2 and 0.5~mJy~beam$^{-1}$. The light blue star marks the position of pulsar J004856--731744 (Carli et al. (in prep.)).}
 \label{fig:smc?}
\end{figure}

In Figs.~\ref{fig:snr1} (marked with `?') and \ref{fig:smc?}, we point to an interesting extended feature of size $\sim$100~arcsec (29\,pc at the distance to the \ac{SMC}) and position just south from the N\,19 region. We cannot associate this feature with any specific source type. Specifically, we investigated possible \ac{SNR} or \ac{AGN} nature but the morphology of this object is different. It is also exclusively seen in radio continuum frequencies suggesting non-thermal origin. Its elongated shape and vicinity to other well-known \ac{SMC} sources indicate possible \ac{SMC} membership. However, it is somewhat similar to bent and twisted radio continuum jets around the peculiar galaxy pair ESO\,295-IG022 \citep{2001A&A...369..467R,2010SerAJ.181...31F}. 
It also coincides with an \ac{AGN} candidate seen in the VISTA Magellanic Clouds (VMC) Survey \citep[][and reference therein]{2011A&A...527A.116C} image (red, probably fairly high redshift). There is a faint \ac{AGN} candidate at RA(J2000)=00$^{\rm h}$48$^{\rm m}$56$^{\rm s}$ and Dec(J2000)=--73\D18\arcmin34\arcsec which has a position consistent with being the central source of the radio jets/lobes. 
Most interestingly, Carli et al. (in prep.) report a young pulsar at RA(J2000)=00$^{\rm h}$48$^{\rm m}$56$^{\rm s}$ and Dec(J2000)=--73\D17\arcmin44'' (with 2~arcsec positional error) that is $\sim$5~arcsec from the radio point-like source J004857--731747 located at the northern tip of this feature (see Fig.~\ref{fig:smc?}). As seen in the case of the DEM\,S5 SNR, the pulsar and radio continuum emission could be slightly displaced and still be related if the latter includes emission from a \ac{PWN}. The radio feature could be a bow-shock \ac{PWN} with  the emission south of the pulsar representing the tail and the radio point source as the \ac{PWN} ``head''. There is no obvious nearby SNR that could be associated with such a putative PWN, but that would not be exceptional ---  several  Galactic bow-shock \ac{PWN}e have no parent \ac{SNR}, such as for example the ``Potoroo'' PWN (Lazarevi\'c et al. submitted). We tentatively classify this source as a \ac{PWN}, but further study is required to establish its nature.

\paragraph{Radio ring galaxy ESO~029--G034}

\begin{figure*}
		\centering
			\includegraphics[width=\textwidth]{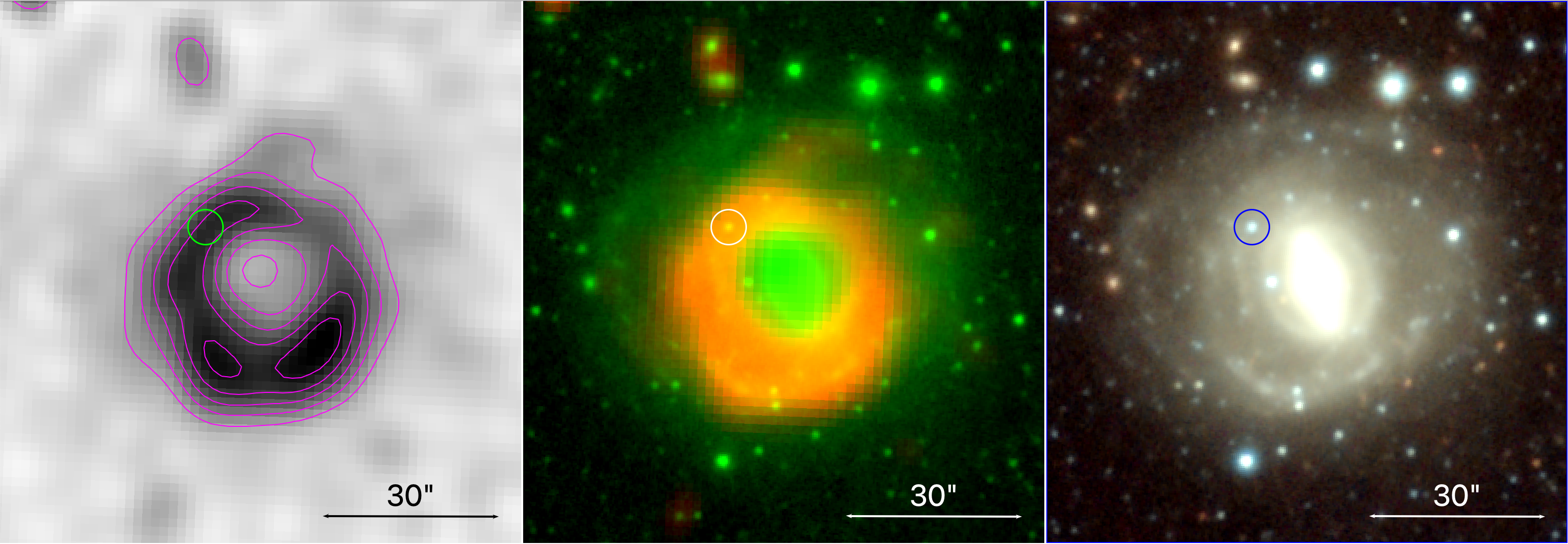}
        \caption{Spiral galaxy ESO~029--G034 (a.k.a. karakond\v zula) where left is MeerKAT image with contours of 50, 75, 100, 125 and 150~$\mu$Jy~beam$^{-1}$, the middle image is a red-green mix from VMC Ks band and MeerKAT (red). On the right is VMC Ks, J and Y bands composite RGB image where the north is up. The circles in each panel indicate the position of \ac{SN} OGLE-2014-SN-046. }
        \label{fig:bckg1}
\end{figure*}

Among the most interesting of the extragalactic discoveries in this survey is the radio ring galaxy (RaRiGx) ESO~029--G034 at a distance of 155~Mpc ($z$=0.036) and diameter of $\sim$65~kpc. This could be a new type of object. As an optical source, it is classified as an interacting double and the largest in the group \citep{1987cspg.book.....A}. However, in the VMC Survey optical images, we can clearly see a face-on spiral galaxy with the dominant optical bar and bulge in the centre. Interestingly, its radio continuum emission is anti-correlated with optical emission as can be seen in Fig.~\ref{fig:bckg1}. Also, \cite{2014ATel.6399....1E}\footnote{Also see \url{http://ogle.astrouw.edu.pl/ogle4/transients/2014/transients.html}.} reported the detection of \ac{SN} OGLE-2014-SN-046 at RA(J2000)=01$^{\rm h}$09$^{\rm m}$06.73$^{\rm s}$ and Dec(J2000)=--74\D30\arcmin46.4\arcsec. A point-like optical source linked to this Type-IIP \ac{SN} is located on the radio ring (see Fig.~\ref{fig:bckg1}). 

\paragraph{Nearby spirals}

\begin{figure}
		\centering
    \includegraphics[width=1.0\columnwidth,trim=0 0 0 0,clip]{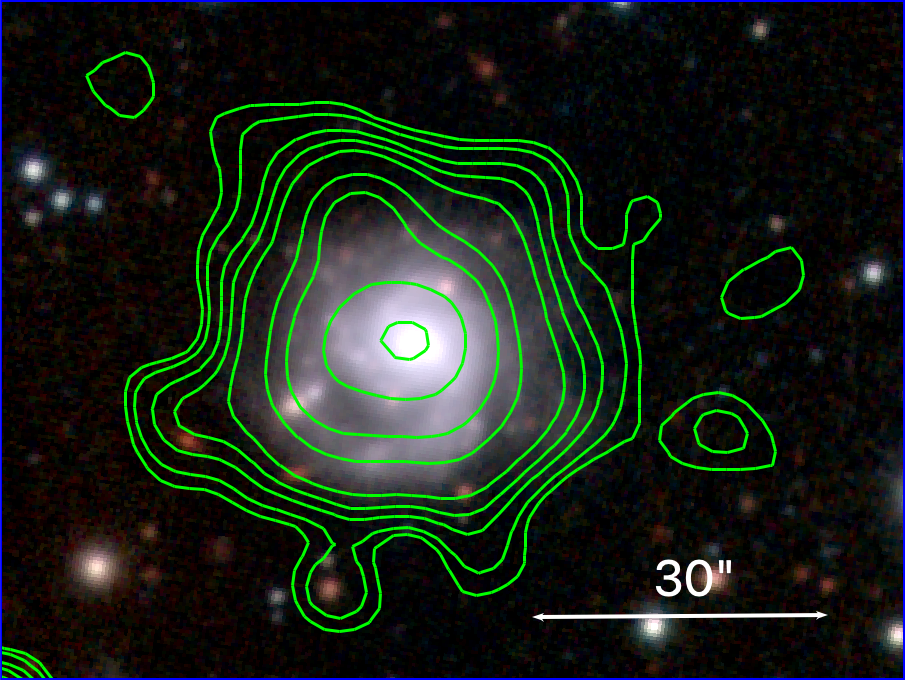}		
    \includegraphics[width=1.0\columnwidth,trim=0 0 0 0,clip]{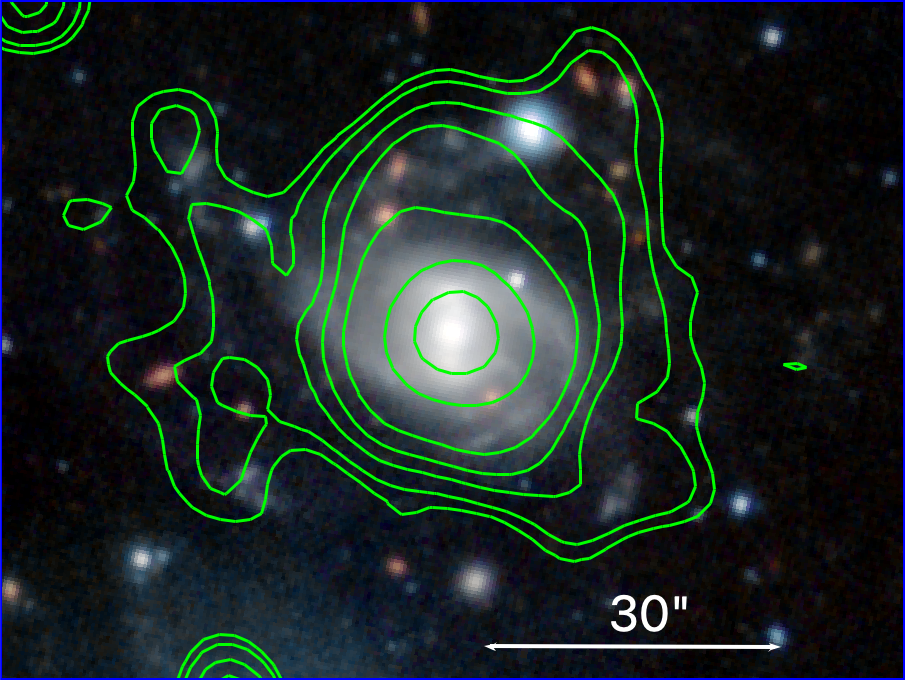}
    \includegraphics[width=1.0\columnwidth,trim=0 0 0 0,clip]{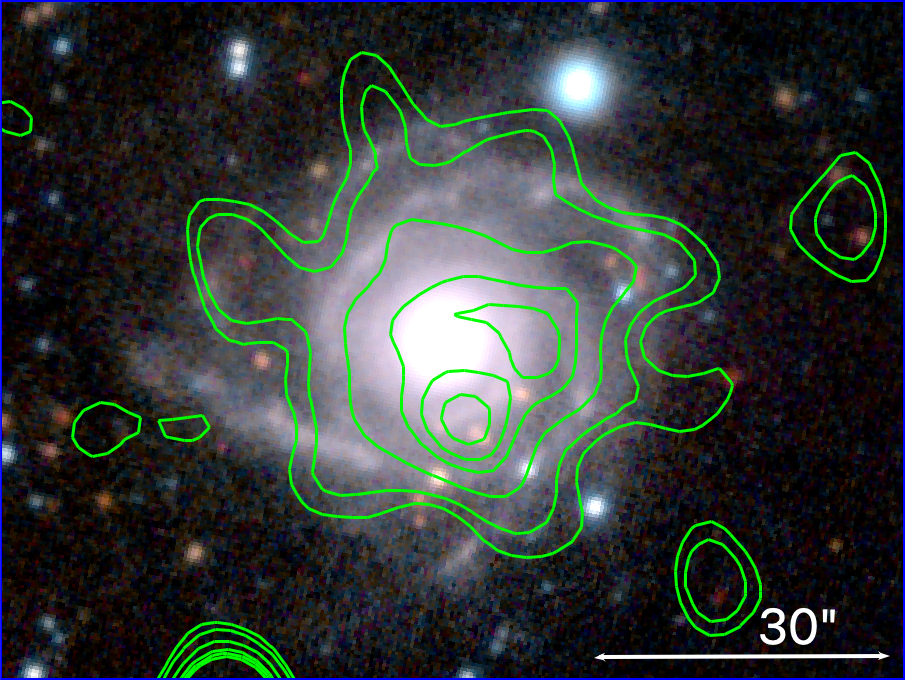}
    \caption{MeerKAT images of three nearby spiral galaxies: 2MASX~J01215707--7516248 (top; a.k.a. Stribog), ESO~029--G047 (middle; a.k.a. Simargl) and ESO~029--G049 (bottom; a.k.a. Da\v zbog). MeerKAT (green) contours are 20, 30, 50, 70, 100, 200, 300, 500, 700 and 1000~$\mu$Jy~beam$^{-1}$. The background RGB image is from the VMC survey where R=Ks, G=J and B=Y band. Note that several radio point sources (some overlapping with the nearby galaxy) are clearly identified with much more distant \ac{AGN} or radio galaxies.        
        }
        \label{fig:bckg3}
\end{figure}

As in other modern radio surveys \citep[e.g., see][]{2022A&A...657A..56K}, continuum emission is detected in our survey from nearby face-on spiral galaxies. In Fig.~\ref{fig:bckg3} we show three such visually striking objects: 2MASX~J01215707--7516248, ESO~029--G047 and ESO~029--G049 at distances of 99~Mpc, 300~Mpc and 174~Mpc, respectively.
We measure their spectral index as --0.11$\pm$0.10, --0.23$\pm$0.10 and --0.29$\pm$0.10, respectively, indicating that thermal emission dominates across these  spiral galaxies.

\section{Summary and Conclusions}
We have imaged the SMC at 0.9--1.7\,GHz using the MeerKAT radio telescope with a median Stokes I rms of $\sim$11\,$\mu$Jy\,beam$^{-1}$. A variety of products are included in the data release associated with this paper. These include a catalogue of point and compact extended radio continuum sources across a $\sim$7\D$\times$7\D\ FoV. The catalogue consists of \TOTALNUMBERCHZEROPEAKFLUX\ point sources and \TOTALBCE\ compact extended sources above 5$\times$ local \ac{RMS} in the broad-band image (a.k.a. channel~0 or total intensity/full band; frequency $\nu$=1283.8~MHz; bandwidth of 800~MHz; beam size of 8~arcsec). The vast majority of catalogued point radio sources are unrelated to the \ac{SMC}, with only a small number of intrinsic \ac{PNe}, compact \HII~regions and \ac{YSOs} as well as foreground stars. However, we expect that about 50~per~cent of \TOTALBCE\ bright compact extended sources are extragalactic objects such as \ac{AGN} with clearly visible radio jets while others could be intrinsic \ac{SMC} objects such as complex \HII~regions. We report each source position (RA, Dec and errors), peak flux densities (with corresponding local \ac{RMS} noise), integrated flux densities (with corresponding errors), and spectral index values that we judge to be reliable for a subset of the sources.  

We used our imaging products to investigate the population of \ac{SMC} \acp{SNR}, pulsars, and \ac{PNe}, while also considering some foreground and background objects. Our main results include:
\begin{enumerate}
    \item Detection of all previously established \ac{SMC} \acp{SNR} and \ac{SNR} candidates; identification of 10 new  \ac{SNR} candidates;   confirmation of MCSNR~J0100--7211 as a bona fide \ac{SNR} with the X-ray magnetar CXOU\,J0110043.1$-$721134 at its centre;
    \item Firm detection of five radio pulsars and two weak radio sources potentially associated with known pulsars, out of 10 known pulsars in the FoV. An X-ray magnetar is not detected as a radio source but is found to be in the centre of a newly established \ac{SNR}; 
    \item Detection of 38 out of 102 optically known \ac{PNe}, of which half are new detections. Almost all of them have a typical \ac{PNe} spectrum but a few such as SMP~9 (LIN66) unexpectedly show a steep spectrum;
    \item Detection of three nearby radio stars at a distance of $\sim$80~pc, among a total of 11 sources that are circularly polarised;
    \item A number of interesting background objects, of which RaRiGx ESO~029--G034 at a distance of 155~Mpc and diameter of $\sim$65~kpc might represent a new type of radio object.  
\end{enumerate}


\section*{Acknowledgements}
The MeerKAT telescope is operated by the South African Radio Astronomy Observatory, which is a facility of the National Research Foundation, an agency of the Department of Science and Innovation.
The National Radio Astronomy Observatory is a facility of the National Science Foundation, operated under a cooperative agreement by Associated Universities, Inc.

\section*{Data Availability}

Raw visibility data is available from the \ac{SARAO} archive (https://archive.sarao.ac.za/) under project codes SSV-20190715-FC-02 and SSV-20200131-SP-01.
Image data products are available using DOI {https://doi.org/10.48479/1fdb-r233}.
The L-band mosaic FITS images are as described in \cite{MFImage}.
The mosaics in Stokes~I, Q, U and V are $7.08^\circ\ \times\ 7.08^\circ$ with 1.5~arcsec pixels centred at J2000 position RA=01:00:0.0 and Dec=--73:00:00.0.
The mosaic cubes contain 16 planes.
The first plane is the band-weighted average flux density at the effective frequency of 1310.89~MHz, the second plane (Stokes~I only) is the fitted spectral index or blanked if no fit could be made, planes 3--16 are the frequency bins centred at the frequencies given in Table~\ref{tab:subchannumbers}.
Two planes are totally blanked due to \ac{RFI}.
The second plane of the Q, U and V images are not physically meaningful and are blanked.
For Stokes~I, a second cube is provided in which the first plane is the broad-band flux density, the second plane is the spectral index where fitted, the third plane is the least squares error estimate for the intensity, the fourth plane is the error estimate for spectral index and the fifth plane the $\chi ^2$ of the fit.
For Stokes Q, U and V, single-plane images of the broad-band average are provided.
A single pointing image at UHF band (544--1088 MHz) with effective frequency of 870.0 MHz is centred at RA=00:52:44.0, dec=-72:49:42 and is $5.0^\circ \times 5.0^\circ$ in extent.
The catalogues underlying this article are available on the CDS/\textsc{VizieR} (\url{https://cds.u-strasbg.fr/}) website as well as from the SARAO archive (https://archive.sarao.ac.za/). 
The full point and compact extended source catalogues (Tables~\ref{tab:main} and \ref{tab:bcecat}) are also provided through the CDS/\textsc{VizieR} service.




\bibliographystyle{mnras}
\bibliography{MeerKAT_SMC.bib} 

\begin{thebibliography}{}
\makeatletter
\relax
\def\mn@urlcharsother{\let\do\@makeother \do\$\do\&\do\#\do\^\do\_\do\%\do\~}
\def\mn@doi{\begingroup\mn@urlcharsother \@ifnextchar [ {\mn@doi@}
  {\mn@doi@[]}}
\def\mn@doi@[#1]#2{\def\@tempa{#1}\ifx\@tempa\@empty \href
  {http://dx.doi.org/#2} {doi:#2}\else \href {http://dx.doi.org/#2} {#1}\fi
  \endgroup}
\def\mn@eprint#1#2{\mn@eprint@#1:#2::\@nil}
\def\mn@eprint@arXiv#1{\href {http://arxiv.org/abs/#1} {{\tt arXiv:#1}}}
\def\mn@eprint@dblp#1{\href {http://dblp.uni-trier.de/rec/bibtex/#1.xml}
  {dblp:#1}}
\def\mn@eprint@#1:#2:#3:#4\@nil{\def\@tempa {#1}\def\@tempb {#2}\def\@tempc
  {#3}\ifx \@tempc \@empty \let \@tempc \@tempb \let \@tempb \@tempa \fi \ifx
  \@tempb \@empty \def\@tempb {arXiv}\fi \@ifundefined
  {mn@eprint@\@tempb}{\@tempb:\@tempc}{\expandafter \expandafter \csname
  mn@eprint@\@tempb\endcsname \expandafter{\@tempc}}}

\bibitem[\protect\citeauthoryear{{Alsaberi} et~al.,}{{Alsaberi}
  et~al.}{2019}]{2019MNRAS.486.2507A}
{Alsaberi} R. Z.~E.,  et~al., 2019, \mn@doi [\mnras] {10.1093/mnras/stz971},
  \href {https://ui.adsabs.harvard.edu/abs/2019MNRAS.486.2507A} {486, 2507}

\bibitem[\protect\citeauthoryear{{Arp} \& {Madore}}{{Arp} \&
  {Madore}}{1987}]{1987cspg.book.....A}
{Arp} H.~C.,  {Madore} B.,  1987, {A catalogue of southern peculiar galaxies
  and associations}.
{Cambridge ; New York : Cambridge University Press}

\bibitem[\protect\citeauthoryear{{Bastian}, {Cotton}  \& {Hallinan}}{{Bastian}
  et~al.}{2022}]{2022ApJ...935...99B}
{Bastian} T.~S.,  {Cotton} W.~D.,   {Hallinan} G.,  2022, \mn@doi [\apj]
  {10.3847/1538-4357/ac7d57}, \href
  {https://ui.adsabs.harvard.edu/abs/2022ApJ...935...99B} {935, 99}

\bibitem[\protect\citeauthoryear{{Bell}, {Bessell}, {Stappers}, {Bailes}  \&
  {Kaspi}}{{Bell} et~al.}{1995}]{1995ApJ...447L.117B}
{Bell} J.~F.,  {Bessell} M.~S.,  {Stappers} B.~W.,  {Bailes} M.,   {Kaspi}
  V.~M.,  1995, \mn@doi [\apjl] {10.1086/309565}, \href
  {https://ui.adsabs.harvard.edu/abs/1995ApJ...447L.117B} {447, L117}

\bibitem[\protect\citeauthoryear{{Boji{\v c}i{\'c}}, {Filipovi{\'c}}  \&
  {Crawford}}{{Boji{\v c}i{\'c}} et~al.}{2010}]{2010SerAJ.181...63B}
{Boji{\v c}i{\'c}} I.~S.,  {Filipovi{\'c}} M.~D.,   {Crawford} E.~J.,  2010,
  \mn@doi [Serbian Astronomical Journal] {10.2298/SAJ1081063B}, \href
  {http://adsabs.harvard.edu/abs/2010SerAJ.181...63B} {181, 63}

\bibitem[\protect\citeauthoryear{{Boji{\v{c}}i{\'c}}, {Filipovi{\'c}},
  {Uro{\v{s}}evi{\'c}}, {Parker}  \& {Galvin}}{{Boji{\v{c}}i{\'c}}
  et~al.}{2021}]{2021MNRAS.503.2887BF}
{Boji{\v{c}}i{\'c}} I.~S.,  {Filipovi{\'c}} M.~D.,  {Uro{\v{s}}evi{\'c}} D.,
  {Parker} Q.~A.,   {Galvin} T.~J.,  2021, \mn@doi [\mnras]
  {10.1093/mnras/stab687}, \href
  {https://ui.adsabs.harvard.edu/abs/2021MNRAS.503.2887B} {503, 2887}

\bibitem[\protect\citeauthoryear{{Bozzetto} et~al.,}{{Bozzetto}
  et~al.}{2017}]{2017ApJS..230....2B}
{Bozzetto} L.~M.,  et~al., 2017, \mn@doi [\apjs] {10.3847/1538-4365/aa653c},
  \href {http://adsabs.harvard.edu/abs/2017ApJS..230....2B} {230, 2}

\bibitem[\protect\citeauthoryear{{Bozzetto} et~al.,}{{Bozzetto}
  et~al.}{2023}]{2023MNRAS.518.2574B}
{Bozzetto} L.~M.,  et~al., 2023, \mn@doi [\mnras] {10.1093/mnras/stac2922},
  \href {https://ui.adsabs.harvard.edu/abs/2023MNRAS.518.2574B} {518, 2574}

\bibitem[\protect\citeauthoryear{{Broadbent}, {Haslam}  \&
  {Osborne}}{{Broadbent} et~al.}{1989}]{broadbent89}
{Broadbent} A.,  {Haslam} C.~G.~T.,   {Osborne} J.~I.,  1989, \mn@doi [\mnras]
  {10.1093/mnras/237.2.381}, \href
  {https://ui.adsabs.harvard.edu/abs/1989MNRAS.237..381B} {237, 381}

\bibitem[\protect\citeauthoryear{{Brunthaler} et~al.,}{{Brunthaler}
  et~al.}{2021}]{Brunthaler2021}
{Brunthaler} A.,  et~al., 2021, \mn@doi [A\&A] {10.1051/0004-6361/202039856},
  \href {https://ui.adsabs.harvard.edu/abs/2021A&A...651A..85B} {651, A85}

\bibitem[\protect\citeauthoryear{{Budding}, {Jones}, {Slee}  \&
  {Watson}}{{Budding} et~al.}{1999}]{1999MNRAS.305..966B}
{Budding} E.,  {Jones} K.~L.,  {Slee} O.~B.,   {Watson} L.,  1999, \mn@doi
  [\mnras] {10.1046/j.1365-8711.1999.02503.x}, \href
  {https://ui.adsabs.harvard.edu/abs/1999MNRAS.305..966B} {305, 966}

\bibitem[\protect\citeauthoryear{Camilo, Scholz, Serylak  \& et al.}{Camilo
  et~al.}{2018}]{Camilo2018}
Camilo F.,  Scholz P.,  Serylak M.,   et al. 2018, \mn@doi [ApJ]
  {10.3847/1538-4357/aab35a}, 856, 180

\bibitem[\protect\citeauthoryear{{Carli} et~al.,}{{Carli}
  et~al.}{2022}]{2022MNRAS.517.5406C}
{Carli} E.,  et~al., 2022, \mn@doi [\mnras] {10.1093/mnras/stac2883}, \href
  {https://ui.adsabs.harvard.edu/abs/2022MNRAS.517.5406C} {517, 5406}

\bibitem[\protect\citeauthoryear{{Cioni} et~al.,}{{Cioni}
  et~al.}{2011}]{2011A&A...527A.116C}
{Cioni} M. R.~L.,  et~al., 2011, \mn@doi [\aap] {10.1051/0004-6361/201016137},
  \href {https://ui.adsabs.harvard.edu/abs/2011A&A...527A.116C} {527, A116}

\bibitem[\protect\citeauthoryear{{Clarke}}{{Clarke}}{1976}]{1976MNRAS.174..393C}
{Clarke} J.~N.,  1976, \mn@doi [\mnras] {10.1093/mnras/174.2.393}, \href
  {http://adsabs.harvard.edu/abs/1976MNRAS.174..393C} {174, 393}

\bibitem[\protect\citeauthoryear{{Cohen} et~al.,}{{Cohen}
  et~al.}{2007}]{cohen07}
{Cohen} M.,  et~al., 2007, \mn@doi [\mnras] {10.1111/j.1365-2966.2006.11209.x},
  \href {https://ui.adsabs.harvard.edu/abs/2007MNRAS.374..979C} {374, 979}

\bibitem[\protect\citeauthoryear{{Collier} et~al.,}{{Collier}
  et~al.}{2018}]{2018MNRAS.477..578C}
{Collier} J.~D.,  et~al., 2018, \mn@doi [\mnras] {10.1093/mnras/sty564}, \href
  {http://adsabs.harvard.edu/abs/2018MNRAS.477..578C} {477, 578}

\bibitem[\protect\citeauthoryear{{Cotton}}{{Cotton}}{2008}]{2008PASP..120..439C}
{Cotton} W.~D.,  2008, \mn@doi [\pasp] {10.1086/586754}, \href
  {https://ui.adsabs.harvard.edu/abs/2008PASP..120..439C} {120, 439}

\bibitem[\protect\citeauthoryear{{Cotton}}{{Cotton}}{2017}]{Cotton2017}
{Cotton} W.~D.,  2017, \mn@doi [\pasp] {10.1088/1538-3873/aa793f}, \href
  {https://ui.adsabs.harvard.edu/abs/2017PASP..129i4501C} {129, 094501}

\bibitem[\protect\citeauthoryear{Cotton}{Cotton}{2019}]{MFImage}
Cotton W.~D.,  2019, Obit Memo series, 63, 1

\bibitem[\protect\citeauthoryear{Cotton et~al.,}{Cotton
  et~al.}{2018}]{Cotton2018}
Cotton W.~D.,  et~al., 2018, \mn@doi [ApJ] {10.3847/1538-4357/aaaec4}, 856, 67

\bibitem[\protect\citeauthoryear{Cotton et~al.,}{Cotton et~al.}{2020}]{XGalaxy}
Cotton W.~D.,  et~al., 2020, \mn@doi [MNRAS] {10.1093/mnras/staa1240}, 495,
  1271

\bibitem[\protect\citeauthoryear{{Crawford}, {Kaspi}, {Manchester}, {Lyne},
  {Camilo}  \& {D'Amico}}{{Crawford} et~al.}{2001}]{2001ApJ...553..367C}
{Crawford} F.,  {Kaspi} V.~M.,  {Manchester} R.~N.,  {Lyne} A.~G.,  {Camilo}
  F.,   {D'Amico} N.,  2001, \mn@doi [\apj] {10.1086/320635}, \href
  {https://ui.adsabs.harvard.edu/abs/2001ApJ...553..367C} {553, 367}

\bibitem[\protect\citeauthoryear{{Crawford}, {Filipovi{\'c}}, {de Horta},
  {Wong}, {Tothill}, {Draskovic}, {Collier}  \& {Galvin}}{{Crawford}
  et~al.}{2011}]{2011SerAJ.183...95C}
{Crawford} E.~J.,  {Filipovi{\'c}} M.~D.,  {de Horta} A.~Y.,  {Wong} G.~F.,
  {Tothill} N.~F.~H.,  {Draskovic} D.,  {Collier} J.~D.,   {Galvin} T.~J.,
  2011, \mn@doi [Serbian Astronomical Journal] {10.2298/SAJ1183095C}, \href
  {http://adsabs.harvard.edu/abs/2011SerAJ.183...95C} {183, 95}

\bibitem[\protect\citeauthoryear{{Crawford}, {Filipovi{\'c}}, {McEntaffer},
  {Brantseg}, {Heitritter}, {Roper}, {Haberl}  \&
  {Uro{\v{s}}evi{\'c}}}{{Crawford} et~al.}{2014}]{2014AJ....148...99C}
{Crawford} E.~J.,  {Filipovi{\'c}} M.~D.,  {McEntaffer} R.~L.,  {Brantseg} T.,
  {Heitritter} K.,  {Roper} Q.,  {Haberl} F.,   {Uro{\v{s}}evi{\'c}} D.,  2014,
  \mn@doi [\aj] {10.1088/0004-6256/148/5/99}, \href
  {https://ui.adsabs.harvard.edu/abs/2014AJ....148...99C} {148, 99}

\bibitem[\protect\citeauthoryear{{Creevey} et~al.,}{{Creevey}
  et~al.}{2022}]{2022arXiv220605864C}
{Creevey} O.~L.,  et~al., 2022, \mn@doi [arXiv e-prints]
  {10.48550/arXiv.2206.05864}, \href
  {https://ui.adsabs.harvard.edu/abs/2022arXiv220605864C} {p. arXiv:2206.05864}

\bibitem[\protect\citeauthoryear{{Dai} et~al.,}{{Dai}
  et~al.}{2015}]{2015MNRAS.449.3223D}
{Dai} S.,  et~al., 2015, \mn@doi [\mnras] {10.1093/mnras/stv508}, \href
  {https://ui.adsabs.harvard.edu/abs/2015MNRAS.449.3223D} {449, 3223}

\bibitem[\protect\citeauthoryear{{Davies}, {Elliott}  \& {Meaburn}}{{Davies}
  et~al.}{1976}]{1976MmRAS..81...89D}
{Davies} R.~D.,  {Elliott} K.~H.,   {Meaburn} J.,  1976, \memras, \href
  {http://adsabs.harvard.edu/abs/1976MmRAS..81...89D} {81, 89}

\bibitem[\protect\citeauthoryear{{Dempsey} et~al.,}{{Dempsey}
  et~al.}{2022}]{2022PASA...39...34D}
{Dempsey} J.,  et~al., 2022, \mn@doi [\pasa] {10.1017/pasa.2022.18}, \href
  {https://ui.adsabs.harvard.edu/abs/2022PASA...39...34D} {39, e034}

\bibitem[\protect\citeauthoryear{{Dra{\v{s}}kovi{\'c}}, {Parker}, {Reid}  \&
  {Stupar}}{{Dra{\v{s}}kovi{\'c}} et~al.}{2016}]{2016JPhCS.728g2008D}
{Dra{\v{s}}kovi{\'c}} D.,  {Parker} Q.~A.,  {Reid} W.~A.,   {Stupar} M.,  2016,
  in Journal of Physics Conference Series. p. 072008 (\mn@eprint {arXiv}
  {1603.07955}), \mn@doi{10.1088/1742-6596/728/7/072008}

\bibitem[\protect\citeauthoryear{{Elias-Rosa} et~al.,}{{Elias-Rosa}
  et~al.}{2014}]{2014ATel.6399....1E}
{Elias-Rosa} N.,  et~al., 2014, The Astronomer's Telegram, \href
  {https://ui.adsabs.harvard.edu/abs/2014ATel.6399....1E} {6399, 1}

\bibitem[\protect\citeauthoryear{{Esposito}, {Rea}  \& {Israel}}{{Esposito}
  et~al.}{2021}]{2021ASSL..461...97E}
{Esposito} P.,  {Rea} N.,   {Israel} G.~L.,  2021, in {Belloni} T.~M.,
  {M{\'e}ndez} M.,   {Zhang} C.,  eds,  Astrophysics and Space Science Library
  Vol. 461, Astrophysics and Space Science Library. pp 97--142 (\mn@eprint
  {arXiv} {1803.05716}), \mn@doi{10.1007/978-3-662-62110-3_3}

\bibitem[\protect\citeauthoryear{Filipovi{\'c} \& Tothill}{Filipovi{\'c} \&
  Tothill}{2021a}]{book2}
Filipovi{\'c} M.~D.,  Tothill N. F.~H.,  eds, 2021a, Multimessenger Astronomy
  in Practice.
2514-3433, IOP Publishing, \mn@doi{10.1088/2514-3433/ac2256}, \url
  {https://dx.doi.org/10.1088/2514-3433/ac2256}

\bibitem[\protect\citeauthoryear{Filipovi{\'c} \& Tothill}{Filipovi{\'c} \&
  Tothill}{2021b}]{book1}
Filipovi{\'c} M.~D.,  Tothill N. F.~H.,  2021b, Principles of Multimessenger
  Astronomy.
2514-3433, IOP Publishing, \mn@doi{10.1088/2514-3433/ac087e}, \url
  {https://dx.doi.org/10.1088/2514-3433/ac087e}

\bibitem[\protect\citeauthoryear{{Filipovic}, {White}, {Jones}, {Haynes},
  {Pietsch}, {Wielebinski}  \& {Klein}}{{Filipovic}
  et~al.}{1996}]{1996ASPC..112...91F}
{Filipovic} M.~D.,  {White} G.~L.,  {Jones} P.~A.,  {Haynes} R.~F.,  {Pietsch}
  W.~N.,  {Wielebinski} R.,   {Klein} U.,  1996, in {Burkert} A.,  {Hartmann}
  D.~H.,   {Majewski} S.~A.,  eds,  Astronomical Society of the Pacific
  Conference Series Vol. 112, The History of the Milky Way and Its Satellite
  System. p.~91

\bibitem[\protect\citeauthoryear{{Filipovi{\'c}}, {Jones}, {White}, {Haynes},
  {Klein}  \& {Wielebinski}}{{Filipovi{\'c}}
  et~al.}{1997}]{1997A&AS..121..321F}
{Filipovi{\'c}} M.~D.,  {Jones} P.~A.,  {White} G.~L.,  {Haynes} R.~F.,
  {Klein} U.,   {Wielebinski} R.,  1997, \mn@doi [\aaps] {10.1051/aas:1997317},
  \href {http://adsabs.harvard.edu/abs/1997A%26AS..121..321F} {121, 321}

\bibitem[\protect\citeauthoryear{Filipovi{\'c}, {Haynes}, {White}  \&
  {Jones}}{Filipovi{\'c} et~al.}{1998}]{1998A&AS..130..421F}
Filipovi{\'c} M.~D.,  {Haynes} R.~F.,  {White} G.~L.,   {Jones} P.~A.,  1998,
  \mn@doi [\aaps] {10.1051/aas:1998417}, \href
  {http://adsabs.harvard.edu/abs/1998A%26AS..130..421F} {130, 421}

\bibitem[\protect\citeauthoryear{{Filipovi{\'c}}, {Haberl}, {Pietsch}  \&
  {Morgan}}{{Filipovi{\'c}} et~al.}{2000}]{2000A&A...353..129F}
{Filipovi{\'c}} M.~D.,  {Haberl} F.,  {Pietsch} W.,   {Morgan} D.~H.,  2000,
  \aap, \href {https://ui.adsabs.harvard.edu/abs/2000A&A...353..129F} {353,
  129}

\bibitem[\protect\citeauthoryear{{Filipovi{\'c}}, {Bohlsen}, {Reid},
  {Staveley-Smith}, {Jones}, {Nohejl}  \& {Goldstein}}{{Filipovi{\'c}}
  et~al.}{2002}]{2002MNRAS.335.1085F}
{Filipovi{\'c}} M.~D.,  {Bohlsen} T.,  {Reid} W.,  {Staveley-Smith} L.,
  {Jones} P.~A.,  {Nohejl} K.,   {Goldstein} G.,  2002, \mn@doi [\mnras]
  {10.1046/j.1365-8711.2002.05702.x}, \href
  {http://adsabs.harvard.edu/abs/2002MNRAS.335.1085F} {335, 1085}

\bibitem[\protect\citeauthoryear{{Filipovi{\'c}}, {Payne}, {Reid}, {Danforth},
  {Staveley-Smith}, {Jones}  \& {White}}{{Filipovi{\'c}}
  et~al.}{2005}]{2005MNRAS.364..217F}
{Filipovi{\'c}} M.~D.,  {Payne} J.~L.,  {Reid} W.,  {Danforth} C.~W.,
  {Staveley-Smith} L.,  {Jones} P.~A.,   {White} G.~L.,  2005, \mn@doi [\mnras]
  {10.1111/j.1365-2966.2005.09554.x}, \href
  {http://adsabs.harvard.edu/abs/2005MNRAS.364..217F} {364, 217}

\bibitem[\protect\citeauthoryear{{Filipovi{\'c}} et~al.,}{{Filipovi{\'c}}
  et~al.}{2008}]{2008A&A...485...63F}
{Filipovi{\'c}} M.~D.,  et~al., 2008, \mn@doi [\aap]
  {10.1051/0004-6361:200809642}, \href
  {https://ui.adsabs.harvard.edu/abs/2008A&A...485...63F} {485, 63}

\bibitem[\protect\citeauthoryear{{Filipovi{\'c}} et~al.,}{{Filipovi{\'c}}
  et~al.}{2009}]{2009MNRAS.399..769F}
{Filipovi{\'c}} M.~D.,  et~al., 2009, \mn@doi [\mnras]
  {10.1111/j.1365-2966.2009.15307.x}, \href
  {http://adsabs.harvard.edu/abs/2009MNRAS.399..769F} {399, 769}

\bibitem[\protect\citeauthoryear{{Filipovic}, {Crawford}, {Jones}  \&
  {White}}{{Filipovic} et~al.}{2010}]{2010SerAJ.181...31F}
{Filipovic} M.~D.,  {Crawford} E.~J.,  {Jones} P.~A.,   {White} G.~L.,  2010,
  \mn@doi [Serbian Astronomical Journal] {10.2298/SAJ1081031F}, \href
  {https://ui.adsabs.harvard.edu/abs/2010SerAJ.181...31F} {181, 31}

\bibitem[\protect\citeauthoryear{{Filipovi{\'c}} et~al.,}{{Filipovi{\'c}}
  et~al.}{2021}]{2021MNRAS.507.2885F}
{Filipovi{\'c}} M.~D.,  et~al., 2021, \mn@doi [MNRAS] {10.1093/mnras/stab2249},
  \href {https://ui.adsabs.harvard.edu/abs/2021MNRAS.507.2885F} {507, 2885}

\bibitem[\protect\citeauthoryear{{Filipovi{\'c}} et~al.,}{{Filipovi{\'c}}
  et~al.}{2022}]{2022MNRAS.512..265F}
{Filipovi{\'c}} M.~D.,  et~al., 2022, \mn@doi [\mnras] {10.1093/mnras/stac210},
  \href {https://ui.adsabs.harvard.edu/abs/2022MNRAS.512..265F} {512, 265}

\bibitem[\protect\citeauthoryear{{Filipovi{\'c}} et~al.,}{{Filipovi{\'c}}
  et~al.}{2023}]{2023AJ....166..149F}
{Filipovi{\'c}} M.~D.,  et~al., 2023, \mn@doi [\aj] {10.3847/1538-3881/acf19c},
  \href {https://ui.adsabs.harvard.edu/abs/2023AJ....166..149F} {166, 149}

\bibitem[\protect\citeauthoryear{{Findlay}}{{Findlay}}{1966}]{1966ARA&A...4...77F}
{Findlay} J.~W.,  1966, \mn@doi [\araa] {10.1146/annurev.aa.04.090166.000453},
  \href {https://ui.adsabs.harvard.edu/abs/1966ARA&A...4...77F} {4, 77}

\bibitem[\protect\citeauthoryear{{Flesch}}{{Flesch}}{2021}]{2021arXiv210512985F}
{Flesch} E.~W.,  2021, arXiv e-prints, \href
  {https://ui.adsabs.harvard.edu/abs/2021arXiv210512985F} {p. arXiv:2105.12985}

\bibitem[\protect\citeauthoryear{{For} et~al.,}{{For}
  et~al.}{2018}]{2018MNRAS.480.2743F}
{For} B.-Q.,  et~al., 2018, \mn@doi [\mnras] {10.1093/mnras/sty1960}, \href
  {http://adsabs.harvard.edu/abs/2018MNRAS.480.2743F} {480, 2743}

\bibitem[\protect\citeauthoryear{{Gaia Collaboration} et~al.,}{{Gaia
  Collaboration} et~al.}{2016}]{2016A&A...595A...2G}
{Gaia Collaboration} et~al., 2016, \mn@doi [\aap]
  {10.1051/0004-6361/201629512}, \href
  {https://ui.adsabs.harvard.edu/abs/2016A&A...595A...2G} {595, A2}

\bibitem[\protect\citeauthoryear{{Gaia Collaboration} et~al.,}{{Gaia
  Collaboration} et~al.}{2022}]{2022arXiv220800211G}
{Gaia Collaboration} et~al., 2022, \mn@doi [arXiv e-prints]
  {10.48550/arXiv.2208.00211}, \href
  {https://ui.adsabs.harvard.edu/abs/2022arXiv220800211G} {p. arXiv:2208.00211}

\bibitem[\protect\citeauthoryear{{Galvin} \& {Filipovic}}{{Galvin} \&
  {Filipovic}}{2014}]{2014SerAJ.189...15G}
{Galvin} T.~J.,  {Filipovic} M.~D.,  2014, \mn@doi [Serbian Astronomical
  Journal] {10.2298/SAJ140505002G}, \href
  {https://ui.adsabs.harvard.edu/abs/2014SerAJ.189...15G} {189, 15}

\bibitem[\protect\citeauthoryear{{Gordon} et~al.,}{{Gordon}
  et~al.}{2009}]{2009ApJ...690L..76G}
{Gordon} K.~D.,  et~al., 2009, \mn@doi [\apjl] {10.1088/0004-637X/690/1/L76},
  \href {https://ui.adsabs.harvard.edu/abs/2009ApJ...690L..76G} {690, L76}

\bibitem[\protect\citeauthoryear{{Graczyk} et~al.,}{{Graczyk}
  et~al.}{2020}]{2020ApJ...904...13G}
{Graczyk} D.,  et~al., 2020, \mn@doi [\apj] {10.3847/1538-4357/abbb2b}, \href
  {https://ui.adsabs.harvard.edu/abs/2020ApJ...904...13G} {904, 13}

\bibitem[\protect\citeauthoryear{{G{\"u}del}}{{G{\"u}del}}{2002}]{Gudel}
{G{\"u}del} M.,  2002, \mn@doi [\araa]
  {10.1146/annurev.astro.40.060401.093806}, \href
  {https://ui.adsabs.harvard.edu/abs/2002ARA&A..40..217G} {40, 217}

\bibitem[\protect\citeauthoryear{{Gvaramadze}, {Kniazev}  \&
  {Oskinova}}{{Gvaramadze} et~al.}{2019}]{2019MNRAS.485L...6G}
{Gvaramadze} V.~V.,  {Kniazev} A.~Y.,   {Oskinova} L.~M.,  2019, \mn@doi
  [\mnras] {10.1093/mnrasl/slz018}, \href
  {http://adsabs.harvard.edu/abs/2019MNRAS.485L...6G} {485, L6}

\bibitem[\protect\citeauthoryear{{Haberl}, {Filipovi{\'c}}, {Pietsch}  \&
  {Kahabka}}{{Haberl} et~al.}{2000}]{haberl2000}
{Haberl} F.,  {Filipovi{\'c}} M.~D.,  {Pietsch} W.,   {Kahabka} P.,  2000,
  \mn@doi [\aaps] {10.1051/aas:2000136}, \href
  {http://adsabs.harvard.edu/abs/2000A%26AS..142...41H} {142, 41}

\bibitem[\protect\citeauthoryear{{Haberl}, {Sturm}, {Filipovi{\'c}}, {Pietsch}
  \& {Crawford}}{{Haberl} et~al.}{2012a}]{2012A&A...537L...1H}
{Haberl} F.,  {Sturm} R.,  {Filipovi{\'c}} M.~D.,  {Pietsch} W.,   {Crawford}
  E.~J.,  2012a, \mn@doi [\aap] {10.1051/0004-6361/201118369}, \href
  {https://ui.adsabs.harvard.edu/abs/2012A&A...537L...1H} {537, L1}

\bibitem[\protect\citeauthoryear{{Haberl} et~al.,}{{Haberl}
  et~al.}{2012b}]{2012A&A...545A.128H}
{Haberl} F.,  et~al., 2012b, \mn@doi [\aap] {10.1051/0004-6361/201219758},
  \href {http://adsabs.harvard.edu/abs/2012A%26A...545A.128H} {545, A128}

\bibitem[\protect\citeauthoryear{{Hales}, {Murphy}, {Curran}, {Middelberg},
  {Gaensler}  \& {Norris}}{{Hales} et~al.}{2012}]{2012MNRAS.425..979H}
{Hales} C.~A.,  {Murphy} T.,  {Curran} J.~R.,  {Middelberg} E.,  {Gaensler}
  B.~M.,   {Norris} R.~P.,  2012, \mn@doi [\mnras]
  {10.1111/j.1365-2966.2012.21373.x}, \href
  {https://ui.adsabs.harvard.edu/abs/2012MNRAS.425..979H} {425, 979}

\bibitem[\protect\citeauthoryear{{Hancock}, {Murphy}, {Gaensler}, {Hopkins}  \&
  {Curran}}{{Hancock} et~al.}{2012}]{2012MNRAS.422.1812H}
{Hancock} P.~J.,  {Murphy} T.,  {Gaensler} B.~M.,  {Hopkins} A.,   {Curran}
  J.~R.,  2012, \mn@doi [\mnras] {10.1111/j.1365-2966.2012.20768.x}, \href
  {https://ui.adsabs.harvard.edu/abs/2012MNRAS.422.1812H} {422, 1812}

\bibitem[\protect\citeauthoryear{{Hancock}, {Trott}  \&
  {Hurley-Walker}}{{Hancock} et~al.}{2018}]{hancock2018}
{Hancock} P.~J.,  {Trott} C.~M.,   {Hurley-Walker} N.,  2018, \mn@doi [PASA]
  {10.1017/pasa.2018.3}, \href
  {https://ui.adsabs.harvard.edu/abs/2018PASA...35...11H} {35, e011}

\bibitem[\protect\citeauthoryear{{Haynes}, {Klein}, {Wielebinski}  \&
  {Murray}}{{Haynes} et~al.}{1986}]{1986A&A...159...22H}
{Haynes} R.~F.,  {Klein} U.,  {Wielebinski} R.,   {Murray} J.~D.,  1986, \aap,
  \href {http://adsabs.harvard.edu/abs/1986A%26A...159...22H} {159, 22}

\bibitem[\protect\citeauthoryear{{Henize}}{{Henize}}{1956}]{1956ApJS....2..315H}
{Henize} K.~G.,  1956, \mn@doi [\apjs] {10.1086/190025}, \href
  {http://adsabs.harvard.edu/abs/1956ApJS....2..315H} {2, 315}

\bibitem[\protect\citeauthoryear{{Heywood} et~al.,}{{Heywood}
  et~al.}{2022}]{2022ApJ...925..165H}
{Heywood} I.,  et~al., 2022, \mn@doi [\apj] {10.3847/1538-4357/ac449a}, \href
  {https://ui.adsabs.harvard.edu/abs/2022ApJ...925..165H} {925, 165}

\bibitem[\protect\citeauthoryear{{Hughes}, {Staveley-Smith}, {Kim}, {Wolleben}
  \& {Filipovi{\'c}}}{{Hughes} et~al.}{2007}]{2007MNRAS.382..543H}
{Hughes} A.,  {Staveley-Smith} L.,  {Kim} S.,  {Wolleben} M.,   {Filipovi{\'c}}
  M.,  2007, \mn@doi [\mnras] {10.1111/j.1365-2966.2007.12466.x}, \href
  {https://ui.adsabs.harvard.edu/abs/2007MNRAS.382..543H} {382, 543}

\bibitem[\protect\citeauthoryear{{Hurley-Walker} et~al.,}{{Hurley-Walker}
  et~al.}{2019a}]{2019PASA...36...45H}
{Hurley-Walker} N.,  et~al., 2019a, \mn@doi [\pasa] {10.1017/pasa.2019.34},
  \href {https://ui.adsabs.harvard.edu/abs/2019PASA...36...45H} {36, e045}

\bibitem[\protect\citeauthoryear{{Hurley-Walker} et~al.,}{{Hurley-Walker}
  et~al.}{2019b}]{2019PASA...36...48H}
{Hurley-Walker} N.,  et~al., 2019b, \mn@doi [\pasa] {10.1017/pasa.2019.33},
  \href {https://ui.adsabs.harvard.edu/abs/2019PASA...36...48H} {36, e048}

\bibitem[\protect\citeauthoryear{{Inoue}, {Koyama}  \& {Tanaka}}{{Inoue}
  et~al.}{1983}]{1983IAUS..101..535I}
{Inoue} H.,  {Koyama} K.,   {Tanaka} Y.,  1983, in {Danziger} J.,  {Gorenstein}
  P.,  eds,  IAU Symposium Vol. 101, Supernova Remnants and their X-ray
  Emission. pp 535--540

\bibitem[\protect\citeauthoryear{{Jacoby}}{{Jacoby}}{1980}]{1980ApJS...42....1J}
{Jacoby} G.~H.,  1980, \mn@doi [\apjs] {10.1086/190642}, \href
  {https://ui.adsabs.harvard.edu/abs/1980ApJS...42....1J} {42, 1}

\bibitem[\protect\citeauthoryear{{Jankowski}, {van Straten}, {Keane}, {Bailes},
  {Barr}, {Johnston}  \& {Kerr}}{{Jankowski}
  et~al.}{2018}]{2018MNRAS.473.4436J}
{Jankowski} F.,  {van Straten} W.,  {Keane} E.~F.,  {Bailes} M.,  {Barr} E.~D.,
   {Johnston} S.,   {Kerr} M.,  2018, \mn@doi [\mnras] {10.1093/mnras/stx2476},
  \href {https://ui.adsabs.harvard.edu/abs/2018MNRAS.473.4436J} {473, 4436}

\bibitem[\protect\citeauthoryear{{Johnston} et~al.,}{{Johnston}
  et~al.}{2020}]{2020MNRAS.493.3608J}
{Johnston} S.,  et~al., 2020, \mn@doi [\mnras] {10.1093/mnras/staa516}, \href
  {https://ui.adsabs.harvard.edu/abs/2020MNRAS.493.3608J} {493, 3608}

\bibitem[\protect\citeauthoryear{Jonas}{Jonas}{2016}]{Jonas2016}
Jonas J.,  2016, ``''.
SARAO, \mn@doi{10.22323/1.277.0001}

\bibitem[\protect\citeauthoryear{{Joseph} et~al.,}{{Joseph}
  et~al.}{2019}]{2019MNRAS.490.1202J}
{Joseph} T.~D.,  et~al., 2019, \mn@doi [MNRAS] {10.1093/mnras/stz2650}, \href
  {https://ui.adsabs.harvard.edu/abs/2019MNRAS.490.1202J} {490, 1202}

\bibitem[\protect\citeauthoryear{{Kaspi}, {Johnston}, {Bell}, {Manchester},
  {Bailes}, {Bessell}, {Lyne}  \& {D'Amico}}{{Kaspi}
  et~al.}{1994}]{1994ApJ...423L..43K}
{Kaspi} V.~M.,  {Johnston} S.,  {Bell} J.~F.,  {Manchester} R.~N.,  {Bailes}
  M.,  {Bessell} M.,  {Lyne} A.~G.,   {D'Amico} N.,  1994, \mn@doi [\apjl]
  {10.1086/187231}, \href
  {https://ui.adsabs.harvard.edu/abs/1994ApJ...423L..43K} {423, L43}

\bibitem[\protect\citeauthoryear{{Kavanagh} et~al.,}{{Kavanagh}
  et~al.}{2019}]{2019A&A...621A.138K}
{Kavanagh} P.~J.,  et~al., 2019, \mn@doi [\aap] {10.1051/0004-6361/201833659},
  \href {https://ui.adsabs.harvard.edu/abs/2019A&A...621A.138K} {621, A138}

\bibitem[\protect\citeauthoryear{{Knowles} et~al.,}{{Knowles}
  et~al.}{2022}]{2022A&A...657A..56K}
{Knowles} K.,  et~al., 2022, \mn@doi [\aap] {10.1051/0004-6361/202141488},
  \href {https://ui.adsabs.harvard.edu/abs/2022A&A...657A..56K} {657, A56}

\bibitem[\protect\citeauthoryear{{Kwok}}{{Kwok}}{2005}]{2005JKAS...38..271K}
{Kwok} S.,  2005, \mn@doi [Journal of Korean Astronomical Society]
  {10.5303/JKAS.2005.38.2.271}, \href
  {http://adsabs.harvard.edu/abs/2005JKAS...38..271K} {38, 271}

\bibitem[\protect\citeauthoryear{{Kwok}}{{Kwok}}{2015}]{2015HiA....16..623K}
{Kwok} S.,  2015, \mn@doi [Highlights of Astronomy]
  {10.1017/S1743921314012526}, \href
  {http://adsabs.harvard.edu/abs/2015HiA....16..623K} {16, 623}

\bibitem[\protect\citeauthoryear{{Lamb}, {Fox}, {Macomb}  \& {Prince}}{{Lamb}
  et~al.}{2002}]{2002ApJ...574L..29L}
{Lamb} R.~C.,  {Fox} D.~W.,  {Macomb} D.~J.,   {Prince} T.~A.,  2002, \mn@doi
  [\apjl] {10.1086/342352}, \href
  {https://ui.adsabs.harvard.edu/abs/2002ApJ...574L..29L} {574, L29}

\bibitem[\protect\citeauthoryear{{Leverenz}, {Filipovi{\'c}}, {Boji{\v
  c}i{\'c}}, {Crawford}, {Collier}, {Grieve}, {Dra{\v s}kovi{\'c}}  \&
  {Reid}}{{Leverenz} et~al.}{2016}]{2016Ap&SS.361..108L}
{Leverenz} H.,  {Filipovi{\'c}} M.~D.,  {Boji{\v c}i{\'c}} I.~S.,  {Crawford}
  E.~J.,  {Collier} J.~D.,  {Grieve} K.,  {Dra{\v s}kovi{\'c}} D.,   {Reid}
  W.~A.,  2016, \mn@doi [\apss] {10.1007/s10509-016-2686-3}, \href
  {http://adsabs.harvard.edu/abs/2016Ap%26SS.361..108L} {361, 108}

\bibitem[\protect\citeauthoryear{{Lindsay}}{{Lindsay}}{1961}]{1961AJ.....66..169L}
{Lindsay} E.~M.,  1961, \mn@doi [\aj] {10.1086/108396}, \href
  {https://ui.adsabs.harvard.edu/abs/1961AJ.....66..169L} {66, 169}

\bibitem[\protect\citeauthoryear{{Luken} et~al.,}{{Luken}
  et~al.}{2020}]{2020MNRAS.492.2606L}
{Luken} K.~J.,  et~al., 2020, \mn@doi [\mnras] {10.1093/mnras/stz3439}, \href
  {https://ui.adsabs.harvard.edu/abs/2020MNRAS.492.2606L} {492, 2606}

\bibitem[\protect\citeauthoryear{{Lyne} et~al.,}{{Lyne}
  et~al.}{1998}]{1998MNRAS.295..743L}
{Lyne} A.~G.,  et~al., 1998, \mn@doi [\mnras]
  {10.1046/j.1365-8711.1998.01144.x}, \href
  {https://ui.adsabs.harvard.edu/abs/1998MNRAS.295..743L} {295, 743}

\bibitem[\protect\citeauthoryear{{Maggi} et~al.,}{{Maggi}
  et~al.}{2016}]{2016A&A...585A.162M}
{Maggi} P.,  et~al., 2016, \mn@doi [\aap] {10.1051/0004-6361/201526932}, \href
  {http://adsabs.harvard.edu/abs/2016A%26A...585A.162M} {585, A162}

\bibitem[\protect\citeauthoryear{{Maggi} et~al.,}{{Maggi}
  et~al.}{2019}]{2019A&A...631A.127M}
{Maggi} P.,  et~al., 2019, \mn@doi [\aap] {10.1051/0004-6361/201936583}, \href
  {https://ui.adsabs.harvard.edu/abs/2019A&A...631A.127M} {631, A127}

\bibitem[\protect\citeauthoryear{{Maitra}, {Ballet}, {Filipovi{\'c}}, {Haberl},
  {Tiengo}, {Grieve}  \& {Roper}}{{Maitra} et~al.}{2015}]{2015A&A...584A..41M}
{Maitra} C.,  {Ballet} J.,  {Filipovi{\'c}} M.~D.,  {Haberl} F.,  {Tiengo} A.,
  {Grieve} K.,   {Roper} Q.,  2015, \mn@doi [\aap]
  {10.1051/0004-6361/201526458}, \href
  {https://ui.adsabs.harvard.edu/abs/2015A&A...584A..41M} {584, A41}

\bibitem[\protect\citeauthoryear{{Maitra}, {Esposito}, {Tiengo}, {Ballet},
  {Haberl}, {Dai}, {Filipovi{\'c}}  \& {Pilia}}{{Maitra}
  et~al.}{2021}]{2021MNRAS.507L...1M}
{Maitra} C.,  {Esposito} P.,  {Tiengo} A.,  {Ballet} J.,  {Haberl} F.,  {Dai}
  S.,  {Filipovi{\'c}} M.~D.,   {Pilia} M.,  2021, \mn@doi [\mnras]
  {10.1093/mnrasl/slab050}, \href
  {https://ui.adsabs.harvard.edu/abs/2021MNRAS.507L...1M} {507, L1}

\bibitem[\protect\citeauthoryear{{Manchester}, {Hobbs}, {Teoh}  \&
  {Hobbs}}{{Manchester} et~al.}{2005}]{2005AJ....129.1993M}
{Manchester} R.~N.,  {Hobbs} G.~B.,  {Teoh} A.,   {Hobbs} M.,  2005, \mn@doi
  [\aj] {10.1086/428488}, \href
  {https://ui.adsabs.harvard.edu/abs/2005AJ....129.1993M} {129, 1993}

\bibitem[\protect\citeauthoryear{{Manchester}, {Fan}, {Lyne}, {Kaspi}  \&
  {Crawford}}{{Manchester} et~al.}{2006}]{2006ApJ...649..235M}
{Manchester} R.~N.,  {Fan} G.,  {Lyne} A.~G.,  {Kaspi} V.~M.,   {Crawford} F.,
  2006, \mn@doi [\apj] {10.1086/505461}, \href
  {https://ui.adsabs.harvard.edu/abs/2006ApJ...649..235M} {649, 235}

\bibitem[\protect\citeauthoryear{{Marocco} et~al.,}{{Marocco}
  et~al.}{2021}]{2021ApJS..253....8M}
{Marocco} F.,  et~al., 2021, \mn@doi [\apjs] {10.3847/1538-4365/abd805}, \href
  {https://ui.adsabs.harvard.edu/abs/2021ApJS..253....8M} {253, 8}

\bibitem[\protect\citeauthoryear{{Mauch}, {Murphy}, {Buttery}, {Curran},
  {Hunstead}, {Piestrzynski}, {Robertson}  \& {Sadler}}{{Mauch}
  et~al.}{2003}]{mau03}
{Mauch} T.,  {Murphy} T.,  {Buttery} H.~J.,  {Curran} J.,  {Hunstead} R.~W.,
  {Piestrzynski} B.,  {Robertson} J.~G.,   {Sadler} E.~M.,  2003, \mn@doi
  [MNRAS] {10.1046/j.1365-8711.2003.06605.x}, \href
  {https://ui.adsabs.harvard.edu/abs/2003MNRAS.342.1117M} {342, 1117}

\bibitem[\protect\citeauthoryear{{Mauch} et~al.,}{{Mauch} et~al.}{2020}]{DEEP2}
{Mauch} T.,  et~al., 2020, \mn@doi [ApJ] {10.3847/1538-4357/ab5d2d}, 888, 61

\bibitem[\protect\citeauthoryear{{McConnell}, {McCulloch}, {Hamilton}, {Ables},
  {Hall}, {Jacka}  \& {Hunt}}{{McConnell} et~al.}{1991}]{1991MNRAS.249..654M}
{McConnell} D.,  {McCulloch} P.~M.,  {Hamilton} P.~A.,  {Ables} J.~G.,  {Hall}
  P.~J.,  {Jacka} C.~E.,   {Hunt} A.~J.,  1991, \mn@doi [\mnras]
  {10.1093/mnras/249.4.654}, \href
  {https://ui.adsabs.harvard.edu/abs/1991MNRAS.249..654M} {249, 654}

\bibitem[\protect\citeauthoryear{{McGee}, {Newton}  \& {Butler}}{{McGee}
  et~al.}{1976}]{1976AuJPh..29..329M}
{McGee} R.~X.,  {Newton} L.~M.,   {Butler} P.~W.,  1976, \mn@doi [Australian
  Journal of Physics] {10.1071/PH760329}, \href
  {http://adsabs.harvard.edu/abs/1976AuJPh..29..329M} {29, 329}

\bibitem[\protect\citeauthoryear{{Meixner} et~al.,}{{Meixner}
  et~al.}{2006}]{meixner_sage}
{Meixner} M.,  et~al., 2006, \mn@doi [\aj] {10.1086/508185}, \href
  {https://ui.adsabs.harvard.edu/abs/2006AJ....132.2268M} {132, 2268}

\bibitem[\protect\citeauthoryear{{Meixner} et~al.,}{{Meixner}
  et~al.}{2013}]{meixner_heritage}
{Meixner} M.,  et~al., 2013, \mn@doi [\aj] {10.1088/0004-6256/146/3/62}, \href
  {https://ui.adsabs.harvard.edu/abs/2013AJ....146...62M} {146, 62}

\bibitem[\protect\citeauthoryear{{Meyssonnier} \& {Azzopardi}}{{Meyssonnier} \&
  {Azzopardi}}{1993}]{1993A&AS..102..451M}
{Meyssonnier} N.,  {Azzopardi} M.,  1993, \aaps, \href
  {https://ui.adsabs.harvard.edu/abs/1993A&AS..102..451M} {102, 451}

\bibitem[\protect\citeauthoryear{{Morgan}}{{Morgan}}{1995}]{1995A&AS..112..445M}
{Morgan} D.~H.,  1995, \aaps, \href
  {https://ui.adsabs.harvard.edu/abs/1995A&AS..112..445M} {112, 445}

\bibitem[\protect\citeauthoryear{{Morgan} \& {Good}}{{Morgan} \&
  {Good}}{1985}]{1985MNRAS.213..491M}
{Morgan} D.~H.,  {Good} A.~R.,  1985, \mn@doi [\mnras]
  {10.1093/mnras/213.3.491}, \href
  {https://ui.adsabs.harvard.edu/abs/1985MNRAS.213..491M} {213, 491}

\bibitem[\protect\citeauthoryear{{Murphy} et~al.,}{{Murphy}
  et~al.}{2017}]{2017PASA...34...20M}
{Murphy} T.,  et~al., 2017, \mn@doi [\pasa] {10.1017/pasa.2017.13}, \href
  {https://ui.adsabs.harvard.edu/abs/2017PASA...34...20M} {34, e020}

\bibitem[\protect\citeauthoryear{{Noordam}}{{Noordam}}{2004}]{Noordam2004}
{Noordam} J.~E.,  2004, in {Oschmann} Jacobus~M. J.,  ed.,  Society of
  Photo-Optical Instrumentation Engineers (SPIE) Conference Series Vol. 5489,
  Ground-based Telescopes. pp 817--825, \mn@doi{10.1117/12.544262}

\bibitem[\protect\citeauthoryear{{Norris} et~al.,}{{Norris}
  et~al.}{2021}]{2021PASA...38...46N}
{Norris} R.~P.,  et~al., 2021, \mn@doi [PASA] {10.1017/pasa.2021.42}, \href
  {https://ui.adsabs.harvard.edu/abs/2021PASA...38...46N} {38, e046}

\bibitem[\protect\citeauthoryear{{Oliveira} et~al.,}{{Oliveira}
  et~al.}{2013}]{2013MNRAS.428.3001O}
{Oliveira} J.~M.,  et~al., 2013, \mn@doi [\mnras] {10.1093/mnras/sts250}, \href
  {http://adsabs.harvard.edu/abs/2013MNRAS.428.3001O} {428, 3001}

\bibitem[\protect\citeauthoryear{{Owen} et~al.,}{{Owen}
  et~al.}{2011}]{2011A&A...530A.132O}
{Owen} R.~A.,  et~al., 2011, \mn@doi [\aap] {10.1051/0004-6361/201116586},
  \href {https://ui.adsabs.harvard.edu/abs/2011A&A...530A.132O} {530, A132}

\bibitem[\protect\citeauthoryear{{Pavlovi{\'c}}, {Uro{\v s}evi{\'c}},
  {Arbutina}, {Orlando}, {Maxted}  \& {Filipovi{\'c}}}{{Pavlovi{\'c}}
  et~al.}{2018}]{2018ApJ...852...84P}
{Pavlovi{\'c}} M.~Z.,  {Uro{\v s}evi{\'c}} D.,  {Arbutina} B.,  {Orlando} S.,
  {Maxted} N.,   {Filipovi{\'c}} M.~D.,  2018, \mn@doi [\apj]
  {10.3847/1538-4357/aaa1e6}, \href
  {http://adsabs.harvard.edu/abs/2018ApJ...852...84P} {852, 84}

\bibitem[\protect\citeauthoryear{{Payne}, {Filipovi{\'c}}, {Reid}, {Jones},
  {Staveley-Smith}  \& {White}}{{Payne} et~al.}{2004}]{2004MNRAS.355...44P}
{Payne} J.~L.,  {Filipovi{\'c}} M.~D.,  {Reid} W.,  {Jones} P.~A.,
  {Staveley-Smith} L.,   {White} G.~L.,  2004, \mn@doi [\mnras]
  {10.1111/j.1365-2966.2004.08287.x}, \href
  {http://adsabs.harvard.edu/abs/2004MNRAS.355...44P} {355, 44}

\bibitem[\protect\citeauthoryear{{Payne}, {White}, {Filipovi{\'c}}  \&
  {Pannuti}}{{Payne} et~al.}{2007}]{2007MNRAS.376.1793P}
{Payne} J.~L.,  {White} G.~L.,  {Filipovi{\'c}} M.~D.,   {Pannuti} T.~G.,
  2007, \mn@doi [\mnras] {10.1111/j.1365-2966.2007.11561.x}, \href
  {http://adsabs.harvard.edu/abs/2007MNRAS.376.1793P} {376, 1793}

\bibitem[\protect\citeauthoryear{{Payne}, {Filipovi{\'c}}, {Crawford}, {de
  Horta}, {White}  \& {Stootman}}{{Payne} et~al.}{2008}]{2008SerAJ.176...65P}
{Payne} J.~L.,  {Filipovi{\'c}} M.~D.,  {Crawford} E.~J.,  {de Horta} A.~Y.,
  {White} G.~L.,   {Stootman} F.~H.,  2008, \mn@doi [Serbian Astronomical
  Journal] {10.2298/SAJ0876065P}, \href
  {http://adsabs.harvard.edu/abs/2008SerAJ.176...65P} {176, 65}

\bibitem[\protect\citeauthoryear{Pedregosa et~al.,}{Pedregosa
  et~al.}{2011}]{scikit-learn}
Pedregosa F.,  et~al., 2011, Journal of Machine Learning Research, 12, 2825

\bibitem[\protect\citeauthoryear{{Pellegrini}, {Oey}, {Winkler}, {Points},
  {Smith}, {Jaskot}  \& {Zastrow}}{{Pellegrini}
  et~al.}{2012}]{2012ApJ...755...40P}
{Pellegrini} E.~W.,  {Oey} M.~S.,  {Winkler} P.~F.,  {Points} S.~D.,  {Smith}
  R.~C.,  {Jaskot} A.~E.,   {Zastrow} J.,  2012, \mn@doi [\apj]
  {10.1088/0004-637X/755/1/40}, \href
  {http://adsabs.harvard.edu/abs/2012ApJ...755...40P} {755, 40}

\bibitem[\protect\citeauthoryear{{Pennock} et~al.,}{{Pennock}
  et~al.}{2021}]{2021MNRAS.506.3540P}
{Pennock} C.~M.,  et~al., 2021, \mn@doi [MNRAS] {10.1093/mnras/stab1858}, \href
  {https://ui.adsabs.harvard.edu/abs/2021MNRAS.506.3540P} {506, 3540}

\bibitem[\protect\citeauthoryear{{Perley}}{{Perley}}{1999}]{Perley1999A}
{Perley} R.~A.,  1999, in {Taylor} G.~B.,  {Carilli} C.~L.,   {Perley} R.~A.,
  eds,  Astronomical Society of the Pacific Conference Series Vol. 180,
  Synthesis Imaging in Radio Astronomy II. pp 383--+

\bibitem[\protect\citeauthoryear{{Pietrzy{\'n}ski} et~al.,}{{Pietrzy{\'n}ski}
  et~al.}{2019}]{2019Natur.567..200P}
{Pietrzy{\'n}ski} G.,  et~al., 2019, \mn@doi [\nat]
  {10.1038/s41586-019-0999-4}, \href
  {http://adsabs.harvard.edu/abs/2019Natur.567..200P} {567, 200}

\bibitem[\protect\citeauthoryear{{Read}, {Filipovi{\'c}}, {Pietsch}  \&
  {Jones}}{{Read} et~al.}{2001}]{2001A&A...369..467R}
{Read} A.~M.,  {Filipovi{\'c}} M.~D.,  {Pietsch} W.,   {Jones} P.~A.,  2001,
  \mn@doi [\aap] {10.1051/0004-6361:20010128}, \href
  {https://ui.adsabs.harvard.edu/abs/2001A&A...369..467R} {369, 467}

\bibitem[\protect\citeauthoryear{{Reid}, {Payne}, {Filipovi{\'c}}, {Danforth},
  {Jones}, {White}  \& {Staveley-Smith}}{{Reid}
  et~al.}{2006}]{2006MNRAS.367.1379R}
{Reid} W.~A.,  {Payne} J.~L.,  {Filipovi{\'c}} M.~D.,  {Danforth} C.~W.,
  {Jones} P.~A.,  {White} G.~L.,   {Staveley-Smith} L.,  2006, \mn@doi [\mnras]
  {10.1111/j.1365-2966.2006.10017.x}, \href
  {http://adsabs.harvard.edu/abs/2006MNRAS.367.1379R} {367, 1379}

\bibitem[\protect\citeauthoryear{{Reynolds}}{{Reynolds}}{1994}]{Reynolds94}
{Reynolds} J.~E.,  1994, {ATNF Memo}, {AT/39.3/040}

\bibitem[\protect\citeauthoryear{{Reynolds}, {Gaensler}  \&
  {Bocchino}}{{Reynolds} et~al.}{2012}]{2012SSRv..166..231R}
{Reynolds} S.~P.,  {Gaensler} B.~M.,   {Bocchino} F.,  2012, \mn@doi [Space
  Science Reviews] {10.1007/s11214-011-9775-y}, \href
  {https://ui.adsabs.harvard.edu/abs/2012SSRv..166..231R} {166, 231}

\bibitem[\protect\citeauthoryear{{Roper}, {McEntaffer}, {DeRoo},
  {Filipovi{\'c}}, {Wong}  \& {Crawford}}{{Roper}
  et~al.}{2015}]{2015ApJ...803..106R}
{Roper} Q.,  {McEntaffer} R.~L.,  {DeRoo} C.,  {Filipovi{\'c}} M.,  {Wong}
  G.~F.,   {Crawford} E.~J.,  2015, \mn@doi [\apj]
  {10.1088/0004-637X/803/2/106}, \href
  {http://adsabs.harvard.edu/abs/2015ApJ...803..106R} {803, 106}

\bibitem[\protect\citeauthoryear{{Sanduleak}, {MacConnell}  \&
  {Philip}}{{Sanduleak} et~al.}{1978}]{1978PASP...90..621S}
{Sanduleak} N.,  {MacConnell} D.~J.,   {Philip} A.~G.~D.,  1978, \mn@doi
  [\pasp] {10.1086/130397}, \href
  {https://ui.adsabs.harvard.edu/abs/1978PASP...90..621S} {90, 621}

\bibitem[\protect\citeauthoryear{{Sano} et~al.,}{{Sano}
  et~al.}{2017}]{2017ApJ...843...61S}
{Sano} H.,  et~al., 2017, \mn@doi [\apj] {10.3847/1538-4357/aa73e0}, \href
  {https://ui.adsabs.harvard.edu/abs/2017ApJ...843...61S} {843, 61}

\bibitem[\protect\citeauthoryear{{Sault}, {Teuben}  \& {Wright}}{{Sault}
  et~al.}{1995}]{1995ASPC...77..433S}
{Sault} R.~J.,  {Teuben} P.~J.,   {Wright} M.~C.~H.,  1995, in {Shaw} R.~A.,
  {Payne} H.~E.,   {Hayes} J.~J.~E.,  eds,  Astronomical Society of the Pacific
  Conference Series Vol. 77, Astronomical Data Analysis Software and Systems
  IV. p.~433 (\mn@eprint {} {astro-ph/0612759})

\bibitem[\protect\citeauthoryear{{Smirnov}}{{Smirnov}}{2011}]{smirnov11}
{Smirnov} O.~M.,  2011, \mn@doi [A\&A] {10.1051/0004-6361/201016082}, \href
  {https://ui.adsabs.harvard.edu/abs/2011A&A...527A.106S} {527, A106}

\bibitem[\protect\citeauthoryear{{Smirnov} \& {Tasse}}{{Smirnov} \&
  {Tasse}}{2015}]{Smirnov2015}
{Smirnov} O.~M.,  {Tasse} C.,  2015, \mn@doi [MNRAS] {10.1093/mnras/stv418},
  \href {https://ui.adsabs.harvard.edu/abs/2015MNRAS.449.2668S} {449, 2668}

\bibitem[\protect\citeauthoryear{{Stanimirovic}, {Staveley-Smith}, {Dickey},
  {Sault}  \& {Snowden}}{{Stanimirovic} et~al.}{1999}]{1999MNRAS.302..417S}
{Stanimirovic} S.,  {Staveley-Smith} L.,  {Dickey} J.~M.,  {Sault} R.~J.,
  {Snowden} S.~L.,  1999, \mn@doi [\mnras] {10.1046/j.1365-8711.1999.02013.x},
  \href {https://ui.adsabs.harvard.edu/abs/1999MNRAS.302..417S} {302, 417}

\bibitem[\protect\citeauthoryear{{Sturm} et~al.,}{{Sturm}
  et~al.}{2013a}]{sturm2013}
{Sturm} R.,  et~al., 2013a, \mn@doi [\aap] {10.1051/0004-6361/201219935}, \href
  {https://ui.adsabs.harvard.edu/abs/2013A&A...558A...3S} {558, A3}

\bibitem[\protect\citeauthoryear{{Sturm} et~al.,}{{Sturm}
  et~al.}{2013b}]{2013A&A...558A.101S}
{Sturm} R.,  et~al., 2013b, \mn@doi [\aap] {10.1051/0004-6361/201220564}, \href
  {https://ui.adsabs.harvard.edu/abs/2013A&A...558A.101S} {558, A101}

\bibitem[\protect\citeauthoryear{{Taylor}}{{Taylor}}{2005}]{2005ASPC..347...29T}
{Taylor} M.~B.,  2005, in {Shopbell} P.,  {Britton} M.,   {Ebert} R.,  eds,
  Astronomical Society of the Pacific Conference Series Vol. 347, Astronomical
  Data Analysis Software and Systems XIV. p.~29

\bibitem[\protect\citeauthoryear{{Titus} et~al.,}{{Titus}
  et~al.}{2019}]{2019MNRAS.487.4332T}
{Titus} N.,  et~al., 2019, \mn@doi [\mnras] {10.1093/mnras/stz1578}, \href
  {https://ui.adsabs.harvard.edu/abs/2019MNRAS.487.4332T} {487, 4332}

\bibitem[\protect\citeauthoryear{{Treumann}}{{Treumann}}{2006}]{2006A&ARv..13..229T}
{Treumann} R.~A.,  2006, \mn@doi [\aapr] {10.1007/s00159-006-0001-y}, \href
  {https://ui.adsabs.harvard.edu/abs/2006A&ARv..13..229T} {13, 229}

\bibitem[\protect\citeauthoryear{{Turtle}, {Ye}, {Amy}  \& {Nicholls}}{{Turtle}
  et~al.}{1998}]{1998PASA...15..280T}
{Turtle} A.~J.,  {Ye} T.,  {Amy} S.~W.,   {Nicholls} J.,  1998, \mn@doi [\pasa]
  {10.1071/AS98280}, \href {http://adsabs.harvard.edu/abs/1998PASA...15..280T}
  {15, 280}

\bibitem[\protect\citeauthoryear{{Vardoulaki} et~al.,}{{Vardoulaki}
  et~al.}{2019}]{2019A&A...627A.142V}
{Vardoulaki} E.,  et~al., 2019, \mn@doi [\aap] {10.1051/0004-6361/201832982},
  \href {https://ui.adsabs.harvard.edu/abs/2019A&A...627A.142V} {627, A142}

\bibitem[\protect\citeauthoryear{Virtanen et~al.,}{Virtanen
  et~al.}{2020}]{2020SciPy-NMeth}
Virtanen P.,  et~al., 2020, \mn@doi [Nature Methods]
  {10.1038/s41592-019-0686-2}, \href {https://rdcu.be/b08Wh} {17, 261}

\bibitem[\protect\citeauthoryear{{Webb} et~al.,}{{Webb}
  et~al.}{2020}]{webb2020}
{Webb} N.~A.,  et~al., 2020, \mn@doi [\aap] {10.1051/0004-6361/201937353},
  \href {https://ui.adsabs.harvard.edu/abs/2020A&A...641A.136W} {641, A136}

\bibitem[\protect\citeauthoryear{{Wenger} et~al.,}{{Wenger}
  et~al.}{2000}]{2000A&AS..143....9W}
{Wenger} M.,  et~al., 2000, \mn@doi [\aaps] {10.1051/aas:2000332}, \href
  {https://ui.adsabs.harvard.edu/abs/2000A&AS..143....9W} {143, 9}

\bibitem[\protect\citeauthoryear{{White} \& {Franciosini}}{{White} \&
  {Franciosini}}{1995}]{1995ApJ...444..342W}
{White} S.~M.,  {Franciosini} E.,  1995, \mn@doi [\apj] {10.1086/175609}, \href
  {https://ui.adsabs.harvard.edu/abs/1995ApJ...444..342W} {444, 342}

\bibitem[\protect\citeauthoryear{{Whiting} \& {Humphreys}}{{Whiting} \&
  {Humphreys}}{2012}]{2012PASA...29..371W}
{Whiting} M.,  {Humphreys} B.,  2012, \mn@doi [\pasa] {10.1071/AS12028}, \href
  {https://ui.adsabs.harvard.edu/abs/2012PASA...29..371W} {29, 371}

\bibitem[\protect\citeauthoryear{{Winkler}, {Smith}, {Points}  \& {MCELS
  Team}}{{Winkler} et~al.}{2015}]{mcels}
{Winkler} P.~F.,  {Smith} R.~C.,  {Points} S.~D.,   {MCELS Team} 2015, in
  {Points} S.,  {Kunder} A.,  eds,  Astronomical Society of the Pacific
  Conference Series Vol. 491, Fifty Years of Wide Field Studies in the Southern
  Hemisphere: Resolved Stellar Populations of the Galactic Bulge and Magellanic
  Clouds. p.~343

\bibitem[\protect\citeauthoryear{{Wong}, {Filipovi{\'c}}, {Crawford}, {de
  Horta}, {Galvin}, {Draskovic}  \& {Payne}}{{Wong}
  et~al.}{2011a}]{2011SerAJ.182...43W}
{Wong} G.~F.,  {Filipovi{\'c}} M.~D.,  {Crawford} E.~J.,  {de Horta} A.~Y.,
  {Galvin} T.,  {Draskovic} D.,   {Payne} J.~L.,  2011a, \mn@doi [Serbian
  Astronomical Journal] {10.2298/SAJ1182043W}, \href
  {http://adsabs.harvard.edu/abs/2011SerAJ.182...43W} {182, 43}

\bibitem[\protect\citeauthoryear{{Wong} et~al.,}{{Wong}
  et~al.}{2011b}]{2011SerAJ.183..103W}
{Wong} G.~F.,  et~al., 2011b, \mn@doi [Serbian Astronomical Journal]
  {10.2298/SAJ1183103W}, \href
  {http://adsabs.harvard.edu/abs/2011SerAJ.183..103W} {183, 103}

\bibitem[\protect\citeauthoryear{{Wong} et~al.,}{{Wong}
  et~al.}{2012a}]{2012SerAJ.184...93W}
{Wong} G.~F.,  et~al., 2012a, \mn@doi [Serbian Astronomical Journal]
  {10.2298/SAJ1284093W}, \href
  {http://adsabs.harvard.edu/abs/2012SerAJ.184...93W} {184, 93}

\bibitem[\protect\citeauthoryear{{Wong}, {Filipovi{\'c}}, {Crawford},
  {Tothill}, {De Horta}  \& {Galvin}}{{Wong}
  et~al.}{2012b}]{2012SerAJ.185...53W}
{Wong} G.~F.,  {Filipovi{\'c}} M.~D.,  {Crawford} E.~J.,  {Tothill} N.~F.~H.,
  {De Horta} A.~Y.,   {Galvin} T.~J.,  2012b, \mn@doi [Serbian Astronomical
  Journal] {10.2298/SAJ1285053W}, \href
  {http://adsabs.harvard.edu/abs/2012SerAJ.185...53W} {185, 53}

\bibitem[\protect\citeauthoryear{{Wright} \& {Otrupcek}}{{Wright} \&
  {Otrupcek}}{1990}]{1990PKS...C......0W}
{Wright} A.,  {Otrupcek} R.,  1990, in PKS Catalog (1990).

\bibitem[\protect\citeauthoryear{{Yamane} et~al.,}{{Yamane}
  et~al.}{2021}]{2021ApJ...918...36Y}
{Yamane} Y.,  et~al., 2021, \mn@doi [\apj] {10.3847/1538-4357/ac0adb}, \href
  {https://ui.adsabs.harvard.edu/abs/2021ApJ...918...36Y} {918, 36}

\bibitem[\protect\citeauthoryear{{Yew} et~al.,}{{Yew}
  et~al.}{2021}]{2021MNRAS.500.2336Y}
{Yew} M.,  et~al., 2021, \mn@doi [\mnras] {10.1093/mnras/staa3382}, \href
  {https://ui.adsabs.harvard.edu/abs/2021MNRAS.500.2336Y} {500, 2336}

\bibitem[\protect\citeauthoryear{{Yu}, {Zijlstra}  \& {Jiang}}{{Yu}
  et~al.}{2021}]{Yu2021}
{Yu} B.,  {Zijlstra} A.,   {Jiang} B.,  2021, \mn@doi [Universe]
  {10.3390/universe7050119}, \href
  {https://ui.adsabs.harvard.edu/abs/2021Univ....7..119Y} {7, 119}

\bibitem[\protect\citeauthoryear{{de Villiers} \& {Cotton}}{{de Villiers} \&
  {Cotton}}{2022}]{2022AJ....163..135D}
{de Villiers} M.~S.,  {Cotton} W.~D.,  2022, \mn@doi [\aj]
  {10.3847/1538-3881/ac460a}, \href
  {https://ui.adsabs.harvard.edu/abs/2022AJ....163..135D} {163, 135}

\makeatother
\end{thebibliography}



\section*{SUPPORTING INFORMATION}

Supplementary data are available at {\it MNRAS} online.

{\bf Table~\ref{tab:main}}. Example of the point source catalogue of \TOTALNUMBERCHZEROPEAKFLUX\ objects in the directions of the \ac{SMC} with its positions, integrated flux densities with associated uncertainty and spectral index.

{\bf Table~\ref{tab:bcecat}}. Example for the compact extended source list (total of \TOTALBCE\ sources) at 1283.8~MHz derived from MeerKAT survey.

Please note: Oxford University Press is not responsible for the content or functionality of any supporting materials supplied by the authors. Any queries (other than missing material) should be directed to the corresponding author for the article.


\bsp	
\label{lastpage}
\end{document}